\documentclass[
reprint,
nofootinbib,
amsmath,amssymb,
aps
]{revtex4-2}

\usepackage{graphicx}
\usepackage{dcolumn}
\usepackage{bm}
\usepackage{hyperref}
\usepackage[version=4]{mhchem}    
\usepackage{csquotes}
\hypersetup{colorlinks=true,linkcolor=red,citecolor=blue}
\usepackage{float}
\usepackage{physics}
\usepackage{xspace}
\usepackage{booktabs} 

\usepackage{lipsum}

\usepackage{prettyref}
\newcommand{\pref}[1]{\prettyref{#1}}
\newrefformat{fig}{Fig.~\ref{#1}}
\newrefformat{tab}{Table~\ref{#1}}
\newrefformat{sec}{Sec.~\ref{#1}}
\newrefformat{ssec}{Sec.~\ref{#1}}
\newrefformat{app}{Appendix~\ref{#1}}
\newrefformat{eqn}{Eq.~(\ref{#1})}

\usepackage{bbding} 

\usepackage[dvipsnames,table]{xcolor}

\newcommand{\lunio}{LuNiO$_3$\xspace}
\newcommand{\lavo}{LaVO$_3$\xspace}
\newcommand{\latio}{LaTiO$_3$\xspace}
\newcommand{\nno}{NdNiO$_3$\xspace}

\begin{document}

\title{Importance of ligand on-site interactions for the description of Mott-insulators in DFT+DMFT}

\author{Alberto Carta}
\author{Anwesha Panda}
\author{Claude Ederer}
\affiliation{Materials Theory, ETH Z\"urich, Wolfgang-Pauli-Strasse 27, 8093 Z\"urich, Switzerland}

\date{\today}

\begin{abstract}
Calculations combining density functional theory (DFT) and dynamical mean-field theory (DMFT) for transition metal (TM) oxides and similar compounds usually focus on improving the description of the TM $d$ states. Here, we emphasize the importance of also accounting for corrections of the ligand $p$ states. 
We demonstrate that focusing exclusively on an improved description of the TM $d$ states results in difficulties to obtain the correct insulating behavior for a variety of materials, and requires to use values for the local interaction parameters that are inconsistent with values obtained using, e.g., the constrained random phase approximation (cRPA).
Importantly, these considerations not only apply to cases where the $p$ states are explicitly included in the DMFT low-energy subspace, but also to cases which only include so-called frontier bands with dominant TM $d$ character. 
We demonstrate that, to a large part, these inconsistencies arise from the use of local/semi-local DFT as starting point for computing interaction parameters within cRPA, and we show that applying a simple empirical correction to the O $p$ and other low energy states not included in the correlated subspace results in improved values for the interaction parameters that then allow to obtain the correct insulating behavior.
For the cases where the ligand states are included in the DMFT subspace, we show that even an approximate but realistic Hartree-Fock-like correction applied to the O $p$ states leads to a correct and, most importantly, also a quantitatively consistent DFT+DMFT description of typical Mott insulators such as \latio, \lavo, or the perovskite rare-earth nickelates, $R$NiO$_3$.
\end{abstract}

\maketitle

\section{\label{sec:Intro}Introduction}

In so-called strongly correlated materials, complex physical behavior arises from the electron-electron interaction~\cite{Fazekas:1999}. Examples are materials exhibiting a metal-insulator transition to a Mott-insulating state, where the strong Coulomb repulsion localizes the electrons~\cite{imada_metalinsulator_1998, roy_mott_2019}. Apart from their intriguing physical properties, such materials also show promise in various technological applications, e.g., as resistive switches in ``Mott'' field-effect transistors~\cite{Newns_et_al:1998, Inoue/Rozenberg:2008, Mannhart/Haensch:2012}, or as photoactive materials with potentially larger quantum efficiency compared to current semiconductor-based photovoltaic materials~\cite{manousakis_photovoltaic_2010a, assmann_oxide_2013, coulter_optoelectronic_2014, wang_device_2015, petocchi_hund_2019}. 

Over the past decades, the combination of density functional theory (DFT) and dynamical mean-field theory (DMFT) has emerged as a powerful tool to describe the behavior of correlated materials~\cite{kotliar_electronic_2006, held_electronic_2007, Kunes_et_al:2010, Paul/Birol:2019}.
DFT+DMFT combines the {\it ab initio} character of DFT with an explicit many-body treatment of the local interaction within a subset of the electronic states -- denoted as \textit{correlated subspace}, while the rest of the electronic degrees of freedom is treated on the level of a standard, typically local or semi-local, DFT functional.

Within the DFT+DMFT formalism, the correlated subspace for a specific material is chosen based on physical and chemical considerations. Typically, a basis of atom-centered orbitals corresponding to the localized $d$ states in a transition metal (TM) oxide, or the $f$ states in a lanthanide or actinide compound are used. However, in some cases it might also be appropriate to use a more unconventional basis, e.g., representing entire molecular orbitals~\cite{Kovacik_et_al:2012, Ferber_et_al:2014, Mlkvik_et_al:2024,  grytsiuk_nb3cl8_2024} or vacancy-centered states~\cite{SoutoCasares/Spaldin/Ederer:2019, souto-casares_oxygen_2021a}.
The localized orbitals can conveniently be defined in terms of Wannier functions~\cite{marzari_maximally_2012, anisimov_full_2005, lechermann_dynamical_2006, beck_charge_2022}, constructed from the Bloch states within a chosen energy window, but other definitions are also in use~\cite{pourovskii_selfconsistency_2007, haule_covalency_2014}. 
In general, the construction of these functions can involve two steps. First, a subset of all electronic bands can be identified, typically by defining a suitable energy window around the Fermi level containing the ``correlated'' degrees of freedom. Then, a set of localized basis orbitals is defined representing the correlated orbitals within these bands. 

\begin{figure}
   \centering
   \includegraphics[width=\columnwidth]{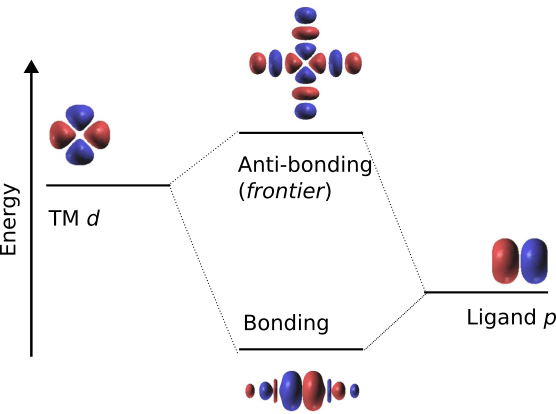}
   \caption{Schematic illustration of the formation of hybridized bonding and antibonding states from atomic TM $d$ and ligand $p$ orbitals for the case with a positive charge transfer energy (i.e., the TM $d$ states are higher in energy than the ligand $p$ states). The antibonding (bonding) bands can be described with a basis of TM (ligand) centered Wannier functions with strong $d$ ($p$) character on the central ion, but with strong admixtures of $p$ ($d$) character on the neighboring ions. In many transition metal oxides and related compounds, the bands corresponding to the antibonding states with dominant TM $d$ character are partially occupied, representing the \emph{frontier} bands immediately around the Fermi energy. 
   }
   \label{fig:hybridization_pd}
\end{figure}

More specifically, in many TM oxides and similar compounds, the correlated subspace can be defined by restricting the energy window to only a few so-called \emph{frontier bands} immediately around the Fermi level, generally formed by states with dominant TM $d$ character (or even by only a subset of $d$ states corresponding to either $e_g$ or $t_{2g}$ states).  
Since, in most cases, these frontier bands are formed from the antibonding combination of hybridized TM $d$ and ligand $p$ states (see \pref{fig:hybridization_pd}), the resulting TM-centered orbitals typically extend to the surrounding ligands and thus have a mixed TM $d$ and ligand $p$ character. 
Nevertheless, it is common to refer to this set of bands simply as the ``$d$ bands''. Consequently, the bands corresponding to the bonding combination of TM $d$ and ligand $p$ states are often referred to as ``$p$ bands''.~\footnote{We note that in systems with a very small or even negative charge transfer energy~\cite{Zaanen/Sawatzky/Allen:1985} the corresponding bands can be strongly mixed and the assignment of bonding and antibonding character to the $p$- and $d$-dominated bands, respectively, is not necessarily valid any more.}
In the following, we will 
use the term \textit{frontier bands} to denote the set of (antibonding) $d$-$p$ hybridized bands around the Fermi level and \emph{frontier orbitals} to denote the corresponding basis orbitals.

Instead of using an energy window restricted only to the frontier bands, it is also common to use a wider energy window, including all bands with dominant TM $d$ and ligand $p$ character. This generally yields more localized basis functions closely resembling the character of the underlying atomic orbitals. In the following, we denote such basis orbitals as \emph{localized orbitals}. 

One could intuitively expect calculations with a larger energy window to be more accurate compared to those restricted to the frontier bands. After all, the corresponding orbitals are more localized and thus the usual approximation of explicitly including only local interactions appears more justified.
Furthermore, it is possible to describe $d$-$p$ charge transfer processes in this case~\cite{hansmann_importance_2014}.
However, while the orbital occupations in the frontier basis generally reflect the nominal (integer) oxidation states of the TM cations, this is not case in the localized basis. Here, due to the strong $d$-$p$ hybridization, the total occupation of the $d$ states on a particular site, $N_d$, can be rather different from the formal charge state.
On the other hand, the emergence of a Mott-insulating state and other correlation effects strongly depend on the filling of the correlated orbitals~\cite{parragh_effective_2013, dang_covalency_2014, dang_density_2014, amaricci_mott_2017}.

It has been shown that this can result in a failure to obtain a Mott-insulating state in DFT+DMFT calculations using a localized basis for materials that are well known to be Mott insulators~\cite{dang_covalency_2014, dang_density_2014}.~\footnote{It should be noted, though, that another DFT+DMFT study, using projections on very localized orbitals to define the correlated subspace (and a different form of the DC correction) did not encounter such problems~\cite{haule_covalency_2014}.} It has then been suggested that adjusting the so-called double-counting (DC) correction can fix the problem to a certain extent~\cite{dang_covalency_2014, dang_density_2014}. 
The role of the DC correction is to subtract the effects of the local interaction within the correlated subspace that are already accounted for on the standard DFT level. 
In practice, the magnitude of the DC correction in the localized basis also controls the $d$-$p$ charge transfer energy, i.e., the energy difference between the TM $d$ and the ligand $p$ states, which in many TM oxides is underestimated by local/semi-local DFT functionals~\cite{dang_covalency_2014, dang_density_2014, haule_covalency_2014}.
Since the charge transfer energy controls the degree of $d$-$p$ hybridization, scaling the DC correction allows to adjust the filling of the localized $d$ orbitals $N_d$~\cite{dang_covalency_2014, dang_density_2014}. 

Other ways of correcting the $d$-$p$ charge transfer energy have also been explored.
Hansmann \textit{et al.} showed that the inclusion of an intersite Coulomb term $U_{pd}$ in combination with an on-site $U_p$ on the ligands recovers the description of the Mott insulating state in cuprates~\cite{hansmann_importance_2014} at the model level. Lechermann \textit{et al.} have tackled the issue of correcting the charge transfer energy in NiO by including a self-interaction correction on the level of the oxygen pseudopotential while treating the Ni $d$ states within DMFT using the standard fully localized limit (FLL) DC correction~\cite{lechermann_interplay_2019}. The amount of self-interaction correction can be chosen to match the experimental $d$-$p$ splitting, which then also results in a good estimate for the bandgap of NiO~\cite{lechermann_interplay_2019, lechermann_oxygen_2024}.

Thus, apart from the explicit treatment of the local interaction within the correlated subspace, a reliable description of the electronic behavior of TM oxides within DFT+DMFT also depends on the correct $d$-$p$ charge transfer energy. 

Furthermore, to arrive at a truly quantitative and predictive description, one also requires to obtain values of the interaction parameters, typically represented in terms of the Hubbard $U$ and the Hund's coupling $J$.
It is worth noting that these values depend strongly on the corresponding orbitals and are therefore fundamentally different for frontier and localized basis sets.
In the DFT+DMFT context, specific values are often obtained using the constrained random-phase approximation (cRPA) \cite{aryasetiawan_frequencydependent_2004,miyake_screened_2008, miyake_initio_2009}, in which the Coulomb interaction felt by the electrons in the correlated subspace is screened by the rest of the ``uncorrelated'' electronic degrees of freedom.
However, it has been pointed out in several studies that this procedure can result in $U$ values that are too small to open a Mott gap \cite{hampel_energetics_2019, kazemi-moridani_strontium_2024, merkel_calculation_2024}.

Here, building on previous studies \cite{dang_covalency_2014, park_computing_2014, lechermann_interplay_2019}, we demonstrate that a better treatment of the on-site interaction of the ligand $p$ orbitals can significantly improve the quantitative description of correlated insulators. Crucially, this holds for both the localized basis, for which we explicitly consider the ligand interaction $U_p$ within the DFT+DMFT framework, as well as in the frontier basis, in which the contribution of the $p$ states is implicitly included in the construction of the basis set.

We consider a selected set of prototypical TM oxides and first show in \pref{sec:frontier_results} that, even though DFT+DMFT calculations using a frontier basis provide a good and intuitive picture of the main physical mechanism, a realistic description requires interaction parameters that appear inconsistent with the corresponding values obtained from cRPA.
We then show in \pref{ssec:localized_results} that similar problems are also encountered in calculations using a localized basis, but that these can be resolved, at least to a large extent, by including a simple explicit treatment of the local interaction between the ligand $p$ states on the level of the Hartree-Fock approximation. 
Finally, in \pref{sec:crpa_results} we perform cRPA calculations where we correct the positions of the O $p$ states (and potentially other states close to the Fermi level) so as to achieve Kohn-Sham bands that more closely match experimental spectroscopic data. We demonstrate that this yields noticeably larger interaction parameters, which turn out to be consistent with the ones used in \pref{sec:frontier_results} and \pref{ssec:localized_results} to reproduce the correct insulating behavior for all materials under consideration. In particular, this holds both for the interaction parameters corresponding to the frontier orbital basis as well as for those corresponding to the $d$-$d$ and $p$-$p$ interaction in the localized basis. 

Our work highlights the importance of not only explicitly considering the TM $d$ states but also of correcting deficiencies in the description of the ligand states. In particular, our work demonstrates the critical role of ligand states also for the calculation of the interaction parameters within cRPA, and that a simple Hartree-Fock-like correction of these states leads to a significantly improved and quantitatively much more consistent description at the DFT+DMFT level. This paves the way towards potential systematic improvements of such calculations, in order to arrive at a truly quantitative and predictive first-principles-based description of correlated materials.

\section{\label{sec:Results} Results}

We are considering two pairs of insulating TM oxides on opposite ends of the 3$d$ series. LaTiO$_3$ and LaVO$_3$, with a $d^1$ and $d^2$ electron configuration respectively, are generally considered as typical Mott insulators with a large charge transfer energy~\cite{dougier_evolution_1976, fujimori_dopinginduced_1992, imada_metalinsulator_1998, maiti_spectroscopic_2000}. 
On the other hand, LuNiO$_3$ and NdNiO$_3$, corresponding to the family of rare earth nickelates, exhibit a rather small or even slightly negative charge transfer energy and a metal insulator transition that involves a charge (or bond) disproportionation leading to two symmetry inequivalent Ni sites~\cite{alonso_hightemperature_2001, medarde_longrange_2008, medarde_charge_2009a}.

\begin{figure}
   \centering
   \includegraphics[width=\columnwidth]{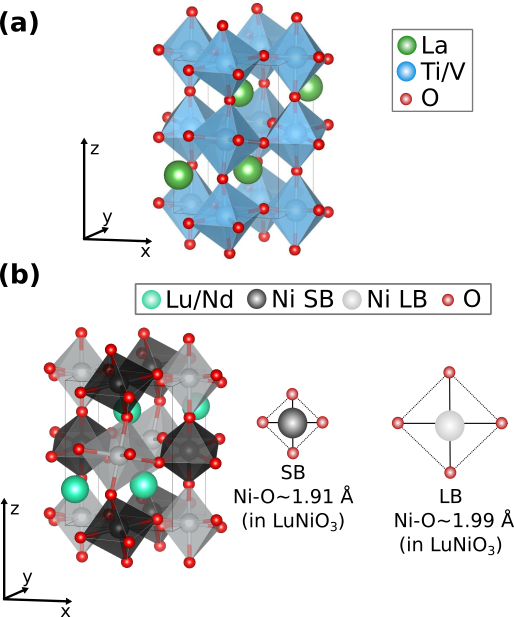}
   \caption{(a) The \textit{Pbnm} crystal structure of \latio and \lavo,  and (b) the distorted $P2_1/n$ structure of \lunio and \nno,  highlighting the difference in the average Ni-O bond length for the short-bond (SB) and long bond (LB) sites for the case of \lunio.  
   }
   \label{fig:materials}
\end{figure}

All four materials crystallize in a distorted perovskite structure with $Pbnm$ space group symmetry, exhibiting GdFeO$_3$-like octahedral tilts (see \pref{fig:materials}(a))~\cite{eitel_high_1986, khan_crystal_2004, alonso_hightemperature_2001}.
In the case of the nickelates, an additional distortion (denoted with the symmetry label $R^+_1$) occurs below the metal-insulator transition temperature, corresponding to a breathing mode of the Ni-O octahedra resulting in long-bond (LB) and short-bond (SB) octahedra arranged in a three-dimensional checkerboard pattern (see \pref{fig:materials}(b)) and $P2_1/n$ space group  symmetry~\cite{alonso_hightemperature_2001, medarde_longrange_2008}.

In all four cases, DFT+DMFT calculations with frontier orbitals have successfully been used to describe the insulating state~\cite{pavarini_mott_2004, deraychaudhury_orbital_2007, dymkowski_straininduced_2014, sclauzero_structural_2015, sclauzero_tuning_2016}, including also the charge disproportionation in the case of the nickelates~\cite{subedi_lowenergy_2015, seth_renormalization_2017, hampel_energetics_2019,  peil_mechanism_2019}. On the other hand, difficulties to obtain an insulating state in DFT+DMFT calculations using the localized basis have been reported, and a scaling of the DC correction has been suggested to remedy this problem~\cite{dang_covalency_2014, dang_density_2014, park_total_2014, park_computing_2014}.
Generally, one could expect that in early TM oxides such as \latio and \lavo, a description in terms of frontier orbitals is well justified, due to the large $d$-$p$ charge transfer energy, whereas in the nickelates a description using a localized $d$-$p$ basis might appear more appropriate.

\subsection{\label{sec:frontier_results} Calculations based on frontier orbitals}

\begin{figure}
   \centering
   \includegraphics[width=\columnwidth]{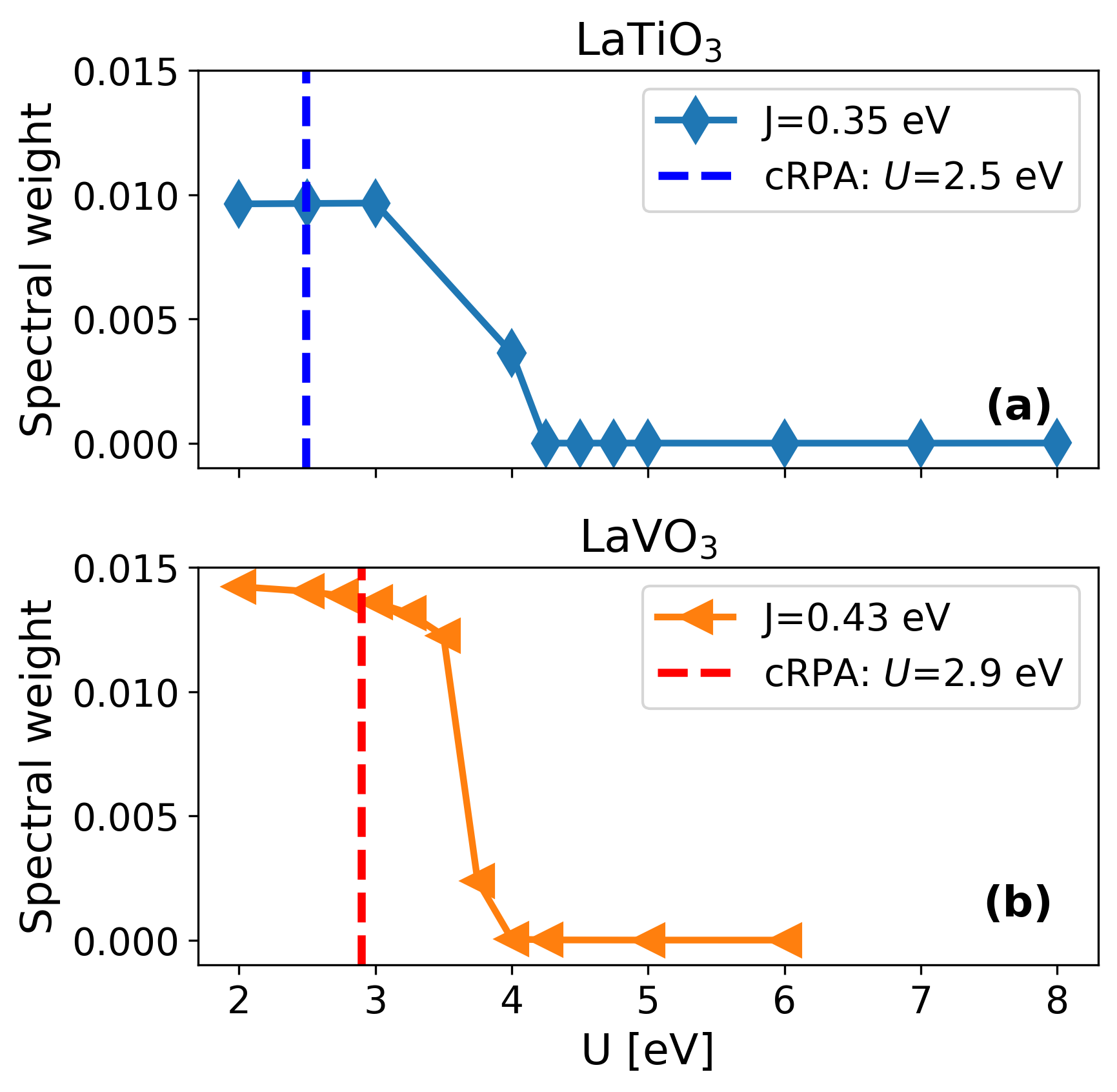}
   \caption{Spectral weight at the Fermi level, $A(\omega=0)$, as a function of the $U$ parameter, obtained from our charge self-consistent DFT+DMFT calculations for (a) \latio and (b) \lavo using the frontier orbital basis. The data represented by the solid lines is obtained with $J$ fixed to the cRPA values from \cite{kim_straininduced_2018} indicated in the legend. The vertical dashed lines indicate the $U$ values from \cite{kim_straininduced_2018}. 
   }
   \label{fig:latio_lavo_csc_frontier}
\end{figure}

%
In both \latio ($d^1$) and \lavo ($d^2$), calculations using frontier orbitals have been used extensively to characterize the Mott-insulating state \cite{pavarini_mott_2004, deraychaudhury_orbital_2007, dymkowski_straininduced_2014, sclauzero_structural_2015, sclauzero_tuning_2016}. Thereby, the local interaction is typically parametrized in the so-called Kanamori form in terms of the intra-orbital Coulomb repulsion $U$ and the Hund's coupling $J$ (see, e.g., \cite{georges_strong_2013}).
Furthermore, values for the corresponding interaction parameters $U$ and $J$ have been obtained from cRPA calculations in Ref.~\cite{kim_straininduced_2018}.

In \pref{fig:latio_lavo_csc_frontier}, we plot the spectral weight at zero frequency, $A(\omega=0)$, obtained from our charge self-consistent DFT+DMFT calculations using $t_{2g}$ frontier orbitals for \latio and \lavo as a function of $U$. Here, $J$ is fixed to the cRPA values of 0.35\,eV for \latio and 0.43\,eV for \lavo reported in Ref.~\cite{kim_straininduced_2018}. 
\pref{app:one-shot} contains further results obtained for a larger range of $U$ and $J$ values but without charge self-consistency. More details on our computational method are given in \pref{sec:Methods}.
We also note that in the frontier basis, by construction, the average occupation on the TM sites corresponds exactly to the formal valence of the TM cation, i.e., one electron for \latio and two for \lavo. 


For both \latio and \lavo, one observes a transition from a metallic state with nonzero spectral weight at low $U$ to a Mott-insulating state with zero spectral weight at large $U$. In both cases, this transition happens roughly around $U\sim4$\,eV.
The cRPA values of $U$ obtained in \cite{kim_straininduced_2018} are indicated by the blue and red vertical dashed lines in \pref{fig:latio_lavo_csc_frontier}. 
It is obvious that these $U$ values are significantly smaller than the critical values for the Mott transition, and therefore result in metallic behavior, in clear disagreement with experimental observations~\cite{fujimori_dopinginduced_1992, maiti_spectroscopic_2000}. 
In order to obtain an insulating state, one needs to increase $U$ relative to the cRPA values by at least 60\% for the case of \latio and by 40\% for the case of \lavo.
Thus, while DFT+DMFT calculations for \latio and \lavo using frontier orbitals can successfully describe their Mott-insulating character, there appears to be a rather substantial quantitative discrepancy regarding the required strength of the interaction $U$ with respect to the corresponding cRPA results. 

\begin{figure}
   \centering
   \includegraphics[width=\columnwidth]{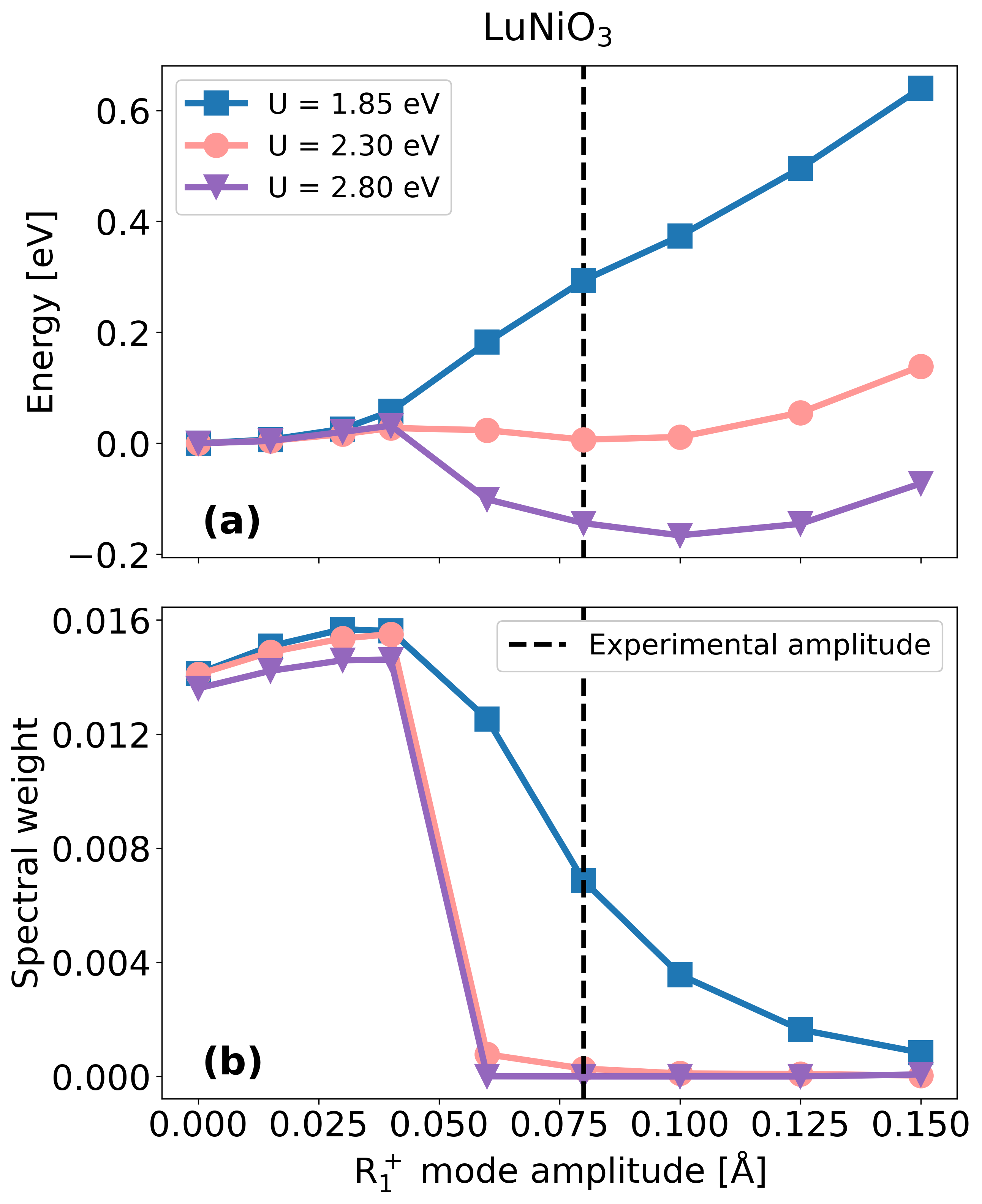}
   \caption{Evolution of (a) the total energy per 20-atom unit cell and (b) the spectral weight $A(\omega = 0)$ of \lunio as a function of the $R_1^+$ breathing mode. Calculations are performed with the frontier orbital basis. We use $J = 0.42$ eV, while the values of $U$ are 1.85\,eV, 2.30\,eV, and 2.80 eV, corresponding to the cRPA value~\cite{hampel_energetics_2019}, and this value scaled by factors of 1.24 and 1.51, respectively. The dashed line represents the experimental amplitude of the $R_1^+$ mode from Ref.~\cite{alonso_hightemperature_2001}.
   \label{fig:lno_energy_curve}
   }
\end{figure}

Next, we show that a very similar situation is also obtained for \lunio using a basis of Ni $e_g$ frontier orbitals. 
Here, we analyze our results as function of the $R_1^+$ breathing mode amplitude, i.e., as function of the structural distortion leading to inequivalent SB and LB Ni sites ({\it cf.} \pref{fig:materials}).
In \pref{fig:lno_energy_curve}(a) we plot the total energy with respect to the undistorted $Pbnm$ structure as a function of the $R^+_{1}$ mode amplitude obtained for different values of $U$ and $J$.
We observe that for $U=1.85$\,eV and $J=0.42$\,eV, which correspond to the interaction parameters obtained from cRPA~\cite{seth_renormalization_2017, hampel_energetics_2019}, the only energy minimum occurs at zero distortion, and the spectral weight, $A(\omega=0)$ (shown in \pref{fig:lno_energy_curve}(b)) indicates a metallic state even for large $R^+_1$ amplitudes. Again, this appears in contradiction to the experimental observation of insulating \lunio with an $R_1^+$ distortion of around 0.08\,\AA.
If, however, the value of $U$ is increased (while keeping $J$ fixed), an energy minimum appears close to the experimentally observed $R_1^+$ amplitude, also corresponding to an insulating state. To achieve this, an increase of $U$ by at least 24\,\% relative to the cRPA value is necessary. 

The results presented so far demonstrate that DFT+DMFT calculations using frontier orbitals can in principle give a good description of the insulating state of the selected materials, but that in all cases the corresponding Hubbard parameters $U$ computed from cRPA are too small to drive the transition into the insulating state.

\subsection{\label{ssec:localized_results} Treating the O $2p$ states in the localized basis set}

We now turn to DFT+DMFT calculations using the localized orbital basis set obtained from a large energy window. We investigate in particular the effect of varying the position of the $p$ bands on the insulating behavior. Here, we follow up on the work of Dang \textit{et al.}~\cite{dang_covalency_2014, dang_density_2014}, but instead of scaling the DC correction (for which we keep the standard FLL form, see \pref{sec:Methods}), we consider an explicit local interaction not only for the TM $d$ states but also for the O $p$ orbitals.

For the latter, we use a simple static Hartree-Fock approximation, resulting in a frequency-independent self-energy $\Sigma^p$. Thereby, we employ a form of the local interaction and the corresponding DC correction analogous to the DFT+$U$ formulation of Dudarev et al.~\cite{dudarev_electronenergyloss_1998}, with a single interaction parameter $U_p$.
The resulting total Hartree-Fock correction to the DFT potential, $\Delta \Sigma^p =  \Sigma^p - \Sigma^p_{DC}$, takes the following simple form:
\begin{equation}
    \label{eqn:dudarev_form}
    \Delta \Sigma^p = \sum_{\lambda} U_p \bigg ( \frac{1}{2} - n_{\lambda} \bigg) \ket{\lambda}\bra{\lambda}.
\end{equation}
Here, $\ket{\lambda}$ and $n_\lambda$ are the eigenvectors and eigenvalues of the local occupation matrix corresponding to the ligand $p$ states. See \pref{sec:Methods} for more details.

The form of $\Delta \Sigma^p$ in \pref{eqn:dudarev_form} favors either completely empty or fully occupied local orbitals. Since the O $p$ states in TM oxides are usually almost completely occupied ($n_\lambda \approx 1$),
$\Delta \Sigma^p$ essentially leads to a downward shift of the O $p$ levels by approximately $-U/2$,  lowering the position of the $p$ states relative to the TM $d$ states and consequently reducing the $d$-$p$ hybridization.

In contrast, the local interaction problem for the $d$ states is still treated dynamically within DMFT, using a numerically exact quantum Monte Carlo solver which provides access to the fully frequency-dependent self-energy $\Sigma^d(\omega)$.
For \latio and \lavo, we consider the interaction only on the TM $t_{2g}$ subshell, with the intra-orbital Hubbard and Hund's interaction parameters denoted as $U_d$ and $J_d$.
For \lunio and \nno, we consider the interaction within the full $d$ shell, using the average Hubbard  and Hund's interaction parameters in the so-called Slater parametrization, denoted as $\mathcal{U}_d$ and $\mathcal{J}_d$. More details are given in \pref{sec:Methods}.

\begin{figure}
   \centering
   \includegraphics[width=\columnwidth]{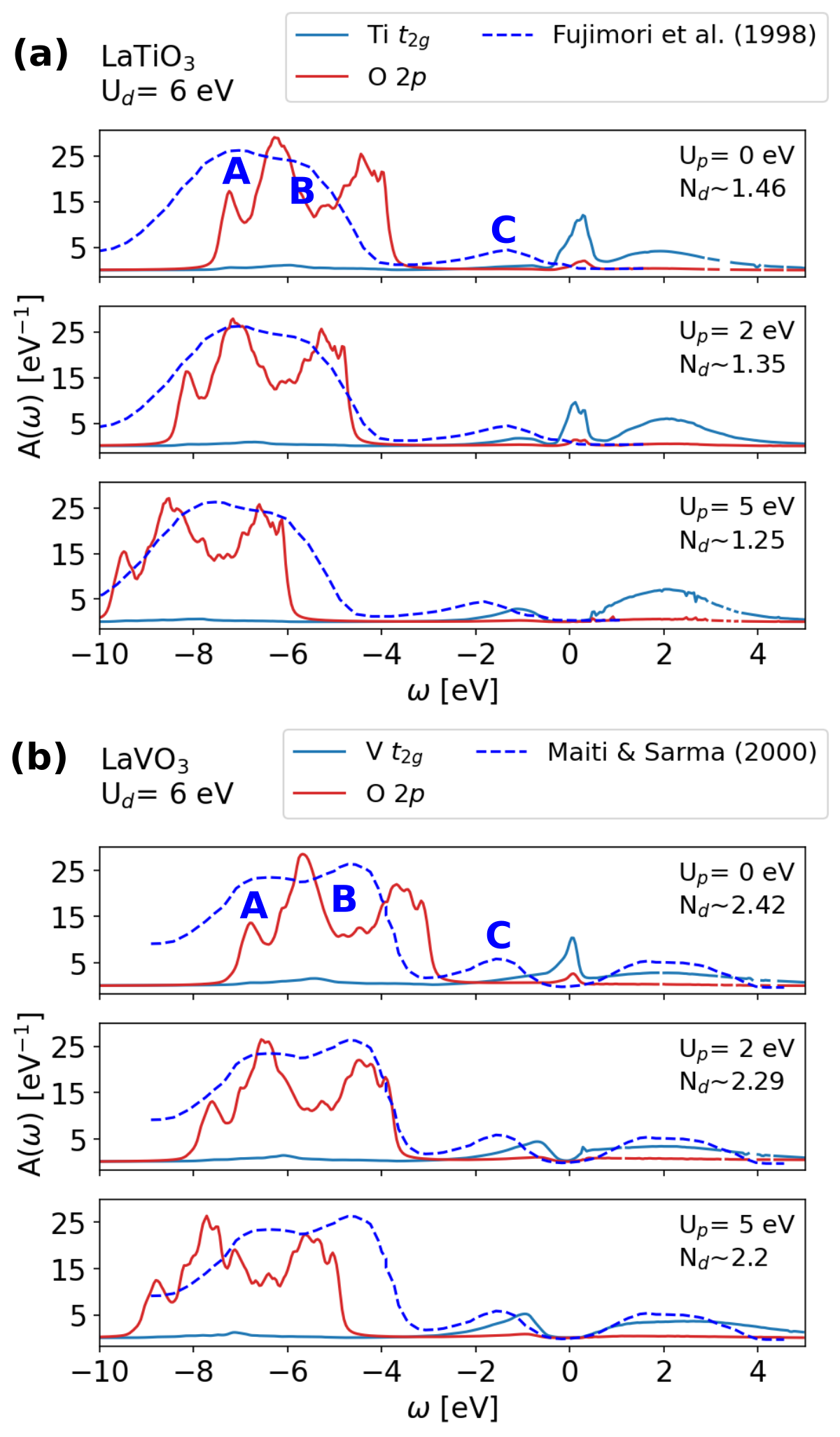}
   \caption{Site resolved spectral functions obtained from localized orbital DFT+DMFT calculations for (a) \latio and (b) \lavo using $U_d=6$\,eV, $J_d=0.65$\,eV and different values for $U_p$. Blue dashed lines correspond to experimental data provided in \cite{fujimori_dopinginduced_1992, maiti_spectroscopic_2000}. A and  B label the two-peak feature dominated by oxygen bands in the experimental spectra, while C indicates the lower Hubbard band corresponding to the TM $d$ states.}
   \label{fig:lanthanates_spectral_functions}
\end{figure}

In \pref{fig:lanthanates_spectral_functions} we plot the site-resolved spectral functions for \latio and \lavo obtained from DFT+DMFT calculations using the localized basis. We vary $U_p$ while fixing the interaction parameters for the TM $d$ states to $U_d=6$\,eV and $J_d=0.65$\,eV. As we show in \pref{ssec:crpa}, these values for $U_d$ and $J_d$ are close to what we obtain from cRPA calculations. Note that these values correspond to localized basis orbitals and are therefore different from the ones corresponding to the frontier orbitals discussed in \pref{sec:frontier_results}. 

\pref{fig:lanthanates_spectral_functions} also includes (as blue dashed lines) experimental data obtained from x-ray valence photoemission spectrosopy (PES) for \latio~\cite{fujimori_dopinginduced_1992} and from a combination of PES and bremsstrahlung isochromat spectrum for \lavo~\cite{maiti_spectroscopic_2000}.
Both experimental spectra show a broad feature with a two peak structure in the energy region around $-6$\,eV, indicated by A and B in the uppermost panels of \pref{fig:lanthanates_spectral_functions}(a) and \pref{fig:lanthanates_spectral_functions}(b), and a weaker peak at around $-2$\,eV (labeled C). While A and B are related to the oxygen bands, C has been identified with the lower TM $t_{2g}$ Hubbard band~\cite{maiti_spectroscopic_2000, fujimori_dopinginduced_1992}.

It is apparent that without explicitly considering the interaction between the $p$ states ($U_p=0$), the $p$ bands for both \latio and \lavo are too close to the $d$ band compared to experiment. Simultaneously, an incorrect metallic solution is obtained for both materials. We also note that, due to the strong $d$-$p$ hybridization, the occupation of the TM $d$ states, $N_d$, deviates significantly from the nominal occupation ({\it cf.} \cite{dang_covalency_2014, dang_density_2014}).

Increasing $U_p$ lowers the position of the O $p$ bands relative to the TM $d$ bands.
The resulting reduction in $d$-$p$ hybridization lowers $N_d$, and, for $U_p=5$\,eV (already for $U_p=2$\,eV in the case of \lavo), a gap opens in the calculated spectral function for fixed $U_d=6$\,eV. 
The gap opens when the deviation from integer occupations is lower than $\sim 0.3$, in qualitative and quantitative agreement with the results presented in \cite{dang_covalency_2014, dang_density_2014}, where the relative position of the $p$ and $d$ bands has been tuned by scaling the DC correction. 
In \pref{app:one-shot}, we present additional data for the spectral weight, $A(\omega=0)$, indicating a metallic or insulating solution, over a wider range of $U_d$ and $U_p$, obtained without charge self-consistency.

It seems that the best agreement with the experimental position of the $p$ bands is obtained around $U_p = 2$\,eV for both \latio and  \lavo. On the other hand the appearance of the gap in \latio and also the size of the gap in \lavo (around 1.1\,eV according to \cite{maiti_spectroscopic_2000}) shows better agreement with experiment for $U_p=5$\,eV. This could indicate that, while introducing $U_p$ already leads to a significant improvement, other effects, such as, e.g., the $d$-$p$ intersite interaction, also need to be considered in order to arrive at an even more accurate quantitative description. 

\begin{figure}
   \centering
   \includegraphics[width=\columnwidth]{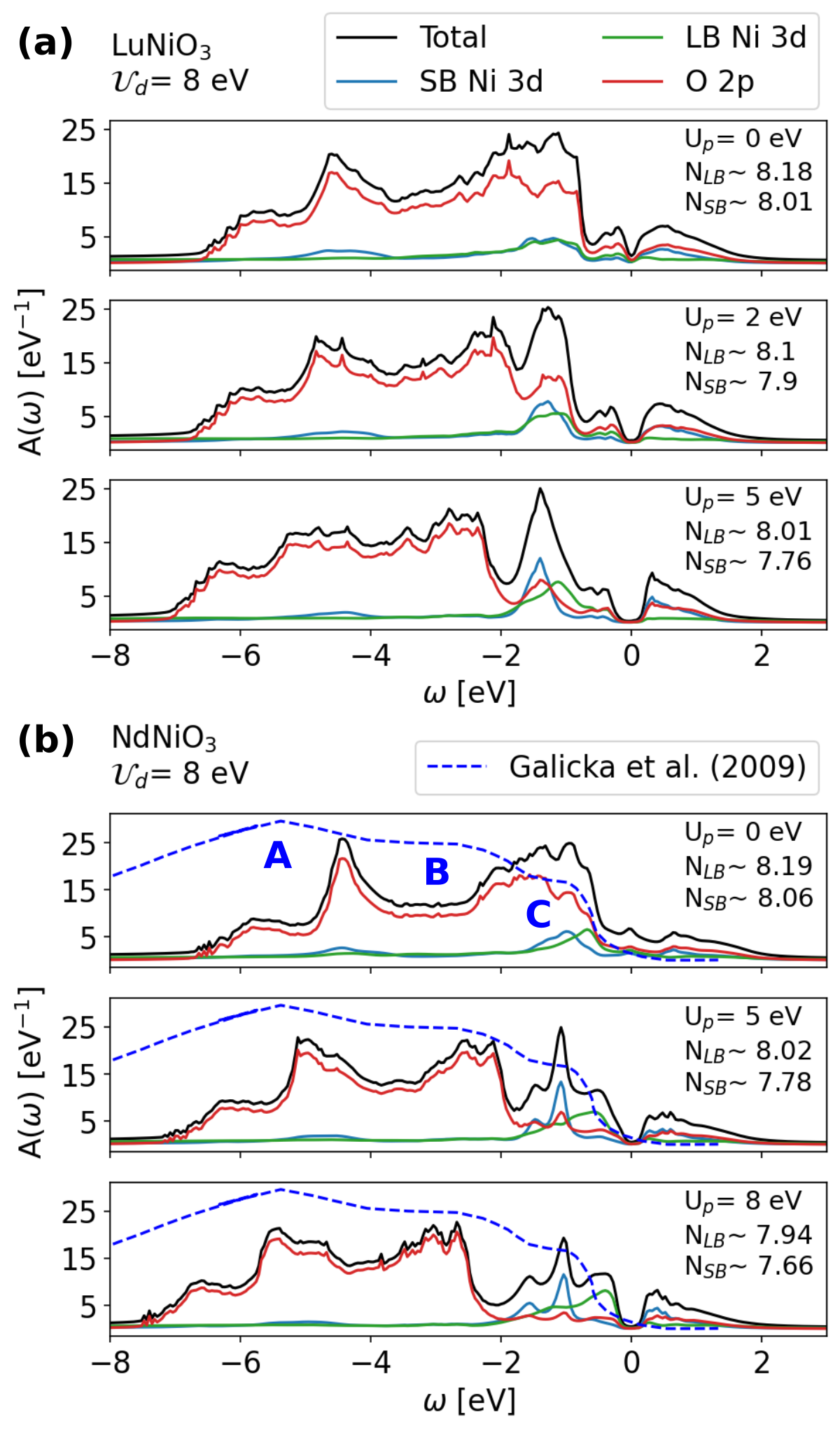}
   \caption{Total and site-resolved spectral functions obtained from DFT+DMFT using localized orbitals for (a) \lunio and (b) \nno. These calculations are performed for $\mathcal{U}_d=8$\,eV, $\mathcal{J}_d=1$\,eV, and different values of $U_p$. The blue dashed lines in (b) correspond to the valence photoemission spectrum for thick films of \nno provided in \cite{galicka_photoemission_2009}. Labels A, B, and C indicate different features in the experimental spectrum (see main text). $N_\text{LB}$ and $N_\text{SB}$ represent the $d$ orbital occupation of the long-bond and short-bond Ni site respectively.}
   \label{fig:nickelates_spectral_functions}
\end{figure}

A similar situation is also observed for the nickelates, for which we use $\mathcal{U}_d=8$\,eV and $\mathcal{J}_d=1$\,eV. This specific choice will also be discussed further in \pref{sec:crpa_results}. 
In \pref{fig:nickelates_spectral_functions} we show the spectral functions for \lunio 
and \nno obtained in the fully distorted low temperature $P2_1/n$ structures.
In previous work, Park \textit{et al.} found that the insulating state of \lunio cannot be obtained in a DFT+DMFT calculation using the localized Wannier basis in combination with the standard FLL DC correction~\cite{park_total_2014, park_siteselective_2012}. However, a small reduction of the DC correction is sufficient to open a gap \cite{park_total_2014, park_siteselective_2012}. 
We obtain a very similar result from our calculations by tuning the local interaction strength of the O $p$ states. For $U_p=0$\,eV [see \pref{fig:nickelates_spectral_functions}(a)], we observe no gap for \lunio, even though the spectral function around the Fermi energy already exhibits a pseudo-gap-like feature. Increasing $U_p$ to 2\,eV then leads to an insulating state.

In \pref{fig:nickelates_spectral_functions} we have also included the occupations $N_\text{LB}$ and $N_\text{SB}$ of the two inequivalent Ni sites, which is closer to $d^8$ than to the nominal $d^7$ configuration of Ni$^{3+}$, with only a small difference between the LB and SB sites. This has also been pointed out in~\cite{haule_mott_2017, park_siteselective_2012} and is discussed in more detail in \pref{app:lunio3}. 
The reduction of the $d$-$p$ hybridization via application of $U_p$ also reduces the overall average $N_d$ and in addition slightly enhances the small occupation difference between the inequivalent Ni sites. However, the average occupation always remains rather close to integer $d^8$, both in the metallic and insulating states, which is somewhat different from the cases of \latio and \lavo, where the Mott gap opens once the occupation $N_d$ gets sufficiently close to the nominal integer occupations of $d^1$ or $d^2$. Further details of the Mott-insulating character of LuNiO$_3$ are discussed in \pref{app:lunio3}.

Next, we turn to \nno [see \pref{fig:nickelates_spectral_functions}(b)], for which experimental PES spectra are available \cite{galicka_photoemission_2009}. The experimental PES spectrum, shown with the blue dashed line in \pref{fig:nickelates_spectral_functions}(b), also exhibits three features, labeled as A, B, and C. Features A and B, around $-5$\,eV and $-3$\,eV, respectively, have been assigned to the O $p$ band, while feature C, at around $-1$\,eV, is assumed to originate from the Ni $d$ states~\cite{galicka_photoemission_2009}.
Similar to the cases of \latio and \lavo, we observe that the spectral function obtained for $U_p=0$\,eV is in qualitative disagreement with the experimental PES spectrum.
In particular, the O $p$ dominated bands appear too close to the Fermi level, resulting in a two-peak structure with no clear Ni $d$ feature at low binding energies.
Increasing the value of $U_p$, shifts the O $p$ bands down in energy, and recovers the qualitative features of the experimental spectrum with two oxygen dominated features at lower energies and an additional peak close to the Fermi level, dominated more by the Ni $d$ states. In addition, a gap opens leading to an insulating spectral function. For $U_p=5$\,eV (which is close to the value obtained by our cRPA calculations in \pref{sec:crpa_results}), we obtain good agreement between the calculated spectral function and the measured PES spectrum.

\subsection{\label{sec:crpa_results} cRPA calculations}

\begin{figure*}
   \centering
   \includegraphics[width=\textwidth]{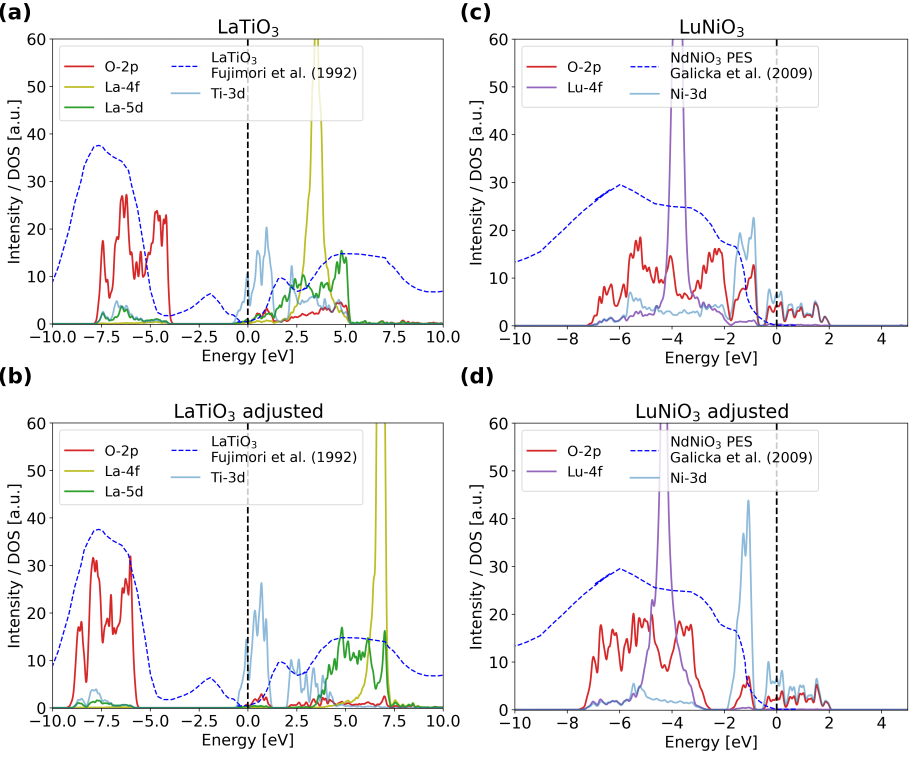}
   \caption{Site (and  shell) resolved densities of states (DOS) corresponding to the Kohn-Sham bandstructure of \latio [panels (1) and (b)] and \lunio [panels (c) and (d)]. The DOS in (a) and (c) are obtained without any adjustment, whereas in (b) and (d) the O $2p$, La $5d$ (and La/Lu $4f$) states are adjusted to better match the experimental data from \cite{fujimori_dopinginduced_1992} for \latio and \cite{galicka_photoemission_2009} for the case of \lunio. Note that the data in \cite{galicka_photoemission_2009} is obtained for thick films of \nno, which we use as reference for \lunio, here, due to lack of experimental data for \lunio.}
   \label{fig:adjusted_lunio_latio}
\end{figure*}

As mentioned in the introduction and demonstrated in \pref{sec:frontier_results}, using the screened interaction parameters obtained from cRPA can lead to qualitative disagreement with experimental results for frontier orbital calculations. To achieve good agreement with experimental observations, a substantial scaling respectively increase of the corresponding $U$ values is required. For the case with localized orbitals, we showed that similar qualitative disagreement is obtained if the local interaction on the ligands is not considered. However, we still need to prove that the corresponding interaction parameters are quantitatively reasonable.

There are in principle several potential explanations why $U$ values obtained from cRPA calculations should not be used ``right out of the box''.
First, cRPA provides a frequency-dependent interaction $U(\omega)$, while in DMFT one typically considers only the value for $U(\omega=0)$ and discards the frequency-dependence of the screening. 
This appears well justified if the frequency dependence of $U(\omega)$ is weak over the energy range of typical excitations within the correlated subspace.
Second, for the  case of the frontier basis, there have been suggestions regarding the importance of non-local interaction terms, which, due to the relatively large spatial extension of these orbitals, are not necessarily negligible~\cite{seth_renormalization_2017}.
Not explicitly considering these non-local interactions might require an adjustment of the local interaction parameters obtained from cRPA.  
However, for the localized basis, intersite interactions are very weak (see, e.g., \cite{haas_incorporating_2024}) and thus are unlikely the cause of the problems described in~\pref{ssec:localized_results}.
Third, 
cRPA is known to overestimate screening, partly due to the omission of higher-order polarization diagrams, which become particularly relevant for low-energy transitions that occur if the screening bands are close in energy to the correlated subspace~\cite{honerkamp_limitations_2018, vanloon_random_2021}.
Finally, as  was pointed out by Werner and Casula~\cite{werner_dynamical_2016}, another reason for an incorrect estimation of screening effects could be related to using the DFT Kohn-Sham band structure as a starting point for the cRPA calculation. The authors remark that this should be particularly relevant in the case of TM oxides, where typical (semi-) local DFT functionals often underestimate the $d$-$p$ splitting~\cite{werner_dynamical_2016}.
This leads to lower excitation energies and, as a consequence, to an overestimation of the screening and an underestimation of $U$ (see the formula for the polarization function in \pref{eqn:crpa_polarization}).

We now explore this last consideration by empirically modifying the Kohn-Sham band structure at the DFT+$U$ level, {\it i.e.}, by applying empirical $+U$ corrections to the O $2p$ and La $5d$ (and La/Lu $4f$) states, in order to better align the corresponding features in the density of states with available experimental photoemission and absorption data. 
Subsequently, we perform cRPA calculations based on this corrected Kohn-Sham band structure to compute the interaction parameters for both the frontier and localized basis sets.  

\pref{fig:adjusted_lunio_latio} illustrates the density of states (DOS) of \latio obtained (a) without and (b) with empirical adjustment of the bands through application of a $+U$ correction to the O $2p$, La $5d$, and La $4f$ states. 
In \pref{fig:adjusted_lunio_latio}(b), the O $2p$ and La $5d$ states are shifted to better reproduce the corresponding features in the experimental spectra reported in~\cite{fujimori_dopinginduced_1992}. The La $4f$ states are shifted towards the top of the La $5d$ band. There is no spectroscopic data available on the position of these states, 
but our test calculations indicate that their contribution to the screening is negligible.
%
The case of \lavo is treated analogously. Thereby, the O $2p$ states are shifted to better reproduce the spectrum reported in~\cite{maiti_spectroscopic_2000}. Due to the lack of spectroscopic information on the empty states in \lavo, the La $5d$ (and $4f$) states are shifted to a similar position as for \latio.
The Ti and V $3d$ states are not shifted, since they are pinned to the Fermi level and a better agreement with the spectroscopic data would require a splitting into lower and upper Hubbard bands which is only achieved in the subsequent DFT+DMFT treatment.

\pref{fig:adjusted_lunio_latio}(c) and (d) show the DOS of \lunio without and with empirical adjustment of the O $2p$ states, respectively.
The filled Lu $4f$ states are not shifted and we note that these states are likely placed significantly too high in energy by the used pseudopotential. However, similar to \latio, we found that the computed interaction parameters exhibit little dependence on the position of these $4f$ states.  


We note that here we use the standard orthonormalized atomic orbitals defined in the QuantumEspresso code to apply the empirical $+U$ corrections. These orbitals are different from the localized Wannier basis used in \pref{ssec:localized_results} and \pref{tab:localized_crpa}. Thus, the corresponding $U$ values cannot be compared directly. Consequently, in order to avoid any confusion, we do not explicitly list the $U$ parameters applied to obtain the corrected densities of states in \pref{fig:adjusted_lunio_latio}(b) and (d). However, they are included in the dataset provided with this work~\cite{MaterialsCloudArchive2024}.

\begin{table}
\centering
\caption{cRPA results for the interaction parameters  $U$ and $J$ in the frontier basis. The first column indicates the different materials, the second column specifies whether the values are obtained with or without correcting the Kohn-Sham bands, or whether the value is taken from the literature. In the cases with empirically corrected Kohn-Sham bands, we also specify which states are shifted.}
\label{tab:frontier_crpa}
\vspace*{3pt}
\begin{tabular}{l l c c}
\toprule
& 
& $U$ (eV) & $J$ (eV) \\
\midrule
LaTiO$_3$           & uncorrected  & 2.88  & 0.38 \\
  & Ref. \cite{kim_straininduced_2018} & 2.49  & 0.35 \\
           & O $2p$, La $5d$, La $4f$ & 5.64  & 0.53 \\
\midrule
LaVO$_3$           & uncorrected & 3.09  & 0.47 \\
   & Ref. \cite{kim_straininduced_2018} & 2.92  & 0.43 \\
           & O $2p$, La $5d$, La $4f$ & 5.26  & 0.56 \\
\midrule
LuNiO$_3$  & uncorrected & 1.80  & 0.42 \\
  & Ref. \cite{hampel_energetics_2019} & 1.83  & 0.37 \\
           & O $2p$  & 2.65  & 0.58 \\
           & O $2p$, Lu $4f$  & 2.70  & 0.59 \\
\bottomrule
\end{tabular}
\end{table}

We first discuss the results of our cRPA calculations obtained for the frontier basis set.
In \pref{tab:frontier_crpa} we report the corresponding values of $U$ and $J$, averaged to the Kanamori parametrization, and obtained either without or with applying the empirical $+U$ corrections to the underlying Kohn-Sham bandstructure. 
The values denoted  by ``uncorrected'' are obtained from standard cRPA calculations based on the uncorrected Kohn-Sham band structures. In this case we obtain $U$ and $J$ values that agree very well with the available literature, which are also included in \pref{tab:localized_crpa} for comparison.
%
As discussed in \pref{sec:frontier_results}, these values of the interaction parameters are too small to obtain the experimentally observed insulating state in the corresponding materials.

If we apply the empirical corrections to the bands surrounding the correlated subspace, we obtain significantly larger values for the $U$ parameters, and also observe an increase in $J$. In all cases the resulting values lead to insulating solutions in the corresponding DFT+DMFT calculations, and thus to a realistic description of the corresponding materials ({\it cf.} \pref{fig:latio_lavo_csc_frontier} and \pref{fig:lno_energy_curve}).
For the case of \lunio, we also compare values obtained without shifting the occupied Lu $4f$ states and values obtained after shifting the $4f$ states to energies below the O $2p$ bands. It can be seen that this indeed has only a very small effect on the resulting interaction parameters.

Next, we discuss our cRPA results for the localized basis orbitals.
Since in this case we are treating both TM $d$ and O $p$ states in the subsequent DFT+DMFT calculations, we remove both of them from the screening subspace and we do not correct the position of the O $2p$ bands in the underlying Kohn-Sham band structure at the DFT+$U$ level.
Since for \lunio, the Lu $f$ states are entangled with the O $p$ bands, we also include these Lu $f$ states in our localized basis, which effectively removes them from the screening subspace.

\begin{table}
\centering
\caption{cRPA results for the interaction parameters in the localized basis. 
Note that for \latio and \lavo the values correspond to the Kanamori parameters $U_d$ and $J_d$ (see, e.g., \cite{georges_strong_2013}), while for \lunio they correspond to the Slater parametrization using $\mathcal{U}_d$ and $\mathcal{J}_d$ (see, e.g., \cite{liechtenstein_densityfunctional_1995}). The second column specifies whether the uncorrected or corrected Kohn-Sham bands is used and to which states the correction is applied.}
\label{tab:localized_crpa}
\vspace*{3pt}
\begin{tabular}{l l c c c}
\toprule
& 
& $U_d$/$\mathcal{U}_d$ (eV) & $J_d$/$\mathcal{J}_d$ (eV) & $U_p$ (eV)  \\
\midrule
LaTiO$_3$  & uncorrected & 3.94  & 0.48  &  4.34  \\
           & La $5d$, La $4f$ & 5.81  & 0.55  &  5.78  \\
\midrule
LaVO$_3$   & uncorrected & 4.18  & 0.58   &  4.06  \\
           & La $5d$, La $4f$ & 5.45  & 0.63  &  5.40  \\
\midrule
LuNiO$_3$  & uncorrected & 8.60  & 1.32  & 5.57  \\
\bottomrule
\end{tabular}
\end{table}

In \pref{tab:localized_crpa} we report the resulting values of the interaction parameters.
Again, one can see that the adjustment of the empty La $5d$ (and $4f$) states for \latio and \lavo leads to a significant increase of the Hubbard parameters, and also to a small increase of the Hund's coupling. Furthermore, in both cases, the values obtained without this adjustment are too small to open a Mott gap, even if the local interaction for both the TM $d$ and the O $p$ states is included in the corresponding DFT+DMFT calculations (see \pref{fig:phase_diagram_localized} in the appendix). 
On the other hand, the interaction parameters obtained with the corrected Kohn-Sham bandstructure are rather similar to the values used in \pref{ssec:localized_results} to  obtain a realistic description of the insulating state. 

For \lunio, the cRPA values obtained without any correction to the Kohn-Sham bandstructure already agree very well with the parameters used to correctly reproduce both the insulating state and the position of the oxygen bands in the corresponding spectra (see \pref{fig:nickelates_spectral_functions}). 
In this case there are no low-lying Lu $d$ or other bands that need to be corrected.

The results presented in this section clearly demonstrate that, for all considered materials, using the uncorrected Kohn-Sham band structure as a starting point for cRPA calculations leads to screened interaction parameters which incorrectly predict metallicity. On the other hand, shifting the position of the ligand states (and potentially other low-lying bands) to better reproduce spectroscopic measurements, leads to larger interaction parameters for which the correct insulating state is obtained in DFT+DMFT calculations. 

\section{Discussion and Summary}
\label{sec:discussion}

In this work we studied several representative correlated insulating TM oxides within DFT+DMFT. Thereby, we used different basis sets to describe the correlated subspace. First, a frontier orbital basis corresponding to only few low-energy bands around the Fermi level, and second, a localized orbital basis describing both TM $d$ and O $p$ dominated bands. 
We show that in both cases, in order to arrive at a consistent description of the insulating state, and to achieve good agreement with experimental spectroscopic data, it is crucial to not only consider the local interaction between the TM $d$ states, but also between the O $p$ states. 
Without explicitly accounting for these ligand interactions, inconsistencies occur in obtaining the correct insulating behavior, which are related to an inaccurate description of the O $2p$ states in the underlying DFT calculation (using standard semi-local functionals). We then show that an explicit treatment of the on-site interaction within the ligand states, even on a simple Hartree-Fock/DFT+$U$ level, leads to a significantly more realistic and quantitatively consistent description of the experimentally observed insulating state.

Thereby, the effect of the ligand states becomes important on two different levels. First, even in the case of frontier orbital calculations, where the ligand states do not explicitly enter the correlated subspace for the DMFT calculations, our results show that an incorrect energetic position of the ligand states strongly affects the screened interaction parameters obtained from cRPA calculations. In particular, without applying any further correction to the DFT Kohn-Sham bandstructure used to perform the cRPA calculation, the resulting interaction parameters are clearly too small to obtain the correct Mott-insulating state in the corresponding DFT+DMFT calculations. Roughly adjusting the Kohn-Sham band structure to match experimental spectroscopic data, results in significantly larger values for the Hubbard $U$, the correct insulating state, and, in the case of \lunio, even in a correct energetic description of the relevant structural distortion.
Thus, independent of the known tendency of cRPA to overscreen~\cite{shinaoka_accuracy_2015, werner_dynamical_2016, honerkamp_limitations_2018, vanloon_random_2021}, an inaccurate description of the electronic degrees of freedom responsible for this screening, 
can lead to an additional ``overscreening'', resulting in an underestimation of the screened interaction parameters. 

For the cases employing a localized basis, such an overscreening of the interaction parameters obtained in cRPA can also be relevant if other low-lying unoccupied bands are present, as in the cases of \latio and \lavo in \pref{tab:localized_crpa}. However, another effect, also related to the incorrect $d$-$p$ splitting in the underlying bandstructure, can become even more important. 
If the O $p$ states remain uncorrected, all materials considered in this work remain metallic in DFT+DMFT calculations using localized orbitals for realistic values of the Hubbard interaction $U_d$.
On the other hand, we show that already a simple Hartree-Fock-like treatment of the on-site interaction of the ligand states, and using interaction parameters $U_d$ and $U_p$ consistent with our cRPA results, leads to the correct insulating behavior and an overall spectrum that is on good agreement with available experimental data. 

This second effect 
has already been pointed out in previous studies, where the $d$-$p$ splitting was adjusted empirically by either shifting or scaling the DC correction~\cite{dang_covalency_2014, dang_density_2014, wang_covalency_2012, park_siteselective_2012, park_computing_2014}. We note that directly correcting the O $p$ states is conceptually much more appealing than adjusting the DC correction. While, ultimately, the exact DC correction remains ill-defined, and improvements compared to the currently used forms are probably difficult to achieve, the explicit treatment of on-site ligand interactions, and potentially also intersite TM-ligand interactions, is conceptually straightforward. Our results show that in many cases a simple and computationally not very demanding Hartree-Fock/DFT+$U$-like treatment of these ``uncorrelated'' states might already provide a significant improvement. Furthermore, the values of the corresponding interaction parameters seem consistent with those obtained from cRPA calculations, provided that similar corrections are incorporated in the band structure used for the cRPA calculations.

We note that our work is of course not yet fully quantitative and self-consistent in all aspect. The shifts applied to the Kohn-Sham bands to ``correct'' the cRPA results are partly phenomenological and do not correspond to the same basis for which the interaction parameters are calculated. Furthermore, a $U_p$ of 5\,eV, consistent with our cRPA results seems to slightly overcorrect the charge transfer splitting in \latio and \lavo. One could speculate whether also incorporating explicit $d$-$p$ or $d$-$d$ intersite interactions, which have also been shown to be important in previous studies~\cite{seth_renormalization_2017, hansmann_importance_2014}, could further improve things. 
In any case, our study points out a clear route for systematic improvements towards more quantitative and predictive DFT+DMFT calculations. We note that it is reasonable to assume that similar considerations as discussed here mainly for correlated insulators are also relevant for a correct description of metallic systems and other correlated materials.

\section{\label{sec:Methods} Details of the underlying methods}

\subsection{\label{ssec:dft_dmft} DFT+DMFT}

Within DFT+DMFT, the electronic degrees of freedom are divided in two subsets: i) the \emph{correlated subspace}, for which the local electron-electron interaction is treated on the DMFT level, and ii) the rest of the electronic degrees of freedom, which is treated at the level of a standard DFT functional~\cite{kotliar_electronic_2006}.
Thus, the Hamiltonian for the correlated subspace can be expressed as:
\begin{equation}
    \label{eqn:general_H}
    H =  H^0_{\text{KS}} + H_{\text{int}} - H_{\text{DC}} \quad ,
\end{equation}
where $H^0_{\text{KS}}$ is the effective single-particle Kohn-Sham Hamiltonian, incorporating the kinetic energy of the Kohn-Sham electrons, the  bare Hartree interaction, and the exchange-correlation contributions associated with the DFT functional,  $H_{\text{int}}$ describes the partially screened local electron-electron interaction confined to the correlated subspace, and $H_{\text{DC}}$ represents a double counting correction, subtracting the portion of the local interaction within the correlated subspace that is already included in $H^0_{\text{KS}}$. 

The self-energy in the correlated subspace basis is then approximated as purely local, neglecting any dependence on wavevector $\mathbf{k}$ while retaining the full dependence on frequency $\omega$, i.e.,  $\Sigma(\omega, \mathbf{k}) \rightarrow \Sigma(\omega)$~\cite{georges_dynamical_1996}. 
Thus, each correlated site can be mapped on an effective impurity problem and the DMFT self-consistency requires  that the corresponding impurity Green's function, $G_{\text{imp}}(\omega)$, matches the local component of the lattice Green's function, $G_{\text{latt}}(\omega, \mathbf{k})$~\cite{georges_dynamical_1996}.


In all frontier orbital calculations, and also in the localized orbital calculations for \latio and \lavo, we parametrize the interaction term $H_{\text{int}}$ on the TM sites using the so-called Kanamori form, specified by parameters $U$, $J$, and $U'=U-J$, as described in~\cite{georges_strong_2013}. 
Whenever ambiguity is present, we will distinguish the interaction on the $d$ orbitals using subscripts, e.g., $U_d$ and $J_d$. 

In the localized orbital calculations for \lunio and \nno, the local interaction on the Ni sites is expressed using the so-called  Slater parametrization, as described in \cite{liechtenstein_densityfunctional_1995}, where $\mathcal{U}_d=F^0$, $\mathcal{J}_d = (F^2+F^4)/14$, and $F^4/F^2 = 0.63$, with $F^n$ representing the Slater integrals~\cite{liechtenstein_densityfunctional_1995}. 
For the local interaction between the O $p$ states in the localized basis, we choose the following simple form:
\begin{equation}
    \label{eqn:H_p}
    H^{p}_{\text{int}} = \sum_{ m m' \sigma \sigma' }U_p n^\sigma_{m}n^{\sigma'}_{m'} (1-\delta_{\sigma\sigma'}\delta_{m m'}).
\end{equation}
Where $m, m'$ and $\sigma, \sigma'$ are orbital and spin indices, respectively, and $n_m^\sigma$ represents the electron number operator for orbital $(m,\sigma)$. We thus use a single parameter $U_p$ representing the average interaction strength in the O $p$ subspace. The mean-field approximation to \pref{eqn:H_p} is equivalent to the simplified DFT+$U$ form suggested by Dudarev {\it et al.} \cite{dudarev_electronenergyloss_1998}.

For the DC correction, we use the standard fully-localized limit (FLL)~\cite{liechtenstein_densityfunctional_1995, karolak_double_2010}, and refrain from applying any additional shifts or scaling factors, as previously done in \cite{dang_covalency_2014, park_computing_2014, park_siteselective_2012, dang_density_2014, wang_covalency_2012}.
For the Kanamori parametrization of the interaction, the DC correction results in an occupation-dependent shift of the on-site potential of the form~\cite{held_electronic_2007}:
\begin{equation}
\label{eqn:dc_kanamori}
\Sigma^{d}_\text{DC, K} = \langle U \rangle \bigg( N_d -\frac{1}{2} \bigg) \quad ,
\end{equation}
with 
\begin{equation}
\langle U \rangle = \frac{U + (U-2J)(M-1) + (U-3J)(M-1)}{2M-1} \quad ,
\end{equation}
where $M$ is the number of orbitals per site.
For the Slater parametrization the corresponding potential shift reads~\cite{liechtenstein_densityfunctional_1995, park_siteselective_2012}:
\begin{equation}
    \label{eqn:dc_slater}
    \Sigma^{d}_\text{DC, S} = \mathcal{U}_d \bigg( N_d -\frac{1}{2} \bigg) - \mathcal{J}_d \bigg( \frac{N_d}{2} -\frac{1}{2} \bigg)
\end{equation}
For our parametrization of the O $p$ interaction, this simplifies to: 
\begin{equation}
    \label{eqn:dc_}
    {\Sigma}^p_{DC} = {U}_p \bigg( N_p -\frac{1}{2} \bigg) \quad ,
\end{equation}
where $N_p$ is the total occupation of the O $p$ states.



\subsection{\label{ssec:correlated_subspace} Definition of the correlated subspace}

\begin{figure}
   \centering
   \includegraphics[width=\columnwidth]{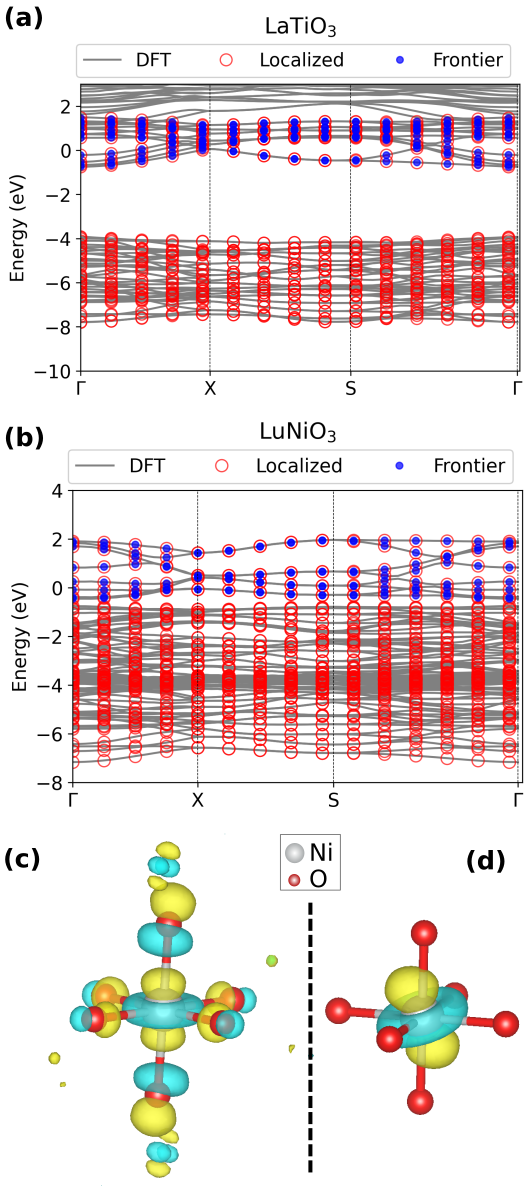}
   \caption{(a-b) Kohn-Sham band structures (grey solid lines) of (a) \latio and (b) \lunio,  together with the bands recalculated from the Wannier functions corresponding to the localized (red empty circles) and frontier (full blue circles) basis sets. (c-d) Isosurface plot of the Ni-centered $d_{z^2}$-like Wannier function in \lunio for the frontier (c) and localized (d) basis sets.}   
   \label{fig:wannier_bands}
\end{figure}

In \pref{fig:wannier_bands}(a-b) we show the Kohn-Sham band-structures of \latio and \lunio and highlight the bands included in the construction of the frontier and localized subspaces. The cases of \lavo and \nno are very similar to that.

In \latio (and \lavo) the frontier subspace describes only the group of bands around the Fermi level with dominant TM $t_{2g}$ character, while for the localized basis this subspace is extended to also include the lower-lying bands with dominant O $p$ character. As a result, the frontier basis consists of three $t_{2g}$-like orbitals centered in each TM site, while the localized basis consists of three much more localized $t_{2g}$ orbitals per TM site plus three $p$ orbitals on each O site.

In the case of \lunio (and \nno), the frontier subspace consists of the bands immediately around the Fermi level with dominant Ni $e_g$ character, resulting in two $e_g$-like frontier orbitals per Ni site, while for the localized subspace we also include the lower-lying bands with Ni $t_{2g}$ and O $p$ character. This results in five $d$ orbitals per Ni and three $p$ orbitals on each O site. 

As an example, \pref{fig:wannier_bands}(c) shows the resulting Ni-centered $d_{z^2}$-like orbitals corresponding to the frontier and localized basis sets in \lunio. It can be seen that the frontier orbital is more extended, with $p$-like ``tails'' on the surrounding O ligands, reflecting the hybridization of the frontier bands ({\it cf.} \pref{fig:hybridization_pd}). In the localized orbital, these tails are absent, since the corresponding orbital character is now represented by the additional O $p$ basis functions.

\subsection{\label{ssec:crpa} Constrained random phase approximation}


In the random-phase approximation (RPA) the screening of the bare Coulomb interaction $V$ is described by simple particle-hole excitations, expressed in terms of the polarization function: 
\begin{align}
\label{eqn:crpa_polarization}
&P(\mathbf{r},\mathbf{r'}, \omega) = \sum^\text{occ}_\nu \sum^{\text{unocc}}_\mu \psi_\nu(\mathbf{r})\psi^*_\nu(\mathbf{r'})\psi^*_\mu(\mathbf{r})\psi_\mu(\mathbf{r'}) \nonumber\\
&\times \bigg[ \frac{1}{\omega - \epsilon_\mu + \epsilon_\nu + i0^+} - \frac{1}{\omega + \epsilon_\mu - \epsilon_\nu - i0^+}   \bigg] \quad ,
\end{align}
where, in the case of RPA performed on top a DFT calculation, $\psi_\nu(\mathbf{r})$ represent the the Kohn-Sham Bloch states with eigenvalues $\epsilon_{\nu}$. 

The idea behind \textit{constrained} RPA (cRPA) is to separate the polarization function into two contributions, $P(\omega) = P_{\text{corr}}(\omega)+ P_r(\omega)$. Here, $P_{\text{corr}}(\omega)$ contains all particle-hole excitations that occur completely within the correlated subspace, while $P_r(\omega)$ contains all other excitations, i.e., excitations happening either entirely outside of the correlated subspace or between the correlated subspace and the rest of the Hilbert space~\cite{aryasetiawan_frequencydependent_2004}.
The partially screened interaction, felt by the electrons within the correlated subspace, is then obtained as $W_r(\omega) = V/[1-P_r(\omega)V]$. Thus, the screening represented by $P_\text{corr}$ is excluded from $W_r$, since the interaction within the correlated subspace is treated explicitly on the DMFT level.

The effective interaction parameters for the correlated subspace ($U$, $J$, etc.) are then obtained by calculating the matrix elements of $W_r(\omega=0)$ within the correlated subspace basis and applying suitable averaging.
%
To obtain the interaction parameters $\mathcal{U}_d$ and $\mathcal{J}_d$ for \lunio, corresponding to the Slater parametrization of the local interaction, we average the full orbital-resolved interaction matrix obtained from cRPA as described  in \cite{merkel_calculation_2024}. 

\subsection{\label{ssec:comp_det} Computational details }

For the DFT part of our DFT+DMFT calculations we use Quantum Espresso version 7.2 \cite{giannozzi_advanced_2017}, employing ultrasoft pseudopotentials from the GBRV library~\cite{garrity_pseudopotentials_2014} and using the PBE exchange-correlation functional~\cite{perdew_generalized_1996}. For all materials, we use a 20 atom unit cell corresponding to a $\sqrt{2} \times \sqrt{2} \times 2$ super-cell of the underlying pseudo-cubic perovskite structure. This cell can accommodate both the GdFeO$_3$-like octahedral rotations and the $R_{1}^{+}$ breathing mode. For \latio and \lavo, we use relaxed crystal structures obtained from spin-degenerate PBE.
For \lunio, we follow \cite{hampel_interplay_2017a} and relax the structure within $Pbnm$ symmetry (using spin-degenerate PBE). We then add the $R^+_1$ breathing mode with varying mode amplitude.
For \nno, we use the experimental structure obtained in~\cite{garcia-munoz_structure_2009}. 
All calculations are performed on a $7 \times 7 \times 5$ $\mathbf{k}$-point grid. 

Wannier90 is used to construct maximally localized Wannier functions as basis orbitals representing the correlated subspace
, as outlined in \pref{ssec:dft_dmft}.
DFT+DMFT calculations are carried out using solid\_dmft \cite{merkel_solid_dmft_2022, beck_charge_2022}, which is part of the TRIQS ecosystem\cite{parcollet_triqs_2015}, for a paramagnetic state at an inverse temperature of $\beta =$ 40 eV$^{-1}$. All DFT+DMFT calculations are fully charge-self-consistent except for the phase diagrams shown in the appendix. 
%
The effective impurity problems on the TM sites are solved using the continuous-time hybridization expansion quantum Monte Carlo impurity solver (CT-HYB)~\cite{seth_triqs_2016}, except for the cases with the five $d$ orbital localized basis on the Ni sites for \lunio and \nno. 
Here, to reduce the computational effort, we only consider the density-density terms of $H_\text{int}$ and solve the Ni impurity problem using the more efficient segment picture continuous-time impurity solver (CT-SEG)~\cite{triqs_ctseg}. 
The O $p$ impurity problem is solved using the more approximate Hartree-Fock impurity solver \cite{triqs_hartree_fock}.
Before solving any impurity problem, we rotate the local Hamiltonian to minimize any off-diagonal elements in the hybridization function, which could cause convergence problems for the Monte Carlo solvers.
Real frequency spectral functions are obtained by analytic continuation of the impurity self-energies using MaxEnt~\cite{levy_implementation_2017}.
Additional details about the specific parameters of our DFT+DMFT calculations will be made available together with the input files in the Materials Cloud database~\cite{talirz_materials_2020}.

For the cRPA calculations we use RESPACK~\cite{nakamura_respack_2021}. Since RESPACK does not allow to use ultrasoft pseudopotentials, we use norm-conserving pseudopotentials from the PseudoDojo library in this case~\cite{vansetten_pseudodojo_2018}. 
The cRPA calculations are performed on a $5 \times 5 \times 3$ $\mathbf{k}$-grid with a polarization function cutoff of 20\,Ry.

\section*{Data availability statement}
The datasets generated and/or analysed during the current study are available in the Materials Cloud Database~\cite{MaterialsCloudArchive2024}. 

\section*{Competing interests}
All authors declare no financial or non-financial competing interests. 

\begin{acknowledgments}
This work was funded by ETH Z\"urich and a grant from the Swiss National Supercomputing Centre (CSCS) under project ID s1128. The computations were carried out on the \enquote{Eiger} cluster at CSCS and the \enquote{Euler} cluster at ETH Zurich. We are grateful to Maximilian E. Merkel, Peter Mlkvik, Alexander Hampel and Sophie Beck for insightful discussions.
\end{acknowledgments}

\appendix

\section{One-shot phase diagrams for \latio and \lavo}
\label{app:one-shot}

\begin{figure*}
   \centering
   \includegraphics[width=0.8\textwidth]{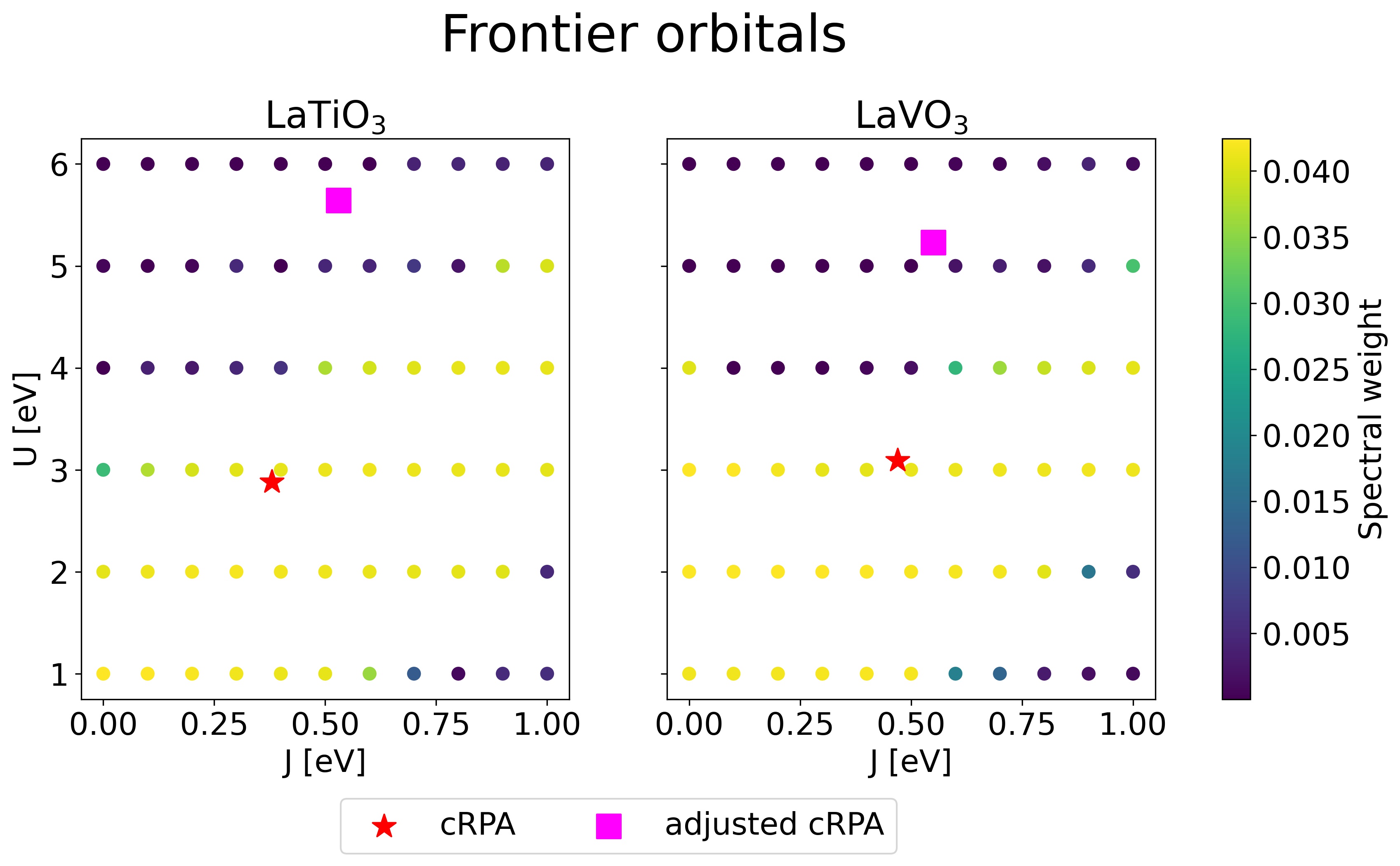}
   \caption{2D Map of the spectral weight, $A(\omega=0)$,  obtained from one-shot frontier basis DFT+DMFT calculations as a function of $U$ and $J$ for \latio and \lavo. The magenta square and the red star correspond to the computed cRPA values of the interaction parameters with and without the empirical correction applied to the surrounding Kohn-Sham states, respectively.  
   }
   \label{fig:phase_diagram_frontier}
\end{figure*}

\begin{figure*}
   \centering
   \includegraphics[width=0.8\textwidth]{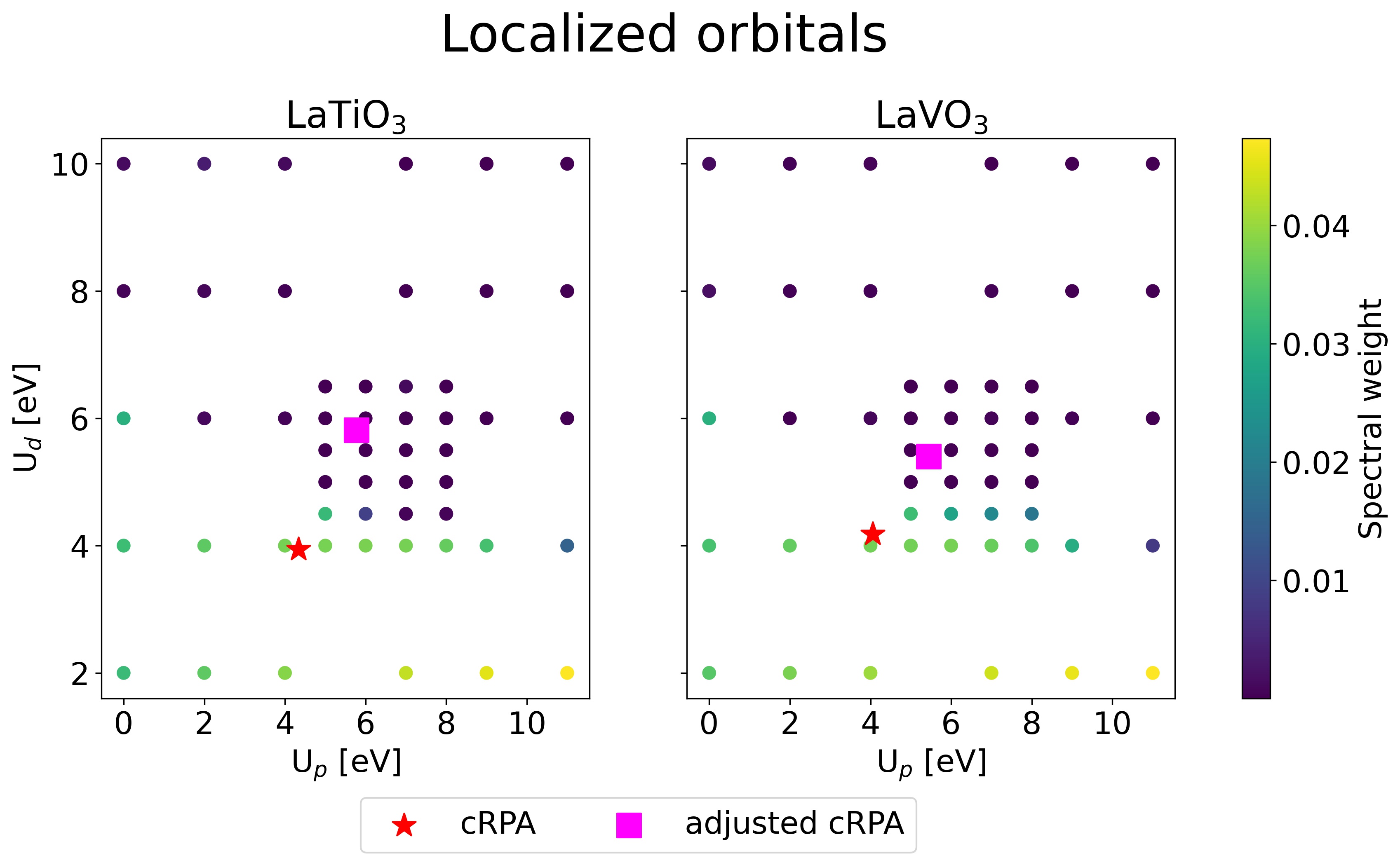}
   \caption{2D Map of the spectral weight, $A(\omega=0)$,  obtained from one-shot localized basis DFT+DMFT calculations as a function of $U_d$ and $U_p$, fixing $J_d$ to 0.65\,eV, for both \latio and \lavo. The magenta square and the red star correspond to the computed cRPA values of the interaction parameters with and without the empirical correction applied to the empty higher-lying Kohn-Sham states, respectively.  
   }
   \label{fig:phase_diagram_localized}
\end{figure*}

Here, we provide additional metal-insulator phase diagrams as function of the interaction parameters for \latio\ and \lavo\ using both the frontier (\pref{fig:phase_diagram_frontier}) and the localized (\pref{fig:phase_diagram_localized}) basis sets. In the frontier basis, we vary $U$ and $J$, while in the localized basis, we vary $U_d$ and $U_p$, fixing $J_d=0.65$\,eV. 
Due to the significant computational effort required, we perform \textit{one-shot} DFT+DMFT calculations rather than fully charge self-consistent (CSC) DFT+DMFT. According to our tests, one-shot and CSC results for \latio\ and \lavo show only minor differences, mostly regarding the onset of the insulating solution, with CSC slightly favoring the metallic state. For example, in \pref{fig:phase_diagram_localized}, \latio\ becomes insulating at $U_d=6$\,eV and $U_p=2$\,eV, whereas in the CSC results presented in \pref{fig:lanthanates_spectral_functions}, the system is still metallic for these interaction parameters. Practically, this implies that both \pref{fig:phase_diagram_frontier} and \pref{fig:phase_diagram_localized} would exhibit only minor variations under CSC conditions, with a slightly larger stability region for the metallic phase.
We note that in the case of the nickelates, the DMFT charge correction is more severe, and CSC and one-shot result (not shown here) can differ significantly.

\section{The insulating state of \lunio from different viewpoints \label{sec:lunio_comparison}}
\label{app:lunio3}

The insulating behavior of LuNiO$_3$ and the other rare earth nickelates has been given different interpretations. In the charge disproportionation picture one considers a charge ordering on the Ni $e_g$ states: $2 e_g^1 \rightarrow (e_g^2)_\text{LB} + (e_g^0)_\text{SB}$, driven by Hund's coupling $J$ \cite{medarde_charge_2009a, medarde_longrange_2008, subedi_lowenergy_2015, hampel_energetics_2019, peil_mechanism_2019}. This leads to a site-selective Mott state, where the half-filled $(e_g^2)_\text{LB}$ site undergoes a Mott transition while the empty $(e_g^0)_\text{SB}$ site remains a band insulator. This picture corresponds to a description in the frontier basis.

\begin{figure}
   \centering
   \includegraphics[width=\columnwidth]{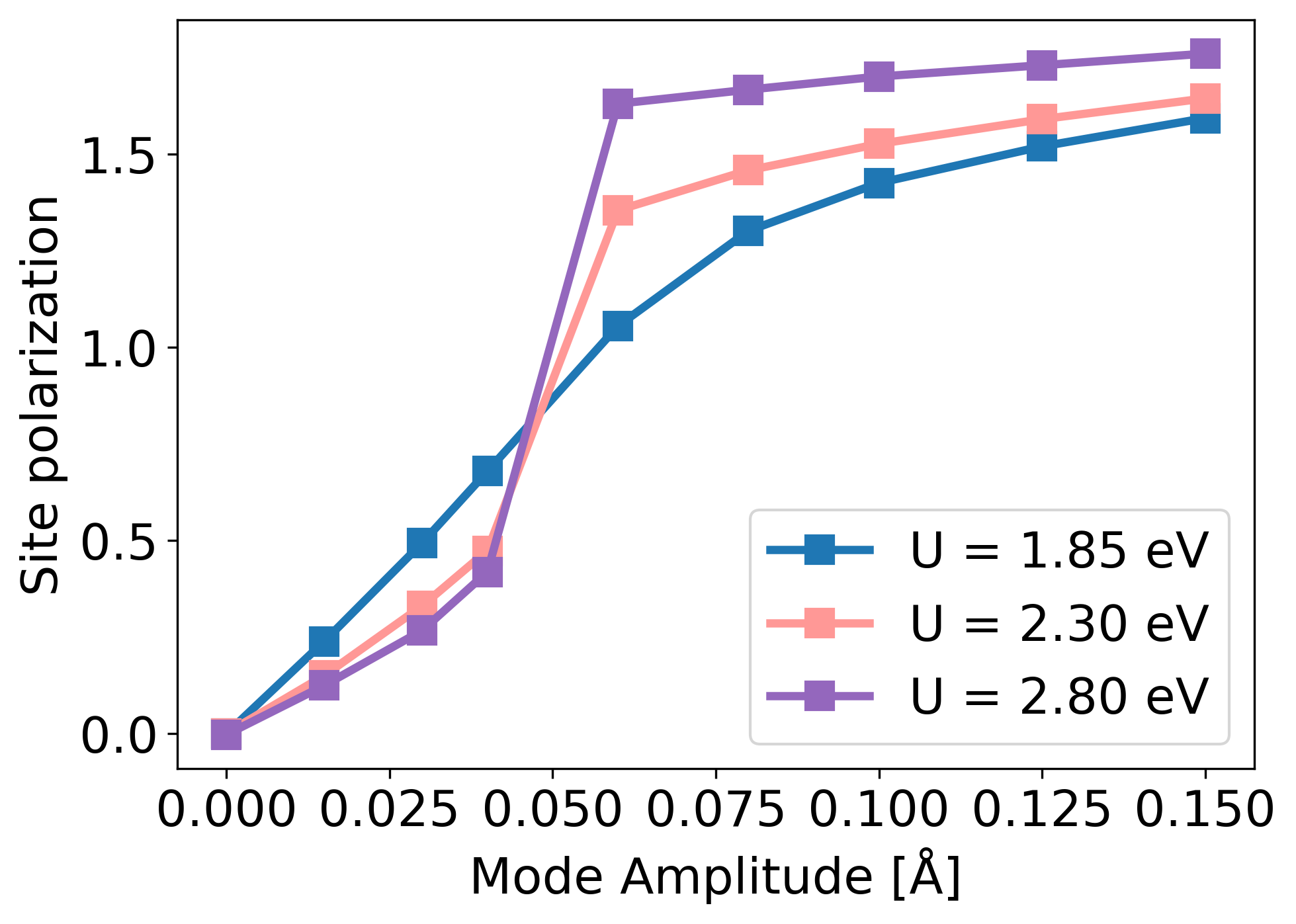}
   \caption{Site polarization (difference in occupation between inequivalent Ni sites) of \lunio as a function of the $R_1^+$ breathing mode, obtained using the frontier orbital basis. The calculations are performed with $J = 0.42$ eV.
   \label{fig:lno_site_pol}
   }
\end{figure}

In \pref{fig:lno_site_pol} we show the ``site polarization'', i.e., the occupation difference between the LB and SB site in the frontier basis. By construction, the average occupation per Ni site corresponds to a formal valence of $d^7$ (i.e., Ni$^{3+}$) or $e_g^1$, and the structural distortion leads to a very pronounced electronic disproportionation.
The metal-insulator transition as shown in \pref{fig:lno_energy_curve}(b) coincides with a jump of the site polarization to $N_\text{LB}-N_\text{SB} \approx 1.5$, corresponding to an LB occupations of about 1.75, which allows a Mott gap to open, while the SB site occupation is severely reduced. These calculations mirror the results described in~\cite{peil_mechanism_2019}, albeit with full charge self-consistency.

Alternatively, the ``ligand-hole'' picture~\cite{mizokawa_spin_2000,  park_siteselective_2012, johnston_charge_2014, haule_mott_2017} acknowledges the potential negative charge transfer character of the nickelates and assumes an average Ni valence of approximately $d^8$, with an additional ligand hole $L$ shared among the oxygen atoms surrounding the Ni sites. During the metal-insulator transition, these ligand-holes order as $2(d^8L) \rightarrow (d^8)_\text{LB} + (d^8L^2)_\text{SB}$, again with a site-selective Mott state on the LB site~\cite{park_computing_2014, park_siteselective_2012, haule_mott_2017} and a simple band-insulating state with singlet formation on the SB site. This picture is consistent with the occupations obtained within the localized Wannier basis set (see \pref{fig:nickelates_spectral_functions} and the corresponding discussion in \pref{ssec:localized_results}).

\begin{figure*}
   \centering
   \includegraphics[width=0.8\textwidth]{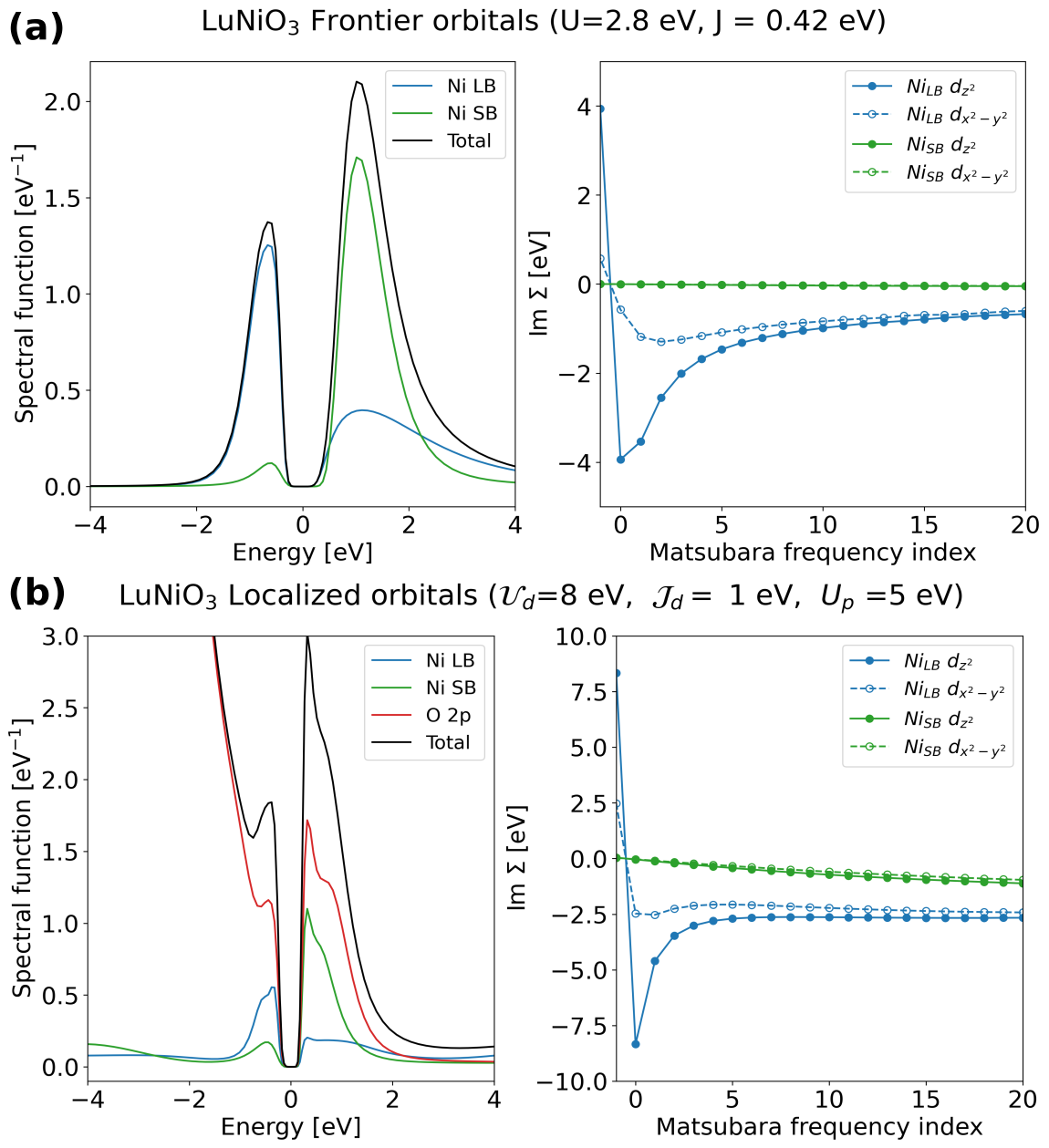}
   \caption{Impurity and total spectral functions and imaginary part of the self energy computed for the insulating state of \lunio at an $R_{1}^+$ mode amplitude of $0.08$\,\AA{} for (a) frontier orbital basis ($U=2.8$\,eV, $J=0.43$\,eV) and (b) localized orbital basis ($\mathcal{U}_d=8$\,eV, $\mathcal{J}_d=1$\,eV, $U_p=5$\,eV).} 
   \label{fig:lno_comparison_frontier_localized}
\end{figure*}

Here, we want to stress that the two apparently different pictures presented above in fact describe essentially the same physics, just from different viewpoints.
To demonstrate this, we plot in \pref{fig:lno_comparison_frontier_localized} the \lunio impurity spectral functions obtained by analytical continuation \cite{levy_implementation_2017} together with the imaginary part of the self energy $\text{Im}[\Sigma(\omega)]$ of the $e_g$ orbitals for both frontier (a) and localized (b) basis. The corresponding interaction parameters are chosen such that an insulating solution is obtained in bith cases.
One can observe that in both cases the qualitative features of the spectral functions and self-energies are rather similar. 
The spectral functions exhibit a similar relative distribution of the Ni LB and SB character in the low energy region of about $\pm 2$\,eV above and below the gap, even though for the case using localized orbitals the corresponding integrated weight is much smaller.
Furthermore, the self-energies of the $e_g$ orbitals on the Ni LB site exhibit pronounced feature with a steep slope of $\text{Im}[\Sigma(i\omega_n)]$ towards low Matsubara frequencies, indicative of a Mott state, while the self energies on the Ni SB site show very little frequency dependence, compatible with an essentially uncorrelated band-insulating character. Thus, the characteristics of the insulating stare are very similar in both basis sets, indicating the general equivalence of the charge disproportionation and ligand hole picture. 

\clearpage



\bibliography{final}

\begin{thebibliography}{91}%
\makeatletter
\providecommand \@ifxundefined [1]{%
 \@ifx{#1\undefined}
}%
\providecommand \@ifnum [1]{%
 \ifnum #1\expandafter \@firstoftwo
 \else \expandafter \@secondoftwo
 \fi
}%
\providecommand \@ifx [1]{%
 \ifx #1\expandafter \@firstoftwo
 \else \expandafter \@secondoftwo
 \fi
}%
\providecommand \natexlab [1]{#1}%
\providecommand \enquote  [1]{``#1''}%
\providecommand \bibnamefont  [1]{#1}%
\providecommand \bibfnamefont [1]{#1}%
\providecommand \citenamefont [1]{#1}%
\providecommand \href@noop [0]{\@secondoftwo}%
\providecommand \href [0]{\begingroup \@sanitize@url \@href}%
\providecommand \@href[1]{\@@startlink{#1}\@@href}%
\providecommand \@@href[1]{\endgroup#1\@@endlink}%
\providecommand \@sanitize@url [0]{\catcode `\\12\catcode `\$12\catcode
  `\&12\catcode `\#12\catcode `\^12\catcode `\_12\catcode `\%12\relax}%
\providecommand \@@startlink[1]{}%
\providecommand \@@endlink[0]{}%
\providecommand \url  [0]{\begingroup\@sanitize@url \@url }%
\providecommand \@url [1]{\endgroup\@href {#1}{\urlprefix }}%
\providecommand \urlprefix  [0]{URL }%
\providecommand \Eprint [0]{\href }%
\providecommand \doibase [0]{https://doi.org/}%
\providecommand \selectlanguage [0]{\@gobble}%
\providecommand \bibinfo  [0]{\@secondoftwo}%
\providecommand \bibfield  [0]{\@secondoftwo}%
\providecommand \translation [1]{[#1]}%
\providecommand \BibitemOpen [0]{}%
\providecommand \bibitemStop [0]{}%
\providecommand \bibitemNoStop [0]{.\EOS\space}%
\providecommand \EOS [0]{\spacefactor3000\relax}%
\providecommand \BibitemShut  [1]{\csname bibitem#1\endcsname}%
\let\auto@bib@innerbib\@empty
\bibitem [{\citenamefont {Fazekas}(1999)}]{Fazekas:1999}%
  \BibitemOpen
  \bibfield  {author} {\bibinfo {author} {\bibfnamefont {P.}~\bibnamefont
  {Fazekas}},\ }\href {https://doi.org/10.1142/2945} {\emph {\bibinfo {title}
  {Lecture {{Notes}} on {{Electron Correlation}} and {{Magnetism}}}}}\
  (\bibinfo  {publisher} {World Scientific},\ \bibinfo {year}
  {1999})\BibitemShut {NoStop}%
\bibitem [{\citenamefont {Imada}\ \emph {et~al.}(1998)\citenamefont {Imada},
  \citenamefont {Fujimori},\ and\ \citenamefont
  {Tokura}}]{imada_metalinsulator_1998}%
  \BibitemOpen
  \bibfield  {author} {\bibinfo {author} {\bibfnamefont {M.}~\bibnamefont
  {Imada}}, \bibinfo {author} {\bibfnamefont {A.}~\bibnamefont {Fujimori}},\
  and\ \bibinfo {author} {\bibfnamefont {Y.}~\bibnamefont {Tokura}},\
  }\bibfield  {title} {\bibinfo {title} {Metal-insulator transitions},\ }\href
  {https://doi.org/10.1103/RevModPhys.70.1039} {\bibfield  {journal} {\bibinfo
  {journal} {Reviews of Modern Physics}\ }\textbf {\bibinfo {volume} {70}},\
  \bibinfo {pages} {1039} (\bibinfo {year} {1998})}\BibitemShut {NoStop}%
\bibitem [{\citenamefont {Roy}(2019)}]{roy_mott_2019}%
  \BibitemOpen
  \bibfield  {author} {\bibinfo {author} {\bibfnamefont {S.~B.}\ \bibnamefont
  {Roy}},\ }\href@noop {} {\emph {\bibinfo {title} {Mott {{Insulators}}:
  {{Physics}} and Applications}}}\ (\bibinfo  {publisher} {IOP Publishing},\
  \bibinfo {year} {2019})\BibitemShut {NoStop}%
\bibitem [{\citenamefont {Newns}\ \emph {et~al.}(1998)\citenamefont {Newns},
  \citenamefont {Misewich}, \citenamefont {Tsuei}, \citenamefont {Gupta},
  \citenamefont {Scott},\ and\ \citenamefont {Schrott}}]{Newns_et_al:1998}%
  \BibitemOpen
  \bibfield  {author} {\bibinfo {author} {\bibfnamefont {D.~M.}\ \bibnamefont
  {Newns}}, \bibinfo {author} {\bibfnamefont {J.~A.}\ \bibnamefont {Misewich}},
  \bibinfo {author} {\bibfnamefont {C.~C.}\ \bibnamefont {Tsuei}}, \bibinfo
  {author} {\bibfnamefont {A.}~\bibnamefont {Gupta}}, \bibinfo {author}
  {\bibfnamefont {B.~A.}\ \bibnamefont {Scott}},\ and\ \bibinfo {author}
  {\bibfnamefont {A.}~\bibnamefont {Schrott}},\ }\bibfield  {title} {\bibinfo
  {title} {Mott transition field effect transistor},\ }\href
  {https://doi.org/10.1063/1.121999} {\bibfield  {journal} {\bibinfo  {journal}
  {Applied Physics Letters}\ }\textbf {\bibinfo {volume} {73}},\ \bibinfo
  {pages} {780} (\bibinfo {year} {1998})}\BibitemShut {NoStop}%
\bibitem [{\citenamefont {Inoue}\ and\ \citenamefont
  {Rozenberg}(2008)}]{Inoue/Rozenberg:2008}%
  \BibitemOpen
  \bibfield  {author} {\bibinfo {author} {\bibfnamefont {I.~H.}\ \bibnamefont
  {Inoue}}\ and\ \bibinfo {author} {\bibfnamefont {M.~J.}\ \bibnamefont
  {Rozenberg}},\ }\bibfield  {title} {\bibinfo {title} {Taming the {{Mott}}
  transition for a novel {{Mott}} transistor},\ }\href
  {https://doi.org/10.1002/adfm.200800558} {\bibfield  {journal} {\bibinfo
  {journal} {Advanced Functional Materials}\ }\textbf {\bibinfo {volume}
  {18}},\ \bibinfo {pages} {2289} (\bibinfo {year} {2008})}\BibitemShut
  {NoStop}%
\bibitem [{\citenamefont {Mannhart}\ and\ \citenamefont
  {Haensch}(2012)}]{Mannhart/Haensch:2012}%
  \BibitemOpen
  \bibfield  {author} {\bibinfo {author} {\bibfnamefont {J.}~\bibnamefont
  {Mannhart}}\ and\ \bibinfo {author} {\bibfnamefont {W.}~\bibnamefont
  {Haensch}},\ }\bibfield  {title} {\bibinfo {title} {Put the pedal to the
  metal},\ }\href {https://doi.org/10.1038/487436a} {\bibfield  {journal}
  {\bibinfo  {journal} {Nature}\ }\textbf {\bibinfo {volume} {487}},\ \bibinfo
  {pages} {436} (\bibinfo {year} {2012})}\BibitemShut {NoStop}%
\bibitem [{\citenamefont {Manousakis}(2010)}]{manousakis_photovoltaic_2010a}%
  \BibitemOpen
  \bibfield  {author} {\bibinfo {author} {\bibfnamefont {E.}~\bibnamefont
  {Manousakis}},\ }\bibfield  {title} {\bibinfo {title} {Photovoltaic effect
  for narrow-gap {{Mott}} insulators},\ }\href
  {https://doi.org/10.1103/PhysRevB.82.125109} {\bibfield  {journal} {\bibinfo
  {journal} {Physical Review B}\ }\textbf {\bibinfo {volume} {82}},\ \bibinfo
  {pages} {125109} (\bibinfo {year} {2010})}\BibitemShut {NoStop}%
\bibitem [{\citenamefont {Assmann}\ \emph {et~al.}(2013)\citenamefont
  {Assmann}, \citenamefont {Blaha}, \citenamefont {Laskowski}, \citenamefont
  {Held}, \citenamefont {Okamoto},\ and\ \citenamefont
  {Sangiovanni}}]{assmann_oxide_2013}%
  \BibitemOpen
  \bibfield  {author} {\bibinfo {author} {\bibfnamefont {E.}~\bibnamefont
  {Assmann}}, \bibinfo {author} {\bibfnamefont {P.}~\bibnamefont {Blaha}},
  \bibinfo {author} {\bibfnamefont {R.}~\bibnamefont {Laskowski}}, \bibinfo
  {author} {\bibfnamefont {K.}~\bibnamefont {Held}}, \bibinfo {author}
  {\bibfnamefont {S.}~\bibnamefont {Okamoto}},\ and\ \bibinfo {author}
  {\bibfnamefont {G.}~\bibnamefont {Sangiovanni}},\ }\bibfield  {title}
  {\bibinfo {title} {Oxide {{Heterostructures}} for {{Efficient Solar
  Cells}}},\ }\href {https://doi.org/10.1103/PhysRevLett.110.078701} {\bibfield
   {journal} {\bibinfo  {journal} {Physical Review Letters}\ }\textbf {\bibinfo
  {volume} {110}},\ \bibinfo {pages} {078701} (\bibinfo {year}
  {2013})}\BibitemShut {NoStop}%
\bibitem [{\citenamefont {Coulter}\ \emph {et~al.}(2014)\citenamefont
  {Coulter}, \citenamefont {Manousakis},\ and\ \citenamefont
  {Gali}}]{coulter_optoelectronic_2014}%
  \BibitemOpen
  \bibfield  {author} {\bibinfo {author} {\bibfnamefont {J.~E.}\ \bibnamefont
  {Coulter}}, \bibinfo {author} {\bibfnamefont {E.}~\bibnamefont
  {Manousakis}},\ and\ \bibinfo {author} {\bibfnamefont {A.}~\bibnamefont
  {Gali}},\ }\bibfield  {title} {\bibinfo {title} {Optoelectronic excitations
  and photovoltaic effect in strongly correlated materials},\ }\href
  {https://doi.org/10.1103/PhysRevB.90.165142} {\bibfield  {journal} {\bibinfo
  {journal} {Physical Review B}\ }\textbf {\bibinfo {volume} {90}},\ \bibinfo
  {pages} {165142} (\bibinfo {year} {2014})}\BibitemShut {NoStop}%
\bibitem [{\citenamefont {Wang}\ \emph {et~al.}(2015)\citenamefont {Wang},
  \citenamefont {Li}, \citenamefont {Bera}, \citenamefont {Ma}, \citenamefont
  {Jin}, \citenamefont {Yuan}, \citenamefont {Yin}, \citenamefont {David},
  \citenamefont {Chen}, \citenamefont {Wu}, \citenamefont {Prellier},
  \citenamefont {Wei},\ and\ \citenamefont {Wu}}]{wang_device_2015}%
  \BibitemOpen
  \bibfield  {author} {\bibinfo {author} {\bibfnamefont {L.}~\bibnamefont
  {Wang}}, \bibinfo {author} {\bibfnamefont {Y.}~\bibnamefont {Li}}, \bibinfo
  {author} {\bibfnamefont {A.}~\bibnamefont {Bera}}, \bibinfo {author}
  {\bibfnamefont {C.}~\bibnamefont {Ma}}, \bibinfo {author} {\bibfnamefont
  {F.}~\bibnamefont {Jin}}, \bibinfo {author} {\bibfnamefont {K.}~\bibnamefont
  {Yuan}}, \bibinfo {author} {\bibfnamefont {W.}~\bibnamefont {Yin}}, \bibinfo
  {author} {\bibfnamefont {A.}~\bibnamefont {David}}, \bibinfo {author}
  {\bibfnamefont {W.}~\bibnamefont {Chen}}, \bibinfo {author} {\bibfnamefont
  {W.}~\bibnamefont {Wu}}, \bibinfo {author} {\bibfnamefont {W.}~\bibnamefont
  {Prellier}}, \bibinfo {author} {\bibfnamefont {S.}~\bibnamefont {Wei}},\ and\
  \bibinfo {author} {\bibfnamefont {T.}~\bibnamefont {Wu}},\ }\bibfield
  {title} {\bibinfo {title} {Device {{Performance}} of the {{Mott Insulator
  LaVO3}} as a {{Photovoltaic Material}}},\ }\href
  {https://doi.org/10.1103/PhysRevApplied.3.064015} {\bibfield  {journal}
  {\bibinfo  {journal} {Physical Review Applied}\ }\textbf {\bibinfo {volume}
  {3}},\ \bibinfo {pages} {064015} (\bibinfo {year} {2015})}\BibitemShut
  {NoStop}%
\bibitem [{\citenamefont {Petocchi}\ \emph {et~al.}(2019)\citenamefont
  {Petocchi}, \citenamefont {Beck}, \citenamefont {Ederer},\ and\ \citenamefont
  {Werner}}]{petocchi_hund_2019}%
  \BibitemOpen
  \bibfield  {author} {\bibinfo {author} {\bibfnamefont {F.}~\bibnamefont
  {Petocchi}}, \bibinfo {author} {\bibfnamefont {S.}~\bibnamefont {Beck}},
  \bibinfo {author} {\bibfnamefont {C.}~\bibnamefont {Ederer}},\ and\ \bibinfo
  {author} {\bibfnamefont {P.}~\bibnamefont {Werner}},\ }\bibfield  {title}
  {\bibinfo {title} {Hund excitations and the efficiency of {{Mott}} solar
  cells},\ }\href {https://doi.org/10.1103/PhysRevB.100.075147} {\bibfield
  {journal} {\bibinfo  {journal} {Physical Review B}\ }\textbf {\bibinfo
  {volume} {100}},\ \bibinfo {pages} {075147} (\bibinfo {year}
  {2019})}\BibitemShut {NoStop}%
\bibitem [{\citenamefont {Kotliar}\ \emph {et~al.}(2006)\citenamefont
  {Kotliar}, \citenamefont {Savrasov}, \citenamefont {Haule}, \citenamefont
  {Oudovenko}, \citenamefont {Parcollet},\ and\ \citenamefont
  {Marianetti}}]{kotliar_electronic_2006}%
  \BibitemOpen
  \bibfield  {author} {\bibinfo {author} {\bibfnamefont {G.}~\bibnamefont
  {Kotliar}}, \bibinfo {author} {\bibfnamefont {S.~Y.}\ \bibnamefont
  {Savrasov}}, \bibinfo {author} {\bibfnamefont {K.}~\bibnamefont {Haule}},
  \bibinfo {author} {\bibfnamefont {V.~S.}\ \bibnamefont {Oudovenko}}, \bibinfo
  {author} {\bibfnamefont {O.}~\bibnamefont {Parcollet}},\ and\ \bibinfo
  {author} {\bibfnamefont {C.~A.}\ \bibnamefont {Marianetti}},\ }\bibfield
  {title} {\bibinfo {title} {Electronic structure calculations with dynamical
  mean-field theory},\ }\href {https://doi.org/irreducib} {\bibfield  {journal}
  {\bibinfo  {journal} {Reviews of Modern Physics}\ }\textbf {\bibinfo {volume}
  {78}},\ \bibinfo {pages} {865} (\bibinfo {year} {2006})}\BibitemShut
  {NoStop}%
\bibitem [{\citenamefont {Held}(2007)}]{held_electronic_2007}%
  \BibitemOpen
  \bibfield  {author} {\bibinfo {author} {\bibfnamefont {K.}~\bibnamefont
  {Held}},\ }\bibfield  {title} {\bibinfo {title} {Electronic structure
  calculations using dynamical mean field theory},\ }\href
  {https://doi.org/10.1080/00018730701619647} {\bibfield  {journal} {\bibinfo
  {journal} {Advances in Physics}\ }\textbf {\bibinfo {volume} {56}},\ \bibinfo
  {pages} {829} (\bibinfo {year} {2007})}\BibitemShut {NoStop}%
\bibitem [{\citenamefont {Kune{\v s}}\ \emph {et~al.}(2010)\citenamefont
  {Kune{\v s}}, \citenamefont {Leonov}, \citenamefont {Kollar}, \citenamefont
  {Byczuk}, \citenamefont {Anisimov},\ and\ \citenamefont
  {Vollhardt}}]{Kunes_et_al:2010}%
  \BibitemOpen
  \bibfield  {author} {\bibinfo {author} {\bibfnamefont {J.}~\bibnamefont
  {Kune{\v s}}}, \bibinfo {author} {\bibfnamefont {I.}~\bibnamefont {Leonov}},
  \bibinfo {author} {\bibfnamefont {M.}~\bibnamefont {Kollar}}, \bibinfo
  {author} {\bibfnamefont {K.}~\bibnamefont {Byczuk}}, \bibinfo {author}
  {\bibfnamefont {V.~I.}\ \bibnamefont {Anisimov}},\ and\ \bibinfo {author}
  {\bibfnamefont {D.}~\bibnamefont {Vollhardt}},\ }\bibfield  {title} {\bibinfo
  {title} {Dynamical mean-field approach to materials with strong electronic
  correlations},\ }\href {https://doi.org/10.1140/epjst/e2010-01209-0}
  {\bibfield  {journal} {\bibinfo  {journal} {The European Physical Journal
  Special Topics}\ }\textbf {\bibinfo {volume} {180}},\ \bibinfo {pages} {5}
  (\bibinfo {year} {2010})}\BibitemShut {NoStop}%
\bibitem [{\citenamefont {Paul}\ and\ \citenamefont
  {Birol}(2019)}]{Paul/Birol:2019}%
  \BibitemOpen
  \bibfield  {author} {\bibinfo {author} {\bibfnamefont {A.}~\bibnamefont
  {Paul}}\ and\ \bibinfo {author} {\bibfnamefont {T.}~\bibnamefont {Birol}},\
  }\bibfield  {title} {\bibinfo {title} {Applications of {{DFT}} + {{DMFT}} in
  {{Materials Science}}},\ }\href
  {https://doi.org/10.1146/annurev-matsci-070218-121825} {\bibfield  {journal}
  {\bibinfo  {journal} {Annual Review of Materials Research}\ }\textbf
  {\bibinfo {volume} {49}},\ \bibinfo {pages} {31} (\bibinfo {year}
  {2019})}\BibitemShut {NoStop}%
\bibitem [{\citenamefont {Kov{\'a}{\v c}ik}\ \emph {et~al.}(2012)\citenamefont
  {Kov{\'a}{\v c}ik}, \citenamefont {Werner}, \citenamefont {Dymkowski},\ and\
  \citenamefont {Ederer}}]{Kovacik_et_al:2012}%
  \BibitemOpen
  \bibfield  {author} {\bibinfo {author} {\bibfnamefont {R.}~\bibnamefont
  {Kov{\'a}{\v c}ik}}, \bibinfo {author} {\bibfnamefont {P.}~\bibnamefont
  {Werner}}, \bibinfo {author} {\bibfnamefont {K.}~\bibnamefont {Dymkowski}},\
  and\ \bibinfo {author} {\bibfnamefont {C.}~\bibnamefont {Ederer}},\
  }\bibfield  {title} {\bibinfo {title} {Rubidium superoxide: {{A}} p-electron
  mott insulator},\ }\href {https://doi.org/10.1103/PhysRevB.86.075130}
  {\bibfield  {journal} {\bibinfo  {journal} {Physical Review B}\ }\textbf
  {\bibinfo {volume} {86}},\ \bibinfo {pages} {075130} (\bibinfo {year}
  {2012})}\BibitemShut {NoStop}%
\bibitem [{\citenamefont {Ferber}\ \emph {et~al.}(2014)\citenamefont {Ferber},
  \citenamefont {Foyevtsova}, \citenamefont {Jeschke},\ and\ \citenamefont
  {Valent{\'i}}}]{Ferber_et_al:2014}%
  \BibitemOpen
  \bibfield  {author} {\bibinfo {author} {\bibfnamefont {J.}~\bibnamefont
  {Ferber}}, \bibinfo {author} {\bibfnamefont {K.}~\bibnamefont {Foyevtsova}},
  \bibinfo {author} {\bibfnamefont {H.~O.}\ \bibnamefont {Jeschke}},\ and\
  \bibinfo {author} {\bibfnamefont {R.}~\bibnamefont {Valent{\'i}}},\
  }\bibfield  {title} {\bibinfo {title} {Unveiling the microscopic nature of
  correlated organic conductors: {{The}} case of
  {$\kappa$}-({{ET}})2{{Cu}}[{{N}}({{CN}})2]{{Br}}$_x${{Cl}}$_{1 - x}$},\
  }\href {https://doi.org/10.1103/PhysRevB.89.205106} {\bibfield  {journal}
  {\bibinfo  {journal} {Physical Review B}\ }\textbf {\bibinfo {volume} {89}},\
  \bibinfo {pages} {205106} (\bibinfo {year} {2014})}\BibitemShut {NoStop}%
\bibitem [{\citenamefont {Mlkvik}\ \emph {et~al.}(2024)\citenamefont {Mlkvik},
  \citenamefont {Merkel}, \citenamefont {Spaldin},\ and\ \citenamefont
  {Ederer}}]{Mlkvik_et_al:2024}%
  \BibitemOpen
  \bibfield  {author} {\bibinfo {author} {\bibfnamefont {P.}~\bibnamefont
  {Mlkvik}}, \bibinfo {author} {\bibfnamefont {M.~E.}\ \bibnamefont {Merkel}},
  \bibinfo {author} {\bibfnamefont {N.~A.}\ \bibnamefont {Spaldin}},\ and\
  \bibinfo {author} {\bibfnamefont {C.}~\bibnamefont {Ederer}},\ }\bibfield
  {title} {\bibinfo {title} {Single-site {{DFT}+{DMFT}} for vanadium dioxide
  using bond-centered orbitals},\ }\href
  {https://doi.org/10.1103/PhysRevResearch.6.033122} {\bibfield  {journal}
  {\bibinfo  {journal} {Phys. Rev. Res.}\ }\textbf {\bibinfo {volume} {6}},\
  \bibinfo {pages} {033122} (\bibinfo {year} {2024})}\BibitemShut {NoStop}%
\bibitem [{\citenamefont {Grytsiuk}\ \emph {et~al.}(2024)\citenamefont
  {Grytsiuk}, \citenamefont {Katsnelson}, \citenamefont {{van Loon}},\ and\
  \citenamefont {Rösner}}]{grytsiuk_nb3cl8_2024}%
  \BibitemOpen
  \bibfield  {author} {\bibinfo {author} {\bibfnamefont {S.}~\bibnamefont
  {Grytsiuk}}, \bibinfo {author} {\bibfnamefont {M.~I.}\ \bibnamefont
  {Katsnelson}}, \bibinfo {author} {\bibfnamefont {E.~G. C.~P.}\ \bibnamefont
  {{van Loon}}},\ and\ \bibinfo {author} {\bibfnamefont {M.}~\bibnamefont
  {Rösner}},\ }\bibfield  {title} {\bibinfo {title} {{{Nb$_3$Cl$_8$}}: A
  prototypical layered {{Mott-Hubbard}} insulator},\ }\href
  {https://doi.org/10.1038/s41535-024-00619-5} {\bibfield  {journal} {\bibinfo
  {journal} {npj Quantum Mater.}\ }\textbf {\bibinfo {volume} {9}},\ \bibinfo
  {pages} {1} (\bibinfo {year} {2024})}\BibitemShut {NoStop}%
\bibitem [{\citenamefont {{Souto-Casares}}\ \emph {et~al.}(2019)\citenamefont
  {{Souto-Casares}}, \citenamefont {Spaldin},\ and\ \citenamefont
  {Ederer}}]{SoutoCasares/Spaldin/Ederer:2019}%
  \BibitemOpen
  \bibfield  {author} {\bibinfo {author} {\bibfnamefont {J.}~\bibnamefont
  {{Souto-Casares}}}, \bibinfo {author} {\bibfnamefont {N.~A.}\ \bibnamefont
  {Spaldin}},\ and\ \bibinfo {author} {\bibfnamefont {C.}~\bibnamefont
  {Ederer}},\ }\bibfield  {title} {\bibinfo {title} {{{DFT}}+{{DMFT}} study of
  oxygen vacancies in a {{Mott}} insulator},\ }\href
  {https://doi.org/10.1103/physrevb.100.085146} {\bibfield  {journal} {\bibinfo
   {journal} {Physical Review B}\ }\textbf {\bibinfo {volume} {100}},\ \bibinfo
  {pages} {085146} (\bibinfo {year} {2019})}\BibitemShut {NoStop}%
\bibitem [{\citenamefont {{Souto-Casares}}\ \emph {et~al.}(2021)\citenamefont
  {{Souto-Casares}}, \citenamefont {Spaldin},\ and\ \citenamefont
  {Ederer}}]{souto-casares_oxygen_2021a}%
  \BibitemOpen
  \bibfield  {author} {\bibinfo {author} {\bibfnamefont {J.}~\bibnamefont
  {{Souto-Casares}}}, \bibinfo {author} {\bibfnamefont {N.~A.}\ \bibnamefont
  {Spaldin}},\ and\ \bibinfo {author} {\bibfnamefont {C.}~\bibnamefont
  {Ederer}},\ }\bibfield  {title} {\bibinfo {title} {Oxygen vacancies in
  strontium titanate: {{A DFT}}+{{DMFT}} study},\ }\href
  {https://doi.org/10.1103/PhysRevResearch.3.023027} {\bibfield  {journal}
  {\bibinfo  {journal} {Physical Review Research}\ }\textbf {\bibinfo {volume}
  {3}},\ \bibinfo {pages} {023027} (\bibinfo {year} {2021})}\BibitemShut
  {NoStop}%
\bibitem [{\citenamefont {Marzari}\ \emph {et~al.}(2012)\citenamefont
  {Marzari}, \citenamefont {Mostofi}, \citenamefont {Yates}, \citenamefont
  {Souza},\ and\ \citenamefont {Vanderbilt}}]{marzari_maximally_2012}%
  \BibitemOpen
  \bibfield  {author} {\bibinfo {author} {\bibfnamefont {N.}~\bibnamefont
  {Marzari}}, \bibinfo {author} {\bibfnamefont {A.~A.}\ \bibnamefont
  {Mostofi}}, \bibinfo {author} {\bibfnamefont {J.~R.}\ \bibnamefont {Yates}},
  \bibinfo {author} {\bibfnamefont {I.}~\bibnamefont {Souza}},\ and\ \bibinfo
  {author} {\bibfnamefont {D.}~\bibnamefont {Vanderbilt}},\ }\bibfield  {title}
  {\bibinfo {title} {Maximally localized {{Wannier}} functions: {{Theory}} and
  applications},\ }\href {https://doi.org/10.1103/revmodphys.84.1419}
  {\bibfield  {journal} {\bibinfo  {journal} {Reviews of Modern Physics}\
  }\textbf {\bibinfo {volume} {84}},\ \bibinfo {pages} {1419} (\bibinfo {year}
  {2012})}\BibitemShut {NoStop}%
\bibitem [{\citenamefont {Anisimov}\ \emph {et~al.}(2005)\citenamefont
  {Anisimov}, \citenamefont {Kondakov}, \citenamefont {Kozhevnikov},
  \citenamefont {Nekrasov}, \citenamefont {Pchelkina}, \citenamefont {Allen},
  \citenamefont {Mo}, \citenamefont {Kim}, \citenamefont {Metcalf},
  \citenamefont {Suga}, \citenamefont {Sekiyama}, \citenamefont {Keller},
  \citenamefont {Leonov}, \citenamefont {Ren},\ and\ \citenamefont
  {Vollhardt}}]{anisimov_full_2005}%
  \BibitemOpen
  \bibfield  {author} {\bibinfo {author} {\bibfnamefont {V.~I.}\ \bibnamefont
  {Anisimov}}, \bibinfo {author} {\bibfnamefont {D.~E.}\ \bibnamefont
  {Kondakov}}, \bibinfo {author} {\bibfnamefont {A.~V.}\ \bibnamefont
  {Kozhevnikov}}, \bibinfo {author} {\bibfnamefont {I.~A.}\ \bibnamefont
  {Nekrasov}}, \bibinfo {author} {\bibfnamefont {Z.~V.}\ \bibnamefont
  {Pchelkina}}, \bibinfo {author} {\bibfnamefont {J.~W.}\ \bibnamefont
  {Allen}}, \bibinfo {author} {\bibfnamefont {S.-K.}\ \bibnamefont {Mo}},
  \bibinfo {author} {\bibfnamefont {H.-D.}\ \bibnamefont {Kim}}, \bibinfo
  {author} {\bibfnamefont {P.}~\bibnamefont {Metcalf}}, \bibinfo {author}
  {\bibfnamefont {S.}~\bibnamefont {Suga}}, \bibinfo {author} {\bibfnamefont
  {A.}~\bibnamefont {Sekiyama}}, \bibinfo {author} {\bibfnamefont
  {G.}~\bibnamefont {Keller}}, \bibinfo {author} {\bibfnamefont
  {I.}~\bibnamefont {Leonov}}, \bibinfo {author} {\bibfnamefont
  {X.}~\bibnamefont {Ren}},\ and\ \bibinfo {author} {\bibfnamefont
  {D.}~\bibnamefont {Vollhardt}},\ }\bibfield  {title} {\bibinfo {title} {Full
  orbital calculation scheme for materials with strongly correlated
  electrons},\ }\bibfield  {journal} {\bibinfo  {journal} {Physical Review B}\
  }\textbf {\bibinfo {volume} {71}},\ \href
  {https://doi.org/10.1103/physrevb.71.125119} {10.1103/physrevb.71.125119}
  (\bibinfo {year} {2005})\BibitemShut {NoStop}%
\bibitem [{\citenamefont {Lechermann}\ \emph {et~al.}(2006)\citenamefont
  {Lechermann}, \citenamefont {Georges}, \citenamefont {Poteryaev},
  \citenamefont {Biermann}, \citenamefont {Posternak}, \citenamefont
  {Yamasaki},\ and\ \citenamefont {Andersen}}]{lechermann_dynamical_2006}%
  \BibitemOpen
  \bibfield  {author} {\bibinfo {author} {\bibfnamefont {F.}~\bibnamefont
  {Lechermann}}, \bibinfo {author} {\bibfnamefont {A.}~\bibnamefont {Georges}},
  \bibinfo {author} {\bibfnamefont {A.}~\bibnamefont {Poteryaev}}, \bibinfo
  {author} {\bibfnamefont {S.}~\bibnamefont {Biermann}}, \bibinfo {author}
  {\bibfnamefont {M.}~\bibnamefont {Posternak}}, \bibinfo {author}
  {\bibfnamefont {A.}~\bibnamefont {Yamasaki}},\ and\ \bibinfo {author}
  {\bibfnamefont {O.~K.}\ \bibnamefont {Andersen}},\ }\bibfield  {title}
  {\bibinfo {title} {Dynamical mean-field theory using {{Wannier}} functions:
  {{A}} flexible route to electronic structure calculations of strongly
  correlated materials},\ }\bibfield  {journal} {\bibinfo  {journal} {Physical
  Review B}\ }\textbf {\bibinfo {volume} {74}},\ \href
  {https://doi.org/10.1103/PhysRevB.74.125120} {10.1103/PhysRevB.74.125120}
  (\bibinfo {year} {2006})\BibitemShut {NoStop}%
\bibitem [{\citenamefont {Beck}\ \emph {et~al.}(2022)\citenamefont {Beck},
  \citenamefont {Hampel}, \citenamefont {Parcollet}, \citenamefont {Ederer},\
  and\ \citenamefont {Georges}}]{beck_charge_2022}%
  \BibitemOpen
  \bibfield  {author} {\bibinfo {author} {\bibfnamefont {S.}~\bibnamefont
  {Beck}}, \bibinfo {author} {\bibfnamefont {A.}~\bibnamefont {Hampel}},
  \bibinfo {author} {\bibfnamefont {O.}~\bibnamefont {Parcollet}}, \bibinfo
  {author} {\bibfnamefont {C.}~\bibnamefont {Ederer}},\ and\ \bibinfo {author}
  {\bibfnamefont {A.}~\bibnamefont {Georges}},\ }\bibfield  {title} {\bibinfo
  {title} {Charge self-consistent electronic structure calculations with
  dynamical mean-field theory using {{Quantum ESPRESSO}}, {{Wannier}} 90 and
  {{TRIQS}}},\ }\href {https://doi.org/10.1088/1361-648X/ac5d1c} {\bibfield
  {journal} {\bibinfo  {journal} {Journal of Physics: Condensed Matter}\
  }\textbf {\bibinfo {volume} {34}},\ \bibinfo {pages} {235601} (\bibinfo
  {year} {2022})}\BibitemShut {NoStop}%
\bibitem [{\citenamefont {Pourovskii}\ \emph {et~al.}(2007)\citenamefont
  {Pourovskii}, \citenamefont {Amadon}, \citenamefont {Biermann},\ and\
  \citenamefont {Georges}}]{pourovskii_selfconsistency_2007}%
  \BibitemOpen
  \bibfield  {author} {\bibinfo {author} {\bibfnamefont {L.~V.}\ \bibnamefont
  {Pourovskii}}, \bibinfo {author} {\bibfnamefont {B.}~\bibnamefont {Amadon}},
  \bibinfo {author} {\bibfnamefont {S.}~\bibnamefont {Biermann}},\ and\
  \bibinfo {author} {\bibfnamefont {A.}~\bibnamefont {Georges}},\ }\bibfield
  {title} {\bibinfo {title} {Self-consistency over the charge density in
  dynamical mean-field theory: {{A}} linear muffin-tin implementation and some
  physical implications},\ }\href {https://doi.org/10.1103/PhysRevB.76.235101}
  {\bibfield  {journal} {\bibinfo  {journal} {Physical Review B}\ }\textbf
  {\bibinfo {volume} {76}},\ \bibinfo {pages} {235101} (\bibinfo {year}
  {2007})}\BibitemShut {NoStop}%
\bibitem [{\citenamefont {Haule}\ \emph {et~al.}(2014)\citenamefont {Haule},
  \citenamefont {Birol},\ and\ \citenamefont {Kotliar}}]{haule_covalency_2014}%
  \BibitemOpen
  \bibfield  {author} {\bibinfo {author} {\bibfnamefont {K.}~\bibnamefont
  {Haule}}, \bibinfo {author} {\bibfnamefont {T.}~\bibnamefont {Birol}},\ and\
  \bibinfo {author} {\bibfnamefont {G.}~\bibnamefont {Kotliar}},\ }\bibfield
  {title} {\bibinfo {title} {Covalency in {{Transition-Metal Oxides}} within
  {{All-Electron Dynamical Mean-Field Theory}}},\ }\href
  {https://doi.org/10.1103/physrevb.90.075136} {\bibfield  {journal} {\bibinfo
  {journal} {Physical Review B}\ }\textbf {\bibinfo {volume} {90}},\ \bibinfo
  {pages} {075136} (\bibinfo {year} {2014})}\BibitemShut {NoStop}%
\bibitem [{\citenamefont {Zaanen}\ \emph {et~al.}(1985)\citenamefont {Zaanen},
  \citenamefont {Sawatzky},\ and\ \citenamefont
  {Allen}}]{Zaanen/Sawatzky/Allen:1985}%
  \BibitemOpen
  \bibfield  {author} {\bibinfo {author} {\bibfnamefont {J.}~\bibnamefont
  {Zaanen}}, \bibinfo {author} {\bibfnamefont {G.~A.}\ \bibnamefont
  {Sawatzky}},\ and\ \bibinfo {author} {\bibfnamefont {J.~W.}\ \bibnamefont
  {Allen}},\ }\bibfield  {title} {\bibinfo {title} {Band gaps and electronic
  structure of transition-metal compounds},\ }\href
  {https://doi.org/10.1103/PhysRevLett.55.418} {\bibfield  {journal} {\bibinfo
  {journal} {Physical Review Letters}\ }\textbf {\bibinfo {volume} {55}},\
  \bibinfo {pages} {418} (\bibinfo {year} {1985})}\BibitemShut {NoStop}%
\bibitem [{\citenamefont {Hansmann}\ \emph {et~al.}(2014)\citenamefont
  {Hansmann}, \citenamefont {Parragh}, \citenamefont {Toschi}, \citenamefont
  {Sangiovanni},\ and\ \citenamefont {Held}}]{hansmann_importance_2014}%
  \BibitemOpen
  \bibfield  {author} {\bibinfo {author} {\bibfnamefont {P.}~\bibnamefont
  {Hansmann}}, \bibinfo {author} {\bibfnamefont {N.}~\bibnamefont {Parragh}},
  \bibinfo {author} {\bibfnamefont {A.}~\bibnamefont {Toschi}}, \bibinfo
  {author} {\bibfnamefont {G.}~\bibnamefont {Sangiovanni}},\ and\ \bibinfo
  {author} {\bibfnamefont {K.}~\bibnamefont {Held}},\ }\bibfield  {title}
  {\bibinfo {title} {Importance of d–p {{Coulomb}} interaction for high
  {{TC}} cuprates and other oxides},\ }\href
  {https://doi.org/10.1088/1367-2630/16/3/033009} {\bibfield  {journal}
  {\bibinfo  {journal} {New Journal of Physics}\ }\textbf {\bibinfo {volume}
  {16}},\ \bibinfo {pages} {033009} (\bibinfo {year} {2014})}\BibitemShut
  {NoStop}%
\bibitem [{\citenamefont {Parragh}\ \emph {et~al.}(2013)\citenamefont
  {Parragh}, \citenamefont {Sangiovanni}, \citenamefont {Hansmann},
  \citenamefont {Hummel}, \citenamefont {Held},\ and\ \citenamefont
  {Toschi}}]{parragh_effective_2013}%
  \BibitemOpen
  \bibfield  {author} {\bibinfo {author} {\bibfnamefont {N.}~\bibnamefont
  {Parragh}}, \bibinfo {author} {\bibfnamefont {G.}~\bibnamefont
  {Sangiovanni}}, \bibinfo {author} {\bibfnamefont {P.}~\bibnamefont
  {Hansmann}}, \bibinfo {author} {\bibfnamefont {S.}~\bibnamefont {Hummel}},
  \bibinfo {author} {\bibfnamefont {K.}~\bibnamefont {Held}},\ and\ \bibinfo
  {author} {\bibfnamefont {A.}~\bibnamefont {Toschi}},\ }\bibfield  {title}
  {\bibinfo {title} {Effective crystal field and {{Fermi}} surface topology:
  {{A}} comparison of d- and dp-orbital models},\ }\href
  {https://doi.org/10.1103/PhysRevB.88.195116} {\bibfield  {journal} {\bibinfo
  {journal} {Physical Review B}\ }\textbf {\bibinfo {volume} {88}},\ \bibinfo
  {pages} {195116} (\bibinfo {year} {2013})}\BibitemShut {NoStop}%
\bibitem [{\citenamefont {Dang}\ \emph
  {et~al.}(2014{\natexlab{a}})\citenamefont {Dang}, \citenamefont {Millis},\
  and\ \citenamefont {Marianetti}}]{dang_covalency_2014}%
  \BibitemOpen
  \bibfield  {author} {\bibinfo {author} {\bibfnamefont {H.~T.}\ \bibnamefont
  {Dang}}, \bibinfo {author} {\bibfnamefont {A.~J.}\ \bibnamefont {Millis}},\
  and\ \bibinfo {author} {\bibfnamefont {C.~A.}\ \bibnamefont {Marianetti}},\
  }\bibfield  {title} {\bibinfo {title} {Covalency and the metal-insulator
  transition in titanate and vanadate perovskites},\ }\href
  {https://doi.org/10.1103/PhysRevB.89.161113} {\bibfield  {journal} {\bibinfo
  {journal} {Physical Review B}\ }\textbf {\bibinfo {volume} {89}},\ \bibinfo
  {pages} {161113} (\bibinfo {year} {2014}{\natexlab{a}})}\BibitemShut
  {NoStop}%
\bibitem [{\citenamefont {Dang}\ \emph
  {et~al.}(2014{\natexlab{b}})\citenamefont {Dang}, \citenamefont {Ai},
  \citenamefont {Millis},\ and\ \citenamefont
  {Marianetti}}]{dang_density_2014}%
  \BibitemOpen
  \bibfield  {author} {\bibinfo {author} {\bibfnamefont {H.~T.}\ \bibnamefont
  {Dang}}, \bibinfo {author} {\bibfnamefont {X.}~\bibnamefont {Ai}}, \bibinfo
  {author} {\bibfnamefont {A.~J.}\ \bibnamefont {Millis}},\ and\ \bibinfo
  {author} {\bibfnamefont {C.~A.}\ \bibnamefont {Marianetti}},\ }\bibfield
  {title} {\bibinfo {title} {Density functional plus dynamical mean-field
  theory of the metal-insulator transition in early transition-metal oxides},\
  }\href {https://doi.org/10.1103/PhysRevB.90.125114} {\bibfield  {journal}
  {\bibinfo  {journal} {Physical Review B}\ }\textbf {\bibinfo {volume} {90}},\
  \bibinfo {pages} {125114} (\bibinfo {year} {2014}{\natexlab{b}})}\BibitemShut
  {NoStop}%
\bibitem [{\citenamefont {Amaricci}\ \emph {et~al.}(2017)\citenamefont
  {Amaricci}, \citenamefont {de’‌ Medici},\ and\ \citenamefont
  {Capone}}]{amaricci_mott_2017}%
  \BibitemOpen
  \bibfield  {author} {\bibinfo {author} {\bibfnamefont {A.}~\bibnamefont
  {Amaricci}}, \bibinfo {author} {\bibfnamefont {L.}~\bibnamefont {de’‌
  Medici}},\ and\ \bibinfo {author} {\bibfnamefont {M.}~\bibnamefont
  {Capone}},\ }\bibfield  {title} {\bibinfo {title} {Mott transitions with
  partially filled correlated orbitals},\ }\href
  {https://doi.org/10.1209/0295-5075/118/17004} {\bibfield  {journal} {\bibinfo
   {journal} {Europhysics Letters}\ }\textbf {\bibinfo {volume} {118}},\
  \bibinfo {pages} {17004} (\bibinfo {year} {2017})}\BibitemShut {NoStop}%
\bibitem [{\citenamefont {Lechermann}\ \emph {et~al.}(2019)\citenamefont
  {Lechermann}, \citenamefont {Körner}, \citenamefont {Urban},\ and\
  \citenamefont {Elsässer}}]{lechermann_interplay_2019}%
  \BibitemOpen
  \bibfield  {author} {\bibinfo {author} {\bibfnamefont {F.}~\bibnamefont
  {Lechermann}}, \bibinfo {author} {\bibfnamefont {W.}~\bibnamefont {Körner}},
  \bibinfo {author} {\bibfnamefont {D.~F.}\ \bibnamefont {Urban}},\ and\
  \bibinfo {author} {\bibfnamefont {C.}~\bibnamefont {Elsässer}},\ }\bibfield
  {title} {\bibinfo {title} {Interplay of charge-transfer and {{Mott-Hubbard}}
  physics approached by an efficient combination of self-interaction correction
  and dynamical mean-field theory},\ }\href
  {https://doi.org/10.1103/PhysRevB.100.115125} {\bibfield  {journal} {\bibinfo
   {journal} {Physical Review B}\ }\textbf {\bibinfo {volume} {100}},\ \bibinfo
  {pages} {115125} (\bibinfo {year} {2019})}\BibitemShut {NoStop}%
\bibitem [{\citenamefont {Lechermann}(2024)}]{lechermann_oxygen_2024}%
  \BibitemOpen
  \bibfield  {author} {\bibinfo {author} {\bibfnamefont {F.}~\bibnamefont
  {Lechermann}},\ }\href {https://doi.org/10.48550/arXiv.2410.06891} {\bibinfo
  {title} {On the oxygen $p$ states in superconducting nickelates}} (\bibinfo
  {year} {2024}),\ \Eprint {https://arxiv.org/abs/2410.06891}
  {arXiv:2410.06891} \BibitemShut {NoStop}%
\bibitem [{\citenamefont {Aryasetiawan}\ \emph {et~al.}(2004)\citenamefont
  {Aryasetiawan}, \citenamefont {Imada}, \citenamefont {Georges}, \citenamefont
  {Kotliar}, \citenamefont {Biermann},\ and\ \citenamefont
  {Lichtenstein}}]{aryasetiawan_frequencydependent_2004}%
  \BibitemOpen
  \bibfield  {author} {\bibinfo {author} {\bibfnamefont {F.}~\bibnamefont
  {Aryasetiawan}}, \bibinfo {author} {\bibfnamefont {M.}~\bibnamefont {Imada}},
  \bibinfo {author} {\bibfnamefont {A.}~\bibnamefont {Georges}}, \bibinfo
  {author} {\bibfnamefont {G.}~\bibnamefont {Kotliar}}, \bibinfo {author}
  {\bibfnamefont {S.}~\bibnamefont {Biermann}},\ and\ \bibinfo {author}
  {\bibfnamefont {A.~I.}\ \bibnamefont {Lichtenstein}},\ }\bibfield  {title}
  {\bibinfo {title} {Frequency-dependent local interactions and low-energy
  effective models from electronic structure calculations},\ }\href
  {https://doi.org/10.1103/PhysRevB.70.195104} {\bibfield  {journal} {\bibinfo
  {journal} {Physical Review B}\ }\textbf {\bibinfo {volume} {70}},\ \bibinfo
  {pages} {195104} (\bibinfo {year} {2004})}\BibitemShut {NoStop}%
\bibitem [{\citenamefont {Miyake}\ and\ \citenamefont
  {Aryasetiawan}(2008)}]{miyake_screened_2008}%
  \BibitemOpen
  \bibfield  {author} {\bibinfo {author} {\bibfnamefont {T.}~\bibnamefont
  {Miyake}}\ and\ \bibinfo {author} {\bibfnamefont {F.}~\bibnamefont
  {Aryasetiawan}},\ }\bibfield  {title} {\bibinfo {title} {Screened {{Coulomb}}
  interaction in the maximally localized {{Wannier}} basis},\ }\href
  {https://doi.org/10.1103/PhysRevB.77.085122} {\bibfield  {journal} {\bibinfo
  {journal} {Physical Review B}\ }\textbf {\bibinfo {volume} {77}},\ \bibinfo
  {pages} {085122} (\bibinfo {year} {2008})}\BibitemShut {NoStop}%
\bibitem [{\citenamefont {Miyake}\ \emph {et~al.}(2009)\citenamefont {Miyake},
  \citenamefont {Aryasetiawan},\ and\ \citenamefont
  {Imada}}]{miyake_initio_2009}%
  \BibitemOpen
  \bibfield  {author} {\bibinfo {author} {\bibfnamefont {T.}~\bibnamefont
  {Miyake}}, \bibinfo {author} {\bibfnamefont {F.}~\bibnamefont
  {Aryasetiawan}},\ and\ \bibinfo {author} {\bibfnamefont {M.}~\bibnamefont
  {Imada}},\ }\bibfield  {title} {\bibinfo {title} {Ab initio procedure for
  constructing effective models of correlated materials with entangled band
  structure},\ }\href {https://doi.org/10.1103/PhysRevB.80.155134} {\bibfield
  {journal} {\bibinfo  {journal} {Physical Review B}\ }\textbf {\bibinfo
  {volume} {80}},\ \bibinfo {pages} {155134} (\bibinfo {year}
  {2009})}\BibitemShut {NoStop}%
\bibitem [{\citenamefont {Hampel}\ \emph {et~al.}(2019)\citenamefont {Hampel},
  \citenamefont {Liu}, \citenamefont {Franchini},\ and\ \citenamefont
  {Ederer}}]{hampel_energetics_2019}%
  \BibitemOpen
  \bibfield  {author} {\bibinfo {author} {\bibfnamefont {A.}~\bibnamefont
  {Hampel}}, \bibinfo {author} {\bibfnamefont {P.}~\bibnamefont {Liu}},
  \bibinfo {author} {\bibfnamefont {C.}~\bibnamefont {Franchini}},\ and\
  \bibinfo {author} {\bibfnamefont {C.}~\bibnamefont {Ederer}},\ }\bibfield
  {title} {\bibinfo {title} {Energetics of the coupled electronic–structural
  transition in the rare-earth nickelates},\ }\href@noop {} {\bibfield
  {journal} {\bibinfo  {journal} {npj Quantum Materials}\ }\textbf {\bibinfo
  {volume} {4}} (\bibinfo {year} {2019})}\BibitemShut {NoStop}%
\bibitem [{\citenamefont {{Kazemi-Moridani}}\ \emph {et~al.}(2024)\citenamefont
  {{Kazemi-Moridani}}, \citenamefont {Beck}, \citenamefont {Hampel},
  \citenamefont {Tremblay}, \citenamefont {Côté},\ and\ \citenamefont
  {Gingras}}]{kazemi-moridani_strontium_2024}%
  \BibitemOpen
  \bibfield  {author} {\bibinfo {author} {\bibfnamefont {A.}~\bibnamefont
  {{Kazemi-Moridani}}}, \bibinfo {author} {\bibfnamefont {S.}~\bibnamefont
  {Beck}}, \bibinfo {author} {\bibfnamefont {A.}~\bibnamefont {Hampel}},
  \bibinfo {author} {\bibfnamefont {A.-M.~S.}\ \bibnamefont {Tremblay}},
  \bibinfo {author} {\bibfnamefont {M.}~\bibnamefont {Côté}},\ and\ \bibinfo
  {author} {\bibfnamefont {O.}~\bibnamefont {Gingras}},\ }\bibfield  {title}
  {\bibinfo {title} {Strontium ferrite under pressure: {{Potential}} analog to
  strontium ruthenate},\ }\href {https://doi.org/10.1103/PhysRevB.109.165146}
  {\bibfield  {journal} {\bibinfo  {journal} {Physical Review B}\ }\textbf
  {\bibinfo {volume} {109}},\ \bibinfo {pages} {165146} (\bibinfo {year}
  {2024})}\BibitemShut {NoStop}%
\bibitem [{\citenamefont {Merkel}\ and\ \citenamefont
  {Ederer}(2024)}]{merkel_calculation_2024}%
  \BibitemOpen
  \bibfield  {author} {\bibinfo {author} {\bibfnamefont {M.~E.}\ \bibnamefont
  {Merkel}}\ and\ \bibinfo {author} {\bibfnamefont {C.}~\bibnamefont
  {Ederer}},\ }\bibfield  {title} {\bibinfo {title} {Calculation of screened
  {{Coulomb}} interaction parameters for the charge-disproportionated insulator
  {{CaFeO3}}},\ }\href {https://doi.org/10.1103/PhysRevResearch.6.013230}
  {\bibfield  {journal} {\bibinfo  {journal} {Physical Review Research}\
  }\textbf {\bibinfo {volume} {6}},\ \bibinfo {pages} {013230} (\bibinfo {year}
  {2024})}\BibitemShut {NoStop}%
\bibitem [{\citenamefont {Park}\ \emph
  {et~al.}(2014{\natexlab{a}})\citenamefont {Park}, \citenamefont {Millis},\
  and\ \citenamefont {Marianetti}}]{park_computing_2014}%
  \BibitemOpen
  \bibfield  {author} {\bibinfo {author} {\bibfnamefont {H.}~\bibnamefont
  {Park}}, \bibinfo {author} {\bibfnamefont {A.~J.}\ \bibnamefont {Millis}},\
  and\ \bibinfo {author} {\bibfnamefont {C.~A.}\ \bibnamefont {Marianetti}},\
  }\bibfield  {title} {\bibinfo {title} {Computing total energies in complex
  materials using charge self-consistent {{DFT}} + {{DMFT}}},\ }\href
  {https://doi.org/10.1103/PhysRevB.90.235103} {\bibfield  {journal} {\bibinfo
  {journal} {Physical Review B}\ }\textbf {\bibinfo {volume} {90}},\ \bibinfo
  {pages} {235103} (\bibinfo {year} {2014}{\natexlab{a}})}\BibitemShut
  {NoStop}%
\bibitem [{\citenamefont {Dougier}\ \emph {et~al.}(1976)\citenamefont
  {Dougier}, \citenamefont {Deglane},\ and\ \citenamefont
  {Hagenmuller}}]{dougier_evolution_1976}%
  \BibitemOpen
  \bibfield  {author} {\bibinfo {author} {\bibfnamefont {P.}~\bibnamefont
  {Dougier}}, \bibinfo {author} {\bibfnamefont {D.}~\bibnamefont {Deglane}},\
  and\ \bibinfo {author} {\bibfnamefont {P.}~\bibnamefont {Hagenmuller}},\
  }\bibfield  {title} {\bibinfo {title} {Evolution des propriétés
  structurales, electriques et magnétiques au sein du système
  la$_{1-x}$ca$_x$vo$_3$},\ }\href
  {https://doi.org/10.1016/0022-4596(76)90160-2} {\bibfield  {journal}
  {\bibinfo  {journal} {Journal of Solid State Chemistry}\ }\textbf {\bibinfo
  {volume} {19}},\ \bibinfo {pages} {135} (\bibinfo {year} {1976})}\BibitemShut
  {NoStop}%
\bibitem [{\citenamefont {Fujimori}\ \emph {et~al.}(1992)\citenamefont
  {Fujimori}, \citenamefont {Hase}, \citenamefont {Nakamura}, \citenamefont
  {Namatame}, \citenamefont {Fujishima}, \citenamefont {Tokura}, \citenamefont
  {Abbate}, \citenamefont {{de Groot}}, \citenamefont {Czyzyk}, \citenamefont
  {Fuggle}, \citenamefont {Strebel}, \citenamefont {Lopez}, \citenamefont
  {Domke},\ and\ \citenamefont {Kaindl}}]{fujimori_dopinginduced_1992}%
  \BibitemOpen
  \bibfield  {author} {\bibinfo {author} {\bibfnamefont {A.}~\bibnamefont
  {Fujimori}}, \bibinfo {author} {\bibfnamefont {I.}~\bibnamefont {Hase}},
  \bibinfo {author} {\bibfnamefont {M.}~\bibnamefont {Nakamura}}, \bibinfo
  {author} {\bibfnamefont {H.}~\bibnamefont {Namatame}}, \bibinfo {author}
  {\bibfnamefont {Y.}~\bibnamefont {Fujishima}}, \bibinfo {author}
  {\bibfnamefont {Y.}~\bibnamefont {Tokura}}, \bibinfo {author} {\bibfnamefont
  {M.}~\bibnamefont {Abbate}}, \bibinfo {author} {\bibfnamefont {F.~M.~F.}\
  \bibnamefont {{de Groot}}}, \bibinfo {author} {\bibfnamefont {M.~T.}\
  \bibnamefont {Czyzyk}}, \bibinfo {author} {\bibfnamefont {J.~C.}\
  \bibnamefont {Fuggle}}, \bibinfo {author} {\bibfnamefont {O.}~\bibnamefont
  {Strebel}}, \bibinfo {author} {\bibfnamefont {F.}~\bibnamefont {Lopez}},
  \bibinfo {author} {\bibfnamefont {M.}~\bibnamefont {Domke}},\ and\ \bibinfo
  {author} {\bibfnamefont {G.}~\bibnamefont {Kaindl}},\ }\bibfield  {title}
  {\bibinfo {title} {Doping-induced changes in the electronic structure of
  {{LaTiO$_3$}}: {{Limitation}} of the one-electron rigid-band model and the
  {{Hubbard}} model},\ }\href {https://doi.org/10.1103/PhysRevB.46.9841}
  {\bibfield  {journal} {\bibinfo  {journal} {Physical Review B}\ }\textbf
  {\bibinfo {volume} {46}},\ \bibinfo {pages} {9841} (\bibinfo {year}
  {1992})}\BibitemShut {NoStop}%
\bibitem [{\citenamefont {Maiti}\ and\ \citenamefont
  {Sarma}(2000)}]{maiti_spectroscopic_2000}%
  \BibitemOpen
  \bibfield  {author} {\bibinfo {author} {\bibfnamefont {K.}~\bibnamefont
  {Maiti}}\ and\ \bibinfo {author} {\bibfnamefont {D.~D.}\ \bibnamefont
  {Sarma}},\ }\bibfield  {title} {\bibinfo {title} {Spectroscopic
  investigations of the electronic structure and metal-insulator transitions in
  a {{Mott-Hubbard}} system la$_{1-x}$ca$_x$vo$_3$},\ }\href
  {https://doi.org/10.1103/PhysRevB.61.2525} {\bibfield  {journal} {\bibinfo
  {journal} {Physical Review B}\ }\textbf {\bibinfo {volume} {61}},\ \bibinfo
  {pages} {2525} (\bibinfo {year} {2000})}\BibitemShut {NoStop}%
\bibitem [{\citenamefont {Alonso}\ \emph {et~al.}(2001)\citenamefont {Alonso},
  \citenamefont {{Martínez-Lope}}, \citenamefont {Casais}, \citenamefont
  {{García-Muñoz}}, \citenamefont {{Fernández-Díaz}},\ and\ \citenamefont
  {Aranda}}]{alonso_hightemperature_2001}%
  \BibitemOpen
  \bibfield  {author} {\bibinfo {author} {\bibfnamefont {J.~A.}\ \bibnamefont
  {Alonso}}, \bibinfo {author} {\bibfnamefont {M.~J.}\ \bibnamefont
  {{Martínez-Lope}}}, \bibinfo {author} {\bibfnamefont {M.~T.}\ \bibnamefont
  {Casais}}, \bibinfo {author} {\bibfnamefont {J.~L.}\ \bibnamefont
  {{García-Muñoz}}}, \bibinfo {author} {\bibfnamefont {M.~T.}\ \bibnamefont
  {{Fernández-Díaz}}},\ and\ \bibinfo {author} {\bibfnamefont {M.~A.~G.}\
  \bibnamefont {Aranda}},\ }\bibfield  {title} {\bibinfo {title}
  {High-temperature structural evolution of {{RNiO$_3$}} ({{R}}={{Ho}}, {{Y}},
  {{Er}}, {{Lu}}) perovskites: {{Charge}} disproportionation and electronic
  localization},\ }\href {https://doi.org/10.1103/PhysRevB.64.094102}
  {\bibfield  {journal} {\bibinfo  {journal} {Physical Review B}\ }\textbf
  {\bibinfo {volume} {64}},\ \bibinfo {pages} {094102} (\bibinfo {year}
  {2001})}\BibitemShut {NoStop}%
\bibitem [{\citenamefont {Medarde}\ \emph {et~al.}(2008)\citenamefont
  {Medarde}, \citenamefont {{Fernández-Díaz}},\ and\ \citenamefont
  {Lacorre}}]{medarde_longrange_2008}%
  \BibitemOpen
  \bibfield  {author} {\bibinfo {author} {\bibfnamefont {M.}~\bibnamefont
  {Medarde}}, \bibinfo {author} {\bibfnamefont {M.~T.}\ \bibnamefont
  {{Fernández-Díaz}}},\ and\ \bibinfo {author} {\bibfnamefont {{\relax
  Ph}.}~\bibnamefont {Lacorre}},\ }\bibfield  {title} {\bibinfo {title}
  {Long-range charge order in the low-temperature insulating phase of
  {{PrNiO$_3$}}},\ }\href {https://doi.org/10.1103/PhysRevB.78.212101}
  {\bibfield  {journal} {\bibinfo  {journal} {Physical Review B}\ }\textbf
  {\bibinfo {volume} {78}},\ \bibinfo {pages} {212101} (\bibinfo {year}
  {2008})}\BibitemShut {NoStop}%
\bibitem [{\citenamefont {Medarde}\ \emph {et~al.}(2009)\citenamefont
  {Medarde}, \citenamefont {Dallera}, \citenamefont {Grioni}, \citenamefont
  {Delley}, \citenamefont {Vernay}, \citenamefont {Mesot}, \citenamefont
  {Sikora}, \citenamefont {Alonso},\ and\ \citenamefont
  {{Martínez-Lope}}}]{medarde_charge_2009a}%
  \BibitemOpen
  \bibfield  {author} {\bibinfo {author} {\bibfnamefont {M.}~\bibnamefont
  {Medarde}}, \bibinfo {author} {\bibfnamefont {C.}~\bibnamefont {Dallera}},
  \bibinfo {author} {\bibfnamefont {M.}~\bibnamefont {Grioni}}, \bibinfo
  {author} {\bibfnamefont {B.}~\bibnamefont {Delley}}, \bibinfo {author}
  {\bibfnamefont {F.}~\bibnamefont {Vernay}}, \bibinfo {author} {\bibfnamefont
  {J.}~\bibnamefont {Mesot}}, \bibinfo {author} {\bibfnamefont
  {M.}~\bibnamefont {Sikora}}, \bibinfo {author} {\bibfnamefont {J.~A.}\
  \bibnamefont {Alonso}},\ and\ \bibinfo {author} {\bibfnamefont {M.~J.}\
  \bibnamefont {{Martínez-Lope}}},\ }\bibfield  {title} {\bibinfo {title}
  {Charge disproportionation in {{RNiO$_3$}} perovskites ({{R}}=rare earth)
  from high-resolution x-ray absorption spectroscopy},\ }\href
  {https://doi.org/10.1103/PhysRevB.80.245105} {\bibfield  {journal} {\bibinfo
  {journal} {Physical Review B}\ }\textbf {\bibinfo {volume} {80}},\ \bibinfo
  {pages} {245105} (\bibinfo {year} {2009})}\BibitemShut {NoStop}%
\bibitem [{\citenamefont {Eitel}\ and\ \citenamefont
  {Greedan}(1986)}]{eitel_high_1986}%
  \BibitemOpen
  \bibfield  {author} {\bibinfo {author} {\bibfnamefont {M.}~\bibnamefont
  {Eitel}}\ and\ \bibinfo {author} {\bibfnamefont {J.~E.}\ \bibnamefont
  {Greedan}},\ }\bibfield  {title} {\bibinfo {title} {A high resolution neutron
  diffraction study of the perovskite {{LaTiO$_3$}}},\ }\href
  {https://doi.org/10.1016/0022-5088(86)90220-1} {\bibfield  {journal}
  {\bibinfo  {journal} {Journal of the Less Common Metals}\ }\textbf {\bibinfo
  {volume} {116}},\ \bibinfo {pages} {95} (\bibinfo {year} {1986})}\BibitemShut
  {NoStop}%
\bibitem [{\citenamefont {Khan}\ \emph {et~al.}(2004)\citenamefont {Khan},
  \citenamefont {Bashir}, \citenamefont {Iqbal},\ and\ \citenamefont
  {Khan}}]{khan_crystal_2004}%
  \BibitemOpen
  \bibfield  {author} {\bibinfo {author} {\bibfnamefont {R.~T.~A.}\
  \bibnamefont {Khan}}, \bibinfo {author} {\bibfnamefont {J.}~\bibnamefont
  {Bashir}}, \bibinfo {author} {\bibfnamefont {N.}~\bibnamefont {Iqbal}},\ and\
  \bibinfo {author} {\bibfnamefont {M.~N.}\ \bibnamefont {Khan}},\ }\bibfield
  {title} {\bibinfo {title} {Crystal structure of {{LaVO$_3$}} by {{Rietveld}}
  refinement method},\ }\href {https://doi.org/10.1016/j.matlet.2003.10.059}
  {\bibfield  {journal} {\bibinfo  {journal} {Materials Letters}\ }\textbf
  {\bibinfo {volume} {58}},\ \bibinfo {pages} {1737} (\bibinfo {year}
  {2004})}\BibitemShut {NoStop}%
\bibitem [{\citenamefont {Pavarini}\ \emph {et~al.}(2004)\citenamefont
  {Pavarini}, \citenamefont {Biermann}, \citenamefont {Poteryaev},
  \citenamefont {Lichtenstein}, \citenamefont {Georges},\ and\ \citenamefont
  {Andersen}}]{pavarini_mott_2004}%
  \BibitemOpen
  \bibfield  {author} {\bibinfo {author} {\bibfnamefont {E.}~\bibnamefont
  {Pavarini}}, \bibinfo {author} {\bibfnamefont {S.}~\bibnamefont {Biermann}},
  \bibinfo {author} {\bibfnamefont {A.}~\bibnamefont {Poteryaev}}, \bibinfo
  {author} {\bibfnamefont {A.~I.}\ \bibnamefont {Lichtenstein}}, \bibinfo
  {author} {\bibfnamefont {A.}~\bibnamefont {Georges}},\ and\ \bibinfo {author}
  {\bibfnamefont {O.~K.}\ \bibnamefont {Andersen}},\ }\bibfield  {title}
  {\bibinfo {title} {Mott {{Transition}} and {{Suppression}} of {{Orbital
  Fluctuations}} in {{Orthorhombic}} 3d1 {{Perovskites}}},\ }\href
  {https://doi.org/10.1103/PhysRevLett.92.176403} {\bibfield  {journal}
  {\bibinfo  {journal} {Physical Review Letters}\ }\textbf {\bibinfo {volume}
  {92}},\ \bibinfo {pages} {176403} (\bibinfo {year} {2004})}\BibitemShut
  {NoStop}%
\bibitem [{\citenamefont {De~Raychaudhury}\ \emph {et~al.}(2007)\citenamefont
  {De~Raychaudhury}, \citenamefont {Pavarini},\ and\ \citenamefont
  {Andersen}}]{deraychaudhury_orbital_2007}%
  \BibitemOpen
  \bibfield  {author} {\bibinfo {author} {\bibfnamefont {M.}~\bibnamefont
  {De~Raychaudhury}}, \bibinfo {author} {\bibfnamefont {E.}~\bibnamefont
  {Pavarini}},\ and\ \bibinfo {author} {\bibfnamefont {O.~K.}\ \bibnamefont
  {Andersen}},\ }\bibfield  {title} {\bibinfo {title} {Orbital {{Fluctuations}}
  in the {{Different Phases}} of {{LaVO3}} and {{YVO3}}},\ }\href
  {https://doi.org/10.1103/PhysRevLett.99.126402} {\bibfield  {journal}
  {\bibinfo  {journal} {Physical Review Letters}\ }\textbf {\bibinfo {volume}
  {99}},\ \bibinfo {pages} {126402} (\bibinfo {year} {2007})}\BibitemShut
  {NoStop}%
\bibitem [{\citenamefont {Dymkowski}\ and\ \citenamefont
  {Ederer}(2014)}]{dymkowski_straininduced_2014}%
  \BibitemOpen
  \bibfield  {author} {\bibinfo {author} {\bibfnamefont {K.}~\bibnamefont
  {Dymkowski}}\ and\ \bibinfo {author} {\bibfnamefont {C.}~\bibnamefont
  {Ederer}},\ }\bibfield  {title} {\bibinfo {title} {Strain-induced
  insulator-to-metal transition in {{LaTiO$_3$}} within {{DFT}} + {{DMFT}}},\
  }\href {https://doi.org/10.1103/PhysRevB.89.161109} {\bibfield  {journal}
  {\bibinfo  {journal} {Physical Review B}\ }\textbf {\bibinfo {volume} {89}},\
  \bibinfo {pages} {161109} (\bibinfo {year} {2014})}\BibitemShut {NoStop}%
\bibitem [{\citenamefont {Sclauzero}\ and\ \citenamefont
  {Ederer}(2015)}]{sclauzero_structural_2015}%
  \BibitemOpen
  \bibfield  {author} {\bibinfo {author} {\bibfnamefont {G.}~\bibnamefont
  {Sclauzero}}\ and\ \bibinfo {author} {\bibfnamefont {C.}~\bibnamefont
  {Ederer}},\ }\bibfield  {title} {\bibinfo {title} {Structural and electronic
  properties of epitaxially strained lavo$_3$ from density functional theory
  and dynamical mean-field theory},\ }\href
  {https://doi.org/10.1103/PhysRevB.92.235112} {\bibfield  {journal} {\bibinfo
  {journal} {Physical Review B}\ }\textbf {\bibinfo {volume} {92}},\ \bibinfo
  {pages} {235112} (\bibinfo {year} {2015})}\BibitemShut {NoStop}%
\bibitem [{\citenamefont {Sclauzero}\ \emph {et~al.}(2016)\citenamefont
  {Sclauzero}, \citenamefont {Dymkowski},\ and\ \citenamefont
  {Ederer}}]{sclauzero_tuning_2016}%
  \BibitemOpen
  \bibfield  {author} {\bibinfo {author} {\bibfnamefont {G.}~\bibnamefont
  {Sclauzero}}, \bibinfo {author} {\bibfnamefont {K.}~\bibnamefont
  {Dymkowski}},\ and\ \bibinfo {author} {\bibfnamefont {C.}~\bibnamefont
  {Ederer}},\ }\bibfield  {title} {\bibinfo {title} {Tuning the metal-insulator
  transition in $d^1$ and $d^2$ perovskites by epitaxial strain: {{A}}
  first-principles-based study},\ }\href
  {https://doi.org/10.1103/PhysRevB.94.245109} {\bibfield  {journal} {\bibinfo
  {journal} {Physical Review B}\ }\textbf {\bibinfo {volume} {94}},\ \bibinfo
  {pages} {245109} (\bibinfo {year} {2016})}\BibitemShut {NoStop}%
\bibitem [{\citenamefont {Subedi}\ \emph {et~al.}(2015)\citenamefont {Subedi},
  \citenamefont {Peil},\ and\ \citenamefont {Georges}}]{subedi_lowenergy_2015}%
  \BibitemOpen
  \bibfield  {author} {\bibinfo {author} {\bibfnamefont {A.}~\bibnamefont
  {Subedi}}, \bibinfo {author} {\bibfnamefont {O.~E.}\ \bibnamefont {Peil}},\
  and\ \bibinfo {author} {\bibfnamefont {A.}~\bibnamefont {Georges}},\
  }\bibfield  {title} {\bibinfo {title} {Low-energy description of the
  metal-insulator transition in the rare-earth nickelates},\ }\href
  {https://doi.org/10.1103/PhysRevB.91.075128} {\bibfield  {journal} {\bibinfo
  {journal} {Physical Review B}\ }\textbf {\bibinfo {volume} {91}},\ \bibinfo
  {pages} {075128} (\bibinfo {year} {2015})}\BibitemShut {NoStop}%
\bibitem [{\citenamefont {Seth}\ \emph {et~al.}(2017)\citenamefont {Seth},
  \citenamefont {Peil}, \citenamefont {Pourovskii}, \citenamefont {Betzinger},
  \citenamefont {Friedrich}, \citenamefont {Parcollet}, \citenamefont
  {Biermann}, \citenamefont {Aryasetiawan},\ and\ \citenamefont
  {Georges}}]{seth_renormalization_2017}%
  \BibitemOpen
  \bibfield  {author} {\bibinfo {author} {\bibfnamefont {P.}~\bibnamefont
  {Seth}}, \bibinfo {author} {\bibfnamefont {O.~E.}\ \bibnamefont {Peil}},
  \bibinfo {author} {\bibfnamefont {L.}~\bibnamefont {Pourovskii}}, \bibinfo
  {author} {\bibfnamefont {M.}~\bibnamefont {Betzinger}}, \bibinfo {author}
  {\bibfnamefont {C.}~\bibnamefont {Friedrich}}, \bibinfo {author}
  {\bibfnamefont {O.}~\bibnamefont {Parcollet}}, \bibinfo {author}
  {\bibfnamefont {S.}~\bibnamefont {Biermann}}, \bibinfo {author}
  {\bibfnamefont {F.}~\bibnamefont {Aryasetiawan}},\ and\ \bibinfo {author}
  {\bibfnamefont {A.}~\bibnamefont {Georges}},\ }\bibfield  {title} {\bibinfo
  {title} {Renormalization of effective interactions in a negative charge
  transfer insulator},\ }\href {https://doi.org/10.1103/PhysRevB.96.205139}
  {\bibfield  {journal} {\bibinfo  {journal} {Physical Review B}\ }\textbf
  {\bibinfo {volume} {96}},\ \bibinfo {pages} {205139} (\bibinfo {year}
  {2017})}\BibitemShut {NoStop}%
\bibitem [{\citenamefont {Peil}\ \emph {et~al.}(2019)\citenamefont {Peil},
  \citenamefont {Hampel}, \citenamefont {Ederer},\ and\ \citenamefont
  {Georges}}]{peil_mechanism_2019}%
  \BibitemOpen
  \bibfield  {author} {\bibinfo {author} {\bibfnamefont {O.~E.}\ \bibnamefont
  {Peil}}, \bibinfo {author} {\bibfnamefont {A.}~\bibnamefont {Hampel}},
  \bibinfo {author} {\bibfnamefont {C.}~\bibnamefont {Ederer}},\ and\ \bibinfo
  {author} {\bibfnamefont {A.}~\bibnamefont {Georges}},\ }\bibfield  {title}
  {\bibinfo {title} {Mechanism and control parameters of the coupled structural
  and metal-insulator transition in nickelates},\ }\href
  {https://doi.org/10.1103/PhysRevB.99.245127} {\bibfield  {journal} {\bibinfo
  {journal} {Physical Review B}\ }\textbf {\bibinfo {volume} {99}},\ \bibinfo
  {pages} {245127} (\bibinfo {year} {2019})}\BibitemShut {NoStop}%
\bibitem [{\citenamefont {Park}\ \emph
  {et~al.}(2014{\natexlab{b}})\citenamefont {Park}, \citenamefont {Millis},\
  and\ \citenamefont {Marianetti}}]{park_total_2014}%
  \BibitemOpen
  \bibfield  {author} {\bibinfo {author} {\bibfnamefont {H.}~\bibnamefont
  {Park}}, \bibinfo {author} {\bibfnamefont {A.~J.}\ \bibnamefont {Millis}},\
  and\ \bibinfo {author} {\bibfnamefont {C.~A.}\ \bibnamefont {Marianetti}},\
  }\bibfield  {title} {\bibinfo {title} {Total {{Energy Calculations Using
  DFT}}+{{DMFT}}: {{Computing}} the {{Pressure Phase Diagram}} of the {{Rare
  Earth Nickelates}}},\ }\href {https://doi.org/10.1103/physrevb.89.245133}
  {\bibfield  {journal} {\bibinfo  {journal} {Physical Review B}\ }\textbf
  {\bibinfo {volume} {89}},\ \bibinfo {pages} {245133} (\bibinfo {year}
  {2014}{\natexlab{b}})}\BibitemShut {NoStop}%
\bibitem [{\citenamefont {Kim}\ \emph {et~al.}(2018)\citenamefont {Kim},
  \citenamefont {Liu}, \citenamefont {Tomczak},\ and\ \citenamefont
  {Franchini}}]{kim_straininduced_2018}%
  \BibitemOpen
  \bibfield  {author} {\bibinfo {author} {\bibfnamefont {B.}~\bibnamefont
  {Kim}}, \bibinfo {author} {\bibfnamefont {P.}~\bibnamefont {Liu}}, \bibinfo
  {author} {\bibfnamefont {J.~M.}\ \bibnamefont {Tomczak}},\ and\ \bibinfo
  {author} {\bibfnamefont {C.}~\bibnamefont {Franchini}},\ }\bibfield  {title}
  {\bibinfo {title} {Strain-induced tuning of the electronic {{Coulomb}}
  interaction in 3d transition metal oxide perovskites},\ }\href
  {https://doi.org/10.1103/PhysRevB.98.075130} {\bibfield  {journal} {\bibinfo
  {journal} {Physical Review B}\ }\textbf {\bibinfo {volume} {98}},\ \bibinfo
  {pages} {075130} (\bibinfo {year} {2018})}\BibitemShut {NoStop}%
\bibitem [{\citenamefont {Georges}\ \emph {et~al.}(2013)\citenamefont
  {Georges}, \citenamefont {de'‌ Medici},\ and\ \citenamefont
  {Mravlje}}]{georges_strong_2013}%
  \BibitemOpen
  \bibfield  {author} {\bibinfo {author} {\bibfnamefont {A.}~\bibnamefont
  {Georges}}, \bibinfo {author} {\bibfnamefont {L.}~\bibnamefont {de'‌
  Medici}},\ and\ \bibinfo {author} {\bibfnamefont {J.}~\bibnamefont
  {Mravlje}},\ }\bibfield  {title} {\bibinfo {title} {Strong {{Correlations}}
  from {{Hund}}'s {{Coupling}}},\ }\href
  {https://doi.org/10.1146/annurev-conmatphys-020911-125045} {\bibfield
  {journal} {\bibinfo  {journal} {Annual Review of Condensed Matter Physics}\
  }\textbf {\bibinfo {volume} {4}},\ \bibinfo {pages} {137} (\bibinfo {year}
  {2013})}\BibitemShut {NoStop}%
\bibitem [{\citenamefont {Dudarev}\ \emph {et~al.}(1998)\citenamefont
  {Dudarev}, \citenamefont {Botton}, \citenamefont {Savrasov}, \citenamefont
  {Humphreys},\ and\ \citenamefont {Sutton}}]{dudarev_electronenergyloss_1998}%
  \BibitemOpen
  \bibfield  {author} {\bibinfo {author} {\bibfnamefont {S.~L.}\ \bibnamefont
  {Dudarev}}, \bibinfo {author} {\bibfnamefont {G.~A.}\ \bibnamefont {Botton}},
  \bibinfo {author} {\bibfnamefont {S.~Y.}\ \bibnamefont {Savrasov}}, \bibinfo
  {author} {\bibfnamefont {C.~J.}\ \bibnamefont {Humphreys}},\ and\ \bibinfo
  {author} {\bibfnamefont {A.~P.}\ \bibnamefont {Sutton}},\ }\bibfield  {title}
  {\bibinfo {title} {Electron-energy-loss spectra and the structural stability
  of nickel oxide: An {{LSDA}}+{{U}} study},\ }\href@noop {} {\bibfield
  {journal} {\bibinfo  {journal} {Physical Review B}\ }\textbf {\bibinfo
  {volume} {57}},\ \bibinfo {pages} {1505} (\bibinfo {year}
  {1998})}\BibitemShut {NoStop}%
\bibitem [{\citenamefont {Galicka}\ \emph {et~al.}(2009)\citenamefont
  {Galicka}, \citenamefont {Szade}, \citenamefont {Ruello}, \citenamefont
  {Laffez},\ and\ \citenamefont {Ratuszna}}]{galicka_photoemission_2009}%
  \BibitemOpen
  \bibfield  {author} {\bibinfo {author} {\bibfnamefont {K.}~\bibnamefont
  {Galicka}}, \bibinfo {author} {\bibfnamefont {J.}~\bibnamefont {Szade}},
  \bibinfo {author} {\bibfnamefont {P.}~\bibnamefont {Ruello}}, \bibinfo
  {author} {\bibfnamefont {P.}~\bibnamefont {Laffez}},\ and\ \bibinfo {author}
  {\bibfnamefont {A.}~\bibnamefont {Ratuszna}},\ }\bibfield  {title} {\bibinfo
  {title} {The photoemission study of {{NdNiO$_3$}}/{{NdGa$O_3$}} thin films,
  through the metal–insulator transition},\ }\href
  {https://doi.org/10.1016/j.apsusc.2008.03.057} {\bibfield  {journal}
  {\bibinfo  {journal} {Applied Surface Science}\ }\textbf {\bibinfo {volume}
  {255}},\ \bibinfo {pages} {4355} (\bibinfo {year} {2009})}\BibitemShut
  {NoStop}%
\bibitem [{\citenamefont {Park}\ \emph {et~al.}(2012)\citenamefont {Park},
  \citenamefont {Millis},\ and\ \citenamefont
  {Marianetti}}]{park_siteselective_2012}%
  \BibitemOpen
  \bibfield  {author} {\bibinfo {author} {\bibfnamefont {H.}~\bibnamefont
  {Park}}, \bibinfo {author} {\bibfnamefont {A.~J.}\ \bibnamefont {Millis}},\
  and\ \bibinfo {author} {\bibfnamefont {C.~A.}\ \bibnamefont {Marianetti}},\
  }\bibfield  {title} {\bibinfo {title} {Site-{{Selective Mott Transition}} in
  {{Rare-Earth-Element Nickelates}}},\ }\href
  {https://doi.org/10.1103/PhysRevLett.109.156402} {\bibfield  {journal}
  {\bibinfo  {journal} {Physical Review Letters}\ }\textbf {\bibinfo {volume}
  {109}},\ \bibinfo {pages} {156402} (\bibinfo {year} {2012})}\BibitemShut
  {NoStop}%
\bibitem [{\citenamefont {Haule}\ and\ \citenamefont
  {Pascut}(2017)}]{haule_mott_2017}%
  \BibitemOpen
  \bibfield  {author} {\bibinfo {author} {\bibfnamefont {K.}~\bibnamefont
  {Haule}}\ and\ \bibinfo {author} {\bibfnamefont {G.~L.}\ \bibnamefont
  {Pascut}},\ }\bibfield  {title} {\bibinfo {title} {Mott {{Transition}} and
  {{Magnetism}} in {{Rare Earth Nickelates}} and its {{Fingerprint}} on the
  {{X-ray Scattering}}},\ }\href@noop {} {\bibfield  {journal} {\bibinfo
  {journal} {Scientific Reports}\ }\textbf {\bibinfo {volume} {7}} (\bibinfo
  {year} {2017})}\BibitemShut {NoStop}%
\bibitem [{\citenamefont {Haas}\ \emph {et~al.}(2024)\citenamefont {Haas},
  \citenamefont {Mlkvik}, \citenamefont {Spaldin},\ and\ \citenamefont
  {Ederer}}]{haas_incorporating_2024}%
  \BibitemOpen
  \bibfield  {author} {\bibinfo {author} {\bibfnamefont {L.}~\bibnamefont
  {Haas}}, \bibinfo {author} {\bibfnamefont {P.}~\bibnamefont {Mlkvik}},
  \bibinfo {author} {\bibfnamefont {N.~A.}\ \bibnamefont {Spaldin}},\ and\
  \bibinfo {author} {\bibfnamefont {C.}~\bibnamefont {Ederer}},\ }\bibfield
  {title} {\bibinfo {title} {Incorporating static intersite correlation effects
  in vanadium dioxide through dft+v},\ }\href
  {https://doi.org/10.1103/PhysRevResearch.6.043177} {\bibfield  {journal}
  {\bibinfo  {journal} {Physical Review Research}\ }\textbf {\bibinfo {volume}
  {6}},\ \bibinfo {pages} {043177} (\bibinfo {year} {2024})}\BibitemShut
  {NoStop}%
\bibitem [{\citenamefont {Honerkamp}\ \emph {et~al.}(2018)\citenamefont
  {Honerkamp}, \citenamefont {Shinaoka}, \citenamefont {Assaad},\ and\
  \citenamefont {Werner}}]{honerkamp_limitations_2018}%
  \BibitemOpen
  \bibfield  {author} {\bibinfo {author} {\bibfnamefont {C.}~\bibnamefont
  {Honerkamp}}, \bibinfo {author} {\bibfnamefont {H.}~\bibnamefont {Shinaoka}},
  \bibinfo {author} {\bibfnamefont {F.~F.}\ \bibnamefont {Assaad}},\ and\
  \bibinfo {author} {\bibfnamefont {P.}~\bibnamefont {Werner}},\ }\bibfield
  {title} {\bibinfo {title} {Limitations of constrained random phase
  approximation downfolding},\ }\href
  {https://doi.org/10.1103/PhysRevB.98.235151} {\bibfield  {journal} {\bibinfo
  {journal} {Physical Review B}\ }\textbf {\bibinfo {volume} {98}},\ \bibinfo
  {pages} {235151} (\bibinfo {year} {2018})}\BibitemShut {NoStop}%
\bibitem [{\citenamefont {{van Loon}}\ \emph {et~al.}(2021)\citenamefont {{van
  Loon}}, \citenamefont {Rösner}, \citenamefont {Katsnelson},\ and\
  \citenamefont {Wehling}}]{vanloon_random_2021}%
  \BibitemOpen
  \bibfield  {author} {\bibinfo {author} {\bibfnamefont {E.~G. C.~P.}\
  \bibnamefont {{van Loon}}}, \bibinfo {author} {\bibfnamefont
  {M.}~\bibnamefont {Rösner}}, \bibinfo {author} {\bibfnamefont {M.~I.}\
  \bibnamefont {Katsnelson}},\ and\ \bibinfo {author} {\bibfnamefont {T.~O.}\
  \bibnamefont {Wehling}},\ }\bibfield  {title} {\bibinfo {title} {Random phase
  approximation for gapped systems: {{Role}} of vertex corrections and
  applicability of the constrained random phase approximation},\ }\href
  {https://doi.org/10.1103/PhysRevB.104.045134} {\bibfield  {journal} {\bibinfo
   {journal} {Physical Review B}\ }\textbf {\bibinfo {volume} {104}},\ \bibinfo
  {pages} {045134} (\bibinfo {year} {2021})}\BibitemShut {NoStop}%
\bibitem [{\citenamefont {Werner}\ and\ \citenamefont
  {Casula}(2016)}]{werner_dynamical_2016}%
  \BibitemOpen
  \bibfield  {author} {\bibinfo {author} {\bibfnamefont {P.}~\bibnamefont
  {Werner}}\ and\ \bibinfo {author} {\bibfnamefont {M.}~\bibnamefont
  {Casula}},\ }\bibfield  {title} {\bibinfo {title} {Dynamical screening in
  correlated electron systems—from lattice models to realistic materials},\
  }\href {https://doi.org/10.1088/0953-8984/28/38/383001} {\bibfield  {journal}
  {\bibinfo  {journal} {Journal of Physics: Condensed Matter}\ }\textbf
  {\bibinfo {volume} {28}},\ \bibinfo {pages} {383001} (\bibinfo {year}
  {2016})}\BibitemShut {NoStop}%
\bibitem [{\citenamefont {Alberto~Carta}\ and\ \citenamefont
  {Ederer}(2024)}]{MaterialsCloudArchive2024}%
  \BibitemOpen
  \bibfield  {author} {\bibinfo {author} {\bibfnamefont {A.~P.}\ \bibnamefont
  {Alberto~Carta}}\ and\ \bibinfo {author} {\bibfnamefont {C.}~\bibnamefont
  {Ederer}},\ }\href@noop {} {\bibinfo {title} {Importance of ligand on-site
  interactions for the description of mott-insulators in dft+dmft}},\ \bibinfo
  {howpublished} {Materials Cloud Archive} (\bibinfo {year} {2024})\BibitemShut
  {NoStop}%
\bibitem [{\citenamefont {Liechtenstein}\ \emph {et~al.}(1995)\citenamefont
  {Liechtenstein}, \citenamefont {Anisimov},\ and\ \citenamefont
  {Zaanen}}]{liechtenstein_densityfunctional_1995}%
  \BibitemOpen
  \bibfield  {author} {\bibinfo {author} {\bibfnamefont {A.~I.}\ \bibnamefont
  {Liechtenstein}}, \bibinfo {author} {\bibfnamefont {V.~I.}\ \bibnamefont
  {Anisimov}},\ and\ \bibinfo {author} {\bibfnamefont {J.}~\bibnamefont
  {Zaanen}},\ }\bibfield  {title} {\bibinfo {title} {Density-functional theory
  and strong interactions: {{Orbital}} ordering in {{Mott-Hubbard}}
  insulators},\ }\href {https://doi.org/10.1103/PhysRevB.52.R5467} {\bibfield
  {journal} {\bibinfo  {journal} {Physical Review B}\ }\textbf {\bibinfo
  {volume} {52}},\ \bibinfo {pages} {R5467} (\bibinfo {year}
  {1995})}\BibitemShut {NoStop}%
\bibitem [{\citenamefont {Shinaoka}\ \emph {et~al.}(2015)\citenamefont
  {Shinaoka}, \citenamefont {Troyer},\ and\ \citenamefont
  {Werner}}]{shinaoka_accuracy_2015}%
  \BibitemOpen
  \bibfield  {author} {\bibinfo {author} {\bibfnamefont {H.}~\bibnamefont
  {Shinaoka}}, \bibinfo {author} {\bibfnamefont {M.}~\bibnamefont {Troyer}},\
  and\ \bibinfo {author} {\bibfnamefont {P.}~\bibnamefont {Werner}},\
  }\bibfield  {title} {\bibinfo {title} {Accuracy of downfolding based on the
  constrained random-phase approximation},\ }\href
  {https://doi.org/10.1103/PhysRevB.91.245156} {\bibfield  {journal} {\bibinfo
  {journal} {Physical Review B}\ }\textbf {\bibinfo {volume} {91}},\ \bibinfo
  {pages} {245156} (\bibinfo {year} {2015})}\BibitemShut {NoStop}%
\bibitem [{\citenamefont {Wang}\ \emph {et~al.}(2012)\citenamefont {Wang},
  \citenamefont {Han}, \citenamefont {{de' Medici}}, \citenamefont {Park},
  \citenamefont {Marianetti},\ and\ \citenamefont
  {Millis}}]{wang_covalency_2012}%
  \BibitemOpen
  \bibfield  {author} {\bibinfo {author} {\bibfnamefont {X.}~\bibnamefont
  {Wang}}, \bibinfo {author} {\bibfnamefont {M.~J.}\ \bibnamefont {Han}},
  \bibinfo {author} {\bibfnamefont {L.}~\bibnamefont {{de' Medici}}}, \bibinfo
  {author} {\bibfnamefont {H.}~\bibnamefont {Park}}, \bibinfo {author}
  {\bibfnamefont {C.~A.}\ \bibnamefont {Marianetti}},\ and\ \bibinfo {author}
  {\bibfnamefont {A.~J.}\ \bibnamefont {Millis}},\ }\bibfield  {title}
  {\bibinfo {title} {Covalency, double-counting, and the metal-insulator phase
  diagram in transition metal oxides},\ }\href
  {https://doi.org/10.1103/PhysRevB.86.195136} {\bibfield  {journal} {\bibinfo
  {journal} {Physical Review B}\ }\textbf {\bibinfo {volume} {86}},\ \bibinfo
  {pages} {195136} (\bibinfo {year} {2012})}\BibitemShut {NoStop}%
\bibitem [{\citenamefont {Georges}\ \emph {et~al.}(1996)\citenamefont
  {Georges}, \citenamefont {Kotliar}, \citenamefont {Krauth},\ and\
  \citenamefont {Rozenberg}}]{georges_dynamical_1996}%
  \BibitemOpen
  \bibfield  {author} {\bibinfo {author} {\bibfnamefont {A.}~\bibnamefont
  {Georges}}, \bibinfo {author} {\bibfnamefont {G.}~\bibnamefont {Kotliar}},
  \bibinfo {author} {\bibfnamefont {W.}~\bibnamefont {Krauth}},\ and\ \bibinfo
  {author} {\bibfnamefont {M.~J.}\ \bibnamefont {Rozenberg}},\ }\bibfield
  {title} {\bibinfo {title} {Dynamical {{Mean-Field Theory}} of {{Strongly
  Correlated Fermion Systems}} and the {{Limit}} of {{Infinite Dimensions}}},\
  }\href {https://doi.org/10.1103/RevModPhys.68.13} {\bibfield  {journal}
  {\bibinfo  {journal} {Reviews of Modern Physics}\ }\textbf {\bibinfo {volume}
  {68}},\ \bibinfo {pages} {13} (\bibinfo {year} {1996})}\BibitemShut {NoStop}%
\bibitem [{\citenamefont {Karolak}\ \emph {et~al.}(2010)\citenamefont
  {Karolak}, \citenamefont {Ulm}, \citenamefont {Wehling}, \citenamefont
  {Mazurenko}, \citenamefont {Poteryaev},\ and\ \citenamefont
  {Lichtenstein}}]{karolak_double_2010}%
  \BibitemOpen
  \bibfield  {author} {\bibinfo {author} {\bibfnamefont {M.}~\bibnamefont
  {Karolak}}, \bibinfo {author} {\bibfnamefont {G.}~\bibnamefont {Ulm}},
  \bibinfo {author} {\bibfnamefont {T.}~\bibnamefont {Wehling}}, \bibinfo
  {author} {\bibfnamefont {V.}~\bibnamefont {Mazurenko}}, \bibinfo {author}
  {\bibfnamefont {A.}~\bibnamefont {Poteryaev}},\ and\ \bibinfo {author}
  {\bibfnamefont {A.}~\bibnamefont {Lichtenstein}},\ }\bibfield  {title}
  {\bibinfo {title} {Double counting in {{LDA}}+{{DMFT}}—{{The}} example of
  {{NiO}}},\ }\href {https://doi.org/10.1016/j.elspec.2010.05.021} {\bibfield
  {journal} {\bibinfo  {journal} {Journal of Electron Spectroscopy and Related
  Phenomena}\ }\textbf {\bibinfo {volume} {181}},\ \bibinfo {pages} {11}
  (\bibinfo {year} {2010})}\BibitemShut {NoStop}%
\bibitem [{\citenamefont {Giannozzi}\ \emph {et~al.}(2017)\citenamefont
  {Giannozzi}, \citenamefont {Andreussi}, \citenamefont {Brumme}, \citenamefont
  {Bunau}, \citenamefont {Nardelli}, \citenamefont {Calandra}, \citenamefont
  {Car}, \citenamefont {Cavazzoni}, \citenamefont {Ceresoli}, \citenamefont
  {Cococcioni}, \citenamefont {Colonna}, \citenamefont {Carnimeo},
  \citenamefont {Corso}, \citenamefont {de~Gironcoli}, \citenamefont {Delugas},
  \citenamefont {DiStasio}, \citenamefont {Ferretti}, \citenamefont {Floris},
  \citenamefont {Fratesi}, \citenamefont {Fugallo}, \citenamefont {Gebauer},
  \citenamefont {Gerstmann}, \citenamefont {Giustino}, \citenamefont {Gorni},
  \citenamefont {Jia}, \citenamefont {Kawamura}, \citenamefont {Ko},
  \citenamefont {Kokalj}, \citenamefont {Küçükbenli}, \citenamefont
  {Lazzeri}, \citenamefont {Marsili}, \citenamefont {Marzari}, \citenamefont
  {Mauri}, \citenamefont {Nguyen}, \citenamefont {Nguyen}, \citenamefont
  {{Otero-de-la-Roza}}, \citenamefont {Paulatto}, \citenamefont {Poncé},
  \citenamefont {Rocca}, \citenamefont {Sabatini}, \citenamefont {Santra},
  \citenamefont {Schlipf}, \citenamefont {Seitsonen}, \citenamefont {Smogunov},
  \citenamefont {Timrov}, \citenamefont {Thonhauser}, \citenamefont {Umari},
  \citenamefont {Vast}, \citenamefont {Wu},\ and\ \citenamefont
  {Baroni}}]{giannozzi_advanced_2017}%
  \BibitemOpen
  \bibfield  {author} {\bibinfo {author} {\bibfnamefont {P.}~\bibnamefont
  {Giannozzi}}, \bibinfo {author} {\bibfnamefont {O.}~\bibnamefont
  {Andreussi}}, \bibinfo {author} {\bibfnamefont {T.}~\bibnamefont {Brumme}},
  \bibinfo {author} {\bibfnamefont {O.}~\bibnamefont {Bunau}}, \bibinfo
  {author} {\bibfnamefont {M.~B.}\ \bibnamefont {Nardelli}}, \bibinfo {author}
  {\bibfnamefont {M.}~\bibnamefont {Calandra}}, \bibinfo {author}
  {\bibfnamefont {R.}~\bibnamefont {Car}}, \bibinfo {author} {\bibfnamefont
  {C.}~\bibnamefont {Cavazzoni}}, \bibinfo {author} {\bibfnamefont
  {D.}~\bibnamefont {Ceresoli}}, \bibinfo {author} {\bibfnamefont
  {M.}~\bibnamefont {Cococcioni}}, \bibinfo {author} {\bibfnamefont
  {N.}~\bibnamefont {Colonna}}, \bibinfo {author} {\bibfnamefont
  {I.}~\bibnamefont {Carnimeo}}, \bibinfo {author} {\bibfnamefont {A.~D.}\
  \bibnamefont {Corso}}, \bibinfo {author} {\bibfnamefont {S.}~\bibnamefont
  {de~Gironcoli}}, \bibinfo {author} {\bibfnamefont {P.}~\bibnamefont
  {Delugas}}, \bibinfo {author} {\bibfnamefont {R.~A.}\ \bibnamefont
  {DiStasio}}, \bibinfo {author} {\bibfnamefont {A.}~\bibnamefont {Ferretti}},
  \bibinfo {author} {\bibfnamefont {A.}~\bibnamefont {Floris}}, \bibinfo
  {author} {\bibfnamefont {G.}~\bibnamefont {Fratesi}}, \bibinfo {author}
  {\bibfnamefont {G.}~\bibnamefont {Fugallo}}, \bibinfo {author} {\bibfnamefont
  {R.}~\bibnamefont {Gebauer}}, \bibinfo {author} {\bibfnamefont
  {U.}~\bibnamefont {Gerstmann}}, \bibinfo {author} {\bibfnamefont
  {F.}~\bibnamefont {Giustino}}, \bibinfo {author} {\bibfnamefont
  {T.}~\bibnamefont {Gorni}}, \bibinfo {author} {\bibfnamefont
  {J.}~\bibnamefont {Jia}}, \bibinfo {author} {\bibfnamefont {M.}~\bibnamefont
  {Kawamura}}, \bibinfo {author} {\bibfnamefont {H.-Y.}\ \bibnamefont {Ko}},
  \bibinfo {author} {\bibfnamefont {A.}~\bibnamefont {Kokalj}}, \bibinfo
  {author} {\bibfnamefont {E.}~\bibnamefont {Küçükbenli}}, \bibinfo {author}
  {\bibfnamefont {M.}~\bibnamefont {Lazzeri}}, \bibinfo {author} {\bibfnamefont
  {M.}~\bibnamefont {Marsili}}, \bibinfo {author} {\bibfnamefont
  {N.}~\bibnamefont {Marzari}}, \bibinfo {author} {\bibfnamefont
  {F.}~\bibnamefont {Mauri}}, \bibinfo {author} {\bibfnamefont {N.~L.}\
  \bibnamefont {Nguyen}}, \bibinfo {author} {\bibfnamefont {H.-V.}\
  \bibnamefont {Nguyen}}, \bibinfo {author} {\bibfnamefont {A.}~\bibnamefont
  {{Otero-de-la-Roza}}}, \bibinfo {author} {\bibfnamefont {L.}~\bibnamefont
  {Paulatto}}, \bibinfo {author} {\bibfnamefont {S.}~\bibnamefont {Poncé}},
  \bibinfo {author} {\bibfnamefont {D.}~\bibnamefont {Rocca}}, \bibinfo
  {author} {\bibfnamefont {R.}~\bibnamefont {Sabatini}}, \bibinfo {author}
  {\bibfnamefont {B.}~\bibnamefont {Santra}}, \bibinfo {author} {\bibfnamefont
  {M.}~\bibnamefont {Schlipf}}, \bibinfo {author} {\bibfnamefont {A.~P.}\
  \bibnamefont {Seitsonen}}, \bibinfo {author} {\bibfnamefont {A.}~\bibnamefont
  {Smogunov}}, \bibinfo {author} {\bibfnamefont {I.}~\bibnamefont {Timrov}},
  \bibinfo {author} {\bibfnamefont {T.}~\bibnamefont {Thonhauser}}, \bibinfo
  {author} {\bibfnamefont {P.}~\bibnamefont {Umari}}, \bibinfo {author}
  {\bibfnamefont {N.}~\bibnamefont {Vast}}, \bibinfo {author} {\bibfnamefont
  {X.}~\bibnamefont {Wu}},\ and\ \bibinfo {author} {\bibfnamefont
  {S.}~\bibnamefont {Baroni}},\ }\bibfield  {title} {\bibinfo {title} {Advanced
  capabilities for materials modelling with {{Quantum ESPRESSO}}},\ }\href
  {https://doi.org/10.1088/1361-648X/aa8f79} {\bibfield  {journal} {\bibinfo
  {journal} {Journal of Physics: Condensed Matter}\ }\textbf {\bibinfo {volume}
  {29}},\ \bibinfo {pages} {465901} (\bibinfo {year} {2017})}\BibitemShut
  {NoStop}%
\bibitem [{\citenamefont {Garrity}\ \emph {et~al.}(2014)\citenamefont
  {Garrity}, \citenamefont {Bennett}, \citenamefont {Rabe},\ and\ \citenamefont
  {Vanderbilt}}]{garrity_pseudopotentials_2014}%
  \BibitemOpen
  \bibfield  {author} {\bibinfo {author} {\bibfnamefont {K.~F.}\ \bibnamefont
  {Garrity}}, \bibinfo {author} {\bibfnamefont {J.~W.}\ \bibnamefont
  {Bennett}}, \bibinfo {author} {\bibfnamefont {K.~M.}\ \bibnamefont {Rabe}},\
  and\ \bibinfo {author} {\bibfnamefont {D.}~\bibnamefont {Vanderbilt}},\
  }\bibfield  {title} {\bibinfo {title} {Pseudopotentials for high-throughput
  {{DFT}} calculations},\ }\href@noop {} {\bibfield  {journal} {\bibinfo
  {journal} {Computational Materials Science}\ }\textbf {\bibinfo {volume}
  {81}},\ \bibinfo {pages} {446} (\bibinfo {year} {2014})}\BibitemShut
  {NoStop}%
\bibitem [{\citenamefont {Perdew}\ \emph {et~al.}(1996)\citenamefont {Perdew},
  \citenamefont {Burke},\ and\ \citenamefont
  {Ernzerhof}}]{perdew_generalized_1996}%
  \BibitemOpen
  \bibfield  {author} {\bibinfo {author} {\bibfnamefont {J.~P.}\ \bibnamefont
  {Perdew}}, \bibinfo {author} {\bibfnamefont {K.}~\bibnamefont {Burke}},\ and\
  \bibinfo {author} {\bibfnamefont {M.}~\bibnamefont {Ernzerhof}},\ }\bibfield
  {title} {\bibinfo {title} {Generalized {{Gradient Approximation Made
  Simple}}},\ }\href {https://doi.org/10.1103/PhysRevLett.77.3865} {\bibfield
  {journal} {\bibinfo  {journal} {Physical Review Letters}\ }\textbf {\bibinfo
  {volume} {77}},\ \bibinfo {pages} {3865} (\bibinfo {year}
  {1996})}\BibitemShut {NoStop}%
\bibitem [{\citenamefont {Hampel}\ and\ \citenamefont
  {Ederer}(2017)}]{hampel_interplay_2017a}%
  \BibitemOpen
  \bibfield  {author} {\bibinfo {author} {\bibfnamefont {A.}~\bibnamefont
  {Hampel}}\ and\ \bibinfo {author} {\bibfnamefont {C.}~\bibnamefont
  {Ederer}},\ }\bibfield  {title} {\bibinfo {title} {Interplay between
  breathing mode distortion and magnetic order in rare-earth nickelates
  {{RNiO$_3$}} within {{DFT}}+{{U}}},\ }\href
  {https://doi.org/10.1103/PhysRevB.96.165130} {\bibfield  {journal} {\bibinfo
  {journal} {Physical Review B}\ }\textbf {\bibinfo {volume} {96}},\ \bibinfo
  {pages} {165130} (\bibinfo {year} {2017})}\BibitemShut {NoStop}%
\bibitem [{\citenamefont {{García-Muñoz}}\ \emph {et~al.}(2009)\citenamefont
  {{García-Muñoz}}, \citenamefont {Aranda}, \citenamefont {Alonso},\ and\
  \citenamefont {{Martínez-Lope}}}]{garcia-munoz_structure_2009}%
  \BibitemOpen
  \bibfield  {author} {\bibinfo {author} {\bibfnamefont {J.~L.}\ \bibnamefont
  {{García-Muñoz}}}, \bibinfo {author} {\bibfnamefont {M.~A.~G.}\
  \bibnamefont {Aranda}}, \bibinfo {author} {\bibfnamefont {J.~A.}\
  \bibnamefont {Alonso}},\ and\ \bibinfo {author} {\bibfnamefont {M.~J.}\
  \bibnamefont {{Martínez-Lope}}},\ }\bibfield  {title} {\bibinfo {title}
  {Structure and charge order in the antiferromagnetic band-insulating phase of
  ndnio$_3$},\ }\href {https://doi.org/10.1103/PhysRevB.79.134432} {\bibfield
  {journal} {\bibinfo  {journal} {Physical Review B}\ }\textbf {\bibinfo
  {volume} {79}},\ \bibinfo {pages} {134432} (\bibinfo {year}
  {2009})}\BibitemShut {NoStop}%
\bibitem [{\citenamefont {Merkel}\ \emph {et~al.}(2022)\citenamefont {Merkel},
  \citenamefont {Carta}, \citenamefont {Beck},\ and\ \citenamefont
  {Hampel}}]{merkel_solid_dmft_2022}%
  \BibitemOpen
  \bibfield  {author} {\bibinfo {author} {\bibfnamefont {M.~E.}\ \bibnamefont
  {Merkel}}, \bibinfo {author} {\bibfnamefont {A.}~\bibnamefont {Carta}},
  \bibinfo {author} {\bibfnamefont {S.}~\bibnamefont {Beck}},\ and\ \bibinfo
  {author} {\bibfnamefont {A.}~\bibnamefont {Hampel}},\ }\bibfield  {title}
  {\bibinfo {title} {Solid\_dmft: Gray-boxing {{DFT}}+{{DMFT}} materials
  simulations with {{TRIQS}}},\ }\href@noop {} {\bibfield  {journal} {\bibinfo
  {journal} {Journal of Open Source Software}\ }\textbf {\bibinfo {volume}
  {7}},\ \bibinfo {pages} {4623} (\bibinfo {year} {2022})}\BibitemShut
  {NoStop}%
\bibitem [{\citenamefont {Parcollet}\ \emph {et~al.}(2015)\citenamefont
  {Parcollet}, \citenamefont {Ferrero}, \citenamefont {Ayral}, \citenamefont
  {Hafermann}, \citenamefont {Krivenko}, \citenamefont {Messio},\ and\
  \citenamefont {Seth}}]{parcollet_triqs_2015}%
  \BibitemOpen
  \bibfield  {author} {\bibinfo {author} {\bibfnamefont {O.}~\bibnamefont
  {Parcollet}}, \bibinfo {author} {\bibfnamefont {M.}~\bibnamefont {Ferrero}},
  \bibinfo {author} {\bibfnamefont {T.}~\bibnamefont {Ayral}}, \bibinfo
  {author} {\bibfnamefont {H.}~\bibnamefont {Hafermann}}, \bibinfo {author}
  {\bibfnamefont {I.}~\bibnamefont {Krivenko}}, \bibinfo {author}
  {\bibfnamefont {L.}~\bibnamefont {Messio}},\ and\ \bibinfo {author}
  {\bibfnamefont {P.}~\bibnamefont {Seth}},\ }\bibfield  {title} {\bibinfo
  {title} {{{TRIQS}}: {{A}} toolbox for research on interacting quantum
  systems},\ }\href@noop {} {\bibfield  {journal} {\bibinfo  {journal}
  {Computer Physics Communications}\ }\textbf {\bibinfo {volume} {196}},\
  \bibinfo {pages} {398} (\bibinfo {year} {2015})}\BibitemShut {NoStop}%
\bibitem [{\citenamefont {Seth}\ \emph {et~al.}(2016)\citenamefont {Seth},
  \citenamefont {Krivenko}, \citenamefont {Ferrero},\ and\ \citenamefont
  {Parcollet}}]{seth_triqs_2016}%
  \BibitemOpen
  \bibfield  {author} {\bibinfo {author} {\bibfnamefont {P.}~\bibnamefont
  {Seth}}, \bibinfo {author} {\bibfnamefont {I.}~\bibnamefont {Krivenko}},
  \bibinfo {author} {\bibfnamefont {M.}~\bibnamefont {Ferrero}},\ and\ \bibinfo
  {author} {\bibfnamefont {O.}~\bibnamefont {Parcollet}},\ }\bibfield  {title}
  {\bibinfo {title} {{{TRIQS}}/{{CTHYB}}: {{A}} continuous-time quantum {{Monte
  Carlo}} hybridisation expansion solver for quantum impurity problems},\
  }\href {https://doi.org/10.1016/j.cpc.2015.10.023} {\bibfield  {journal}
  {\bibinfo  {journal} {Computer Physics Communications}\ }\textbf {\bibinfo
  {volume} {200}},\ \bibinfo {pages} {274} (\bibinfo {year}
  {2016})}\BibitemShut {NoStop}%
\bibitem [{tri(2024{\natexlab{a}})}]{triqs_ctseg}%
  \BibitemOpen
  \href@noop {} {\bibinfo {title} {{{TRIQS}}/ctseg}},\ \bibinfo {howpublished}
  {TRIQS} (\bibinfo {year} {2024}{\natexlab{a}})\BibitemShut {NoStop}%
\bibitem [{tri(2024{\natexlab{b}})}]{triqs_hartree_fock}%
  \BibitemOpen
  \href@noop {} {\bibinfo {title} {{Hartree-Fock: Lattice and impurity solvers
  based on the TRIQS library}}},\ \bibinfo {howpublished}
  {\url{https://github.com/TRIQS/hartree_fock}} (\bibinfo {year}
  {2024}{\natexlab{b}}),\ \bibinfo {note} {accessed: 2024-09-16}\BibitemShut
  {NoStop}%
\bibitem [{\citenamefont {Levy}\ \emph {et~al.}(2017)\citenamefont {Levy},
  \citenamefont {LeBlanc},\ and\ \citenamefont
  {Gull}}]{levy_implementation_2017}%
  \BibitemOpen
  \bibfield  {author} {\bibinfo {author} {\bibfnamefont {R.}~\bibnamefont
  {Levy}}, \bibinfo {author} {\bibfnamefont {J.~P.~F.}\ \bibnamefont
  {LeBlanc}},\ and\ \bibinfo {author} {\bibfnamefont {E.}~\bibnamefont
  {Gull}},\ }\bibfield  {title} {\bibinfo {title} {Implementation of the
  maximum entropy method for analytic continuation},\ }\href
  {https://doi.org/10.1016/j.cpc.2017.01.018} {\bibfield  {journal} {\bibinfo
  {journal} {Computer Physics Communications}\ }\textbf {\bibinfo {volume}
  {215}},\ \bibinfo {pages} {149} (\bibinfo {year} {2017})}\BibitemShut
  {NoStop}%
\bibitem [{\citenamefont {Talirz}\ \emph {et~al.}(2020)\citenamefont {Talirz},
  \citenamefont {Kumbhar}, \citenamefont {Passaro}, \citenamefont {Yakutovich},
  \citenamefont {Granata}, \citenamefont {Gargiulo}, \citenamefont {Borelli},
  \citenamefont {Uhrin}, \citenamefont {Huber}, \citenamefont {Zoupanos},
  \citenamefont {Adorf}, \citenamefont {Andersen}, \citenamefont {Schütt},
  \citenamefont {Pignedoli}, \citenamefont {Passerone}, \citenamefont
  {VandeVondele}, \citenamefont {Schulthess}, \citenamefont {Smit},
  \citenamefont {Pizzi},\ and\ \citenamefont
  {Marzari}}]{talirz_materials_2020}%
  \BibitemOpen
  \bibfield  {author} {\bibinfo {author} {\bibfnamefont {L.}~\bibnamefont
  {Talirz}}, \bibinfo {author} {\bibfnamefont {S.}~\bibnamefont {Kumbhar}},
  \bibinfo {author} {\bibfnamefont {E.}~\bibnamefont {Passaro}}, \bibinfo
  {author} {\bibfnamefont {A.~V.}\ \bibnamefont {Yakutovich}}, \bibinfo
  {author} {\bibfnamefont {V.}~\bibnamefont {Granata}}, \bibinfo {author}
  {\bibfnamefont {F.}~\bibnamefont {Gargiulo}}, \bibinfo {author}
  {\bibfnamefont {M.}~\bibnamefont {Borelli}}, \bibinfo {author} {\bibfnamefont
  {M.}~\bibnamefont {Uhrin}}, \bibinfo {author} {\bibfnamefont {S.~P.}\
  \bibnamefont {Huber}}, \bibinfo {author} {\bibfnamefont {S.}~\bibnamefont
  {Zoupanos}}, \bibinfo {author} {\bibfnamefont {C.~S.}\ \bibnamefont {Adorf}},
  \bibinfo {author} {\bibfnamefont {C.~W.}\ \bibnamefont {Andersen}}, \bibinfo
  {author} {\bibfnamefont {O.}~\bibnamefont {Schütt}}, \bibinfo {author}
  {\bibfnamefont {C.~A.}\ \bibnamefont {Pignedoli}}, \bibinfo {author}
  {\bibfnamefont {D.}~\bibnamefont {Passerone}}, \bibinfo {author}
  {\bibfnamefont {J.}~\bibnamefont {VandeVondele}}, \bibinfo {author}
  {\bibfnamefont {T.~C.}\ \bibnamefont {Schulthess}}, \bibinfo {author}
  {\bibfnamefont {B.}~\bibnamefont {Smit}}, \bibinfo {author} {\bibfnamefont
  {G.}~\bibnamefont {Pizzi}},\ and\ \bibinfo {author} {\bibfnamefont
  {N.}~\bibnamefont {Marzari}},\ }\bibfield  {title} {\bibinfo {title}
  {Materials {{Cloud}}, a platform for open computational science},\ }\href
  {https://doi.org/10.1038/s41597-020-00637-5} {\bibfield  {journal} {\bibinfo
  {journal} {Scientific Data}\ }\textbf {\bibinfo {volume} {7}},\ \bibinfo
  {pages} {299} (\bibinfo {year} {2020})}\BibitemShut {NoStop}%
\bibitem [{\citenamefont {Nakamura}\ \emph {et~al.}(2021)\citenamefont
  {Nakamura}, \citenamefont {Yoshimoto}, \citenamefont {Nomura}, \citenamefont
  {Tadano}, \citenamefont {Kawamura}, \citenamefont {Kosugi}, \citenamefont
  {Yoshimi}, \citenamefont {Misawa},\ and\ \citenamefont
  {Motoyama}}]{nakamura_respack_2021}%
  \BibitemOpen
  \bibfield  {author} {\bibinfo {author} {\bibfnamefont {K.}~\bibnamefont
  {Nakamura}}, \bibinfo {author} {\bibfnamefont {Y.}~\bibnamefont {Yoshimoto}},
  \bibinfo {author} {\bibfnamefont {Y.}~\bibnamefont {Nomura}}, \bibinfo
  {author} {\bibfnamefont {T.}~\bibnamefont {Tadano}}, \bibinfo {author}
  {\bibfnamefont {M.}~\bibnamefont {Kawamura}}, \bibinfo {author}
  {\bibfnamefont {T.}~\bibnamefont {Kosugi}}, \bibinfo {author} {\bibfnamefont
  {K.}~\bibnamefont {Yoshimi}}, \bibinfo {author} {\bibfnamefont
  {T.}~\bibnamefont {Misawa}},\ and\ \bibinfo {author} {\bibfnamefont
  {Y.}~\bibnamefont {Motoyama}},\ }\bibfield  {title} {\bibinfo {title}
  {{{RESPACK}}: {{An}} {\emph{ab initio}} tool for derivation of effective
  low-energy model of material},\ }\href
  {https://doi.org/10.1016/j.cpc.2020.107781} {\bibfield  {journal} {\bibinfo
  {journal} {Computer Physics Communications}\ }\textbf {\bibinfo {volume}
  {261}},\ \bibinfo {pages} {107781} (\bibinfo {year} {2021})}\BibitemShut
  {NoStop}%
\bibitem [{\citenamefont {{van Setten}}\ \emph {et~al.}(2018)\citenamefont
  {{van Setten}}, \citenamefont {Giantomassi}, \citenamefont {Bousquet},
  \citenamefont {Verstraete}, \citenamefont {Hamann}, \citenamefont {Gonze},\
  and\ \citenamefont {Rignanese}}]{vansetten_pseudodojo_2018}%
  \BibitemOpen
  \bibfield  {author} {\bibinfo {author} {\bibfnamefont {M.~J.}\ \bibnamefont
  {{van Setten}}}, \bibinfo {author} {\bibfnamefont {M.}~\bibnamefont
  {Giantomassi}}, \bibinfo {author} {\bibfnamefont {E.}~\bibnamefont
  {Bousquet}}, \bibinfo {author} {\bibfnamefont {M.~J.}\ \bibnamefont
  {Verstraete}}, \bibinfo {author} {\bibfnamefont {D.~R.}\ \bibnamefont
  {Hamann}}, \bibinfo {author} {\bibfnamefont {X.}~\bibnamefont {Gonze}},\ and\
  \bibinfo {author} {\bibfnamefont {G.~M.}\ \bibnamefont {Rignanese}},\
  }\bibfield  {title} {\bibinfo {title} {The {{PseudoDojo}}: {{Training}} and
  grading a 85 element optimized norm-conserving pseudopotential table},\
  }\href {https://doi.org/10.1016/j.cpc.2018.01.012} {\bibfield  {journal}
  {\bibinfo  {journal} {Computer Physics Communications}\ }\textbf {\bibinfo
  {volume} {226}},\ \bibinfo {pages} {39} (\bibinfo {year} {2018})}\BibitemShut
  {NoStop}%
\bibitem [{\citenamefont {Mizokawa}\ \emph {et~al.}(2000)\citenamefont
  {Mizokawa}, \citenamefont {Khomskii},\ and\ \citenamefont
  {Sawatzky}}]{mizokawa_spin_2000}%
  \BibitemOpen
  \bibfield  {author} {\bibinfo {author} {\bibfnamefont {T.}~\bibnamefont
  {Mizokawa}}, \bibinfo {author} {\bibfnamefont {D.~I.}\ \bibnamefont
  {Khomskii}},\ and\ \bibinfo {author} {\bibfnamefont {G.~A.}\ \bibnamefont
  {Sawatzky}},\ }\bibfield  {title} {\bibinfo {title} {Spin and charge ordering
  in self-doped {{Mott}} insulators},\ }\href
  {https://doi.org/10.1103/PhysRevB.61.11263} {\bibfield  {journal} {\bibinfo
  {journal} {Physical Review B}\ }\textbf {\bibinfo {volume} {61}},\ \bibinfo
  {pages} {11263} (\bibinfo {year} {2000})}\BibitemShut {NoStop}%
\bibitem [{\citenamefont {Johnston}\ \emph {et~al.}(2014)\citenamefont
  {Johnston}, \citenamefont {Mukherjee}, \citenamefont {Elfimov}, \citenamefont
  {Berciu},\ and\ \citenamefont {Sawatzky}}]{johnston_charge_2014}%
  \BibitemOpen
  \bibfield  {author} {\bibinfo {author} {\bibfnamefont {S.}~\bibnamefont
  {Johnston}}, \bibinfo {author} {\bibfnamefont {A.}~\bibnamefont {Mukherjee}},
  \bibinfo {author} {\bibfnamefont {I.}~\bibnamefont {Elfimov}}, \bibinfo
  {author} {\bibfnamefont {M.}~\bibnamefont {Berciu}},\ and\ \bibinfo {author}
  {\bibfnamefont {G.~A.}\ \bibnamefont {Sawatzky}},\ }\bibfield  {title}
  {\bibinfo {title} {Charge {{Disproportionation}} without {{Charge Transfer}}
  in the {{Rare-Earth-Element Nickelates}} as a {{Possible Mechanism}} for the
  {{Metal-Insulator Transition}}},\ }\href
  {https://doi.org/10.1103/PhysRevLett.112.106404} {\bibfield  {journal}
  {\bibinfo  {journal} {Physical Review Letters}\ }\textbf {\bibinfo {volume}
  {112}},\ \bibinfo {pages} {106404} (\bibinfo {year} {2014})}\BibitemShut
  {NoStop}%
\end{thebibliography}%
\end{document}